\documentclass[acmsmall, authorversion, nonacm]{acmart}
\usepackage{booktabs}
\usepackage{amsmath} 
\usepackage{amsthm}
\usepackage{enumerate}
\usepackage{enumitem}
\usepackage{hyperref}
\usepackage{subcaption} 

\usepackage{geometry}
\usepackage{stmaryrd}
\usepackage{thmtools}
\usepackage{bm}
\usepackage{footmisc}
\usepackage{fainekos-macros}
\usepackage{rod-macros}
\usepackage[toc,page]{appendix}
\usepackage{multirow}
\usepackage[ruled,vlined]{algorithm2e}
\DeclareMathOperator{\sign}{sign}
\AtBeginDocument{%
  \providecommand\BibTeX{{%
    \normalfont B\kern-0.5em{\scshape i\kern-0.25em b}\kern-0.8em\TeX}}}

\setcopyright{acmcopyright}
\copyrightyear{2022}
\acmYear{2022}
\acmDOI{XXXXXXX.XXXXXXX}

\acmJournal{TECS}
\acmVolume{!!!!!}
\acmNumber{!!!!}
\acmArticle{}
\acmMonth{!!!}

\begin{document}

%%
%% The "title" command has an optional parameter,
%% allowing the author to define a "short title" to be used in page headers.
\title{{Temporal Robustness of Temporal Logic Specifications: Analysis and Control Design}}

%%
%% The "author" command and its associated commands are used to define
%% the authors and their affiliations.
%% Of note is the shared affiliation of the first two authors, and the
%% "authornote" and "authornotemark" commands
%% used to denote shared contribution to the research.
\author{Al\"ena Rodionova}
%\authornote{Both authors contributed equally to this research.}
\email{alena.rodionova@seas.upenn.edu}
\orcid{0000-0001-8455-9917}
\author{Lars Lindemann}
%\authornotemark[1]
\email{larsl@seas.upenn.edu}
\author{Manfred Morari}
\email{morari@seas.upenn.edu}
\author{George J. Pappas}
\email{pappasg@seas.upenn.edu}
\affiliation{%
  \institution{\\University of Pennsylvania}
  \department{Department of Electrical and Systems Engineering}
  \streetaddress{200 South 33rd Street}
  \city{Philadelphia}
  \state{Pennsylvania}
  \country{USA}
  \postcode{19104}
}

%%
%% By default, the full list of authors will be used in the page
%% headers. Often, this list is too long, and will overlap
%% other information printed in the page headers. This command allows
%% the author to define a more concise list
%% of authors' names for this purpose.
%\renewcommand{\shortauthors}{Rodionova, et al.}

%%
%% The abstract is a short summary of the work to be presented in the
%% article.
\begin{abstract}
  We study the temporal robustness of temporal logic specifications and show how to design temporally robust control laws for time-critical control systems. This topic is of particular interest in connected systems and interleaving processes such as multi-robot and human-robot systems where uncertainty in the behavior of individual agents and humans can induce timing uncertainty.  Despite the importance of time-critical systems, temporal robustness of temporal logic specifications has not been studied, especially from a control design point of view. We define synchronous and asynchronous temporal robustness and show that these notions quantify the robustness with respect to synchronous and asynchronous time shifts in the predicates of the temporal logic specification.  It is further shown that the synchronous temporal robustness upper bounds the asynchronous temporal robustness. We then study the control design problem in which we aim  to  design a control law that maximizes the temporal robustness of a dynamical system. Our solution consists of a Mixed-Integer Linear Programming (MILP) encoding that can be used to obtain a sequence of optimal control inputs. While asynchronous temporal robustness is arguably more nuanced than synchronous temporal robustness, we show that control design using synchronous temporal robustness is computationally more efficient. This trade-off can be exploited by the designer depending on the particular application at hand. We conclude the paper with a variety of case studies.
\end{abstract}

%%
%% The code below is generated by the tool at http://dl.acm.org/ccs.cfm.
%% Please copy and paste the code instead of the example below.
%%


%%
%% Keywords. The author(s) should pick words that accurately describe
%% the work being presented. Separate the keywords with commas.

%%
%% This command processes the author and affiliation and title
%% information and builds the first part of the formatted document.
\maketitle

\section{Introduction}

\emph{Time-critical systems} are systems that need to satisfy stringent \emph{real-time constraints}. Examples of time-critical systems include, but are not limited to, medical devices, autonomous driving, automated warehouses, and air traffic control systems. The reliability and safety of such time-critical systems greatly depends on being robust with respect to timing uncertainty. For instance,  Germany's rail network has become prone to frequent train delays which makes scheduling and planning a difficult task \cite{connolly2018we}. It is hence natural to study \emph{temporal robustness} of time-critical systems, i.e.,  to study a system's ability to be robust with respect to timing uncertainty.

\emph{Temporal logics} provide a principled mechanism to express a broad range of real-time constraints that can be imposed on time-critical systems \cite{kress2009temporal,kloetzer2008fully,guo2015multi,kantaros2018sampling}. In fact, there exists a variety of temporal logics, all with their own merits, that have been explored from an analysis and control design point of view. However, the temporal robustness of systems under temporal logic specifications has not yet been studied and used for control design. An initial attempt to define temporal robustness was made in \cite{Donze10STLRob} without, however, further analyzing properties of temporal robustness and without using it for control design. In fact, a natural objective is to design a  system to be robust to previously mentioned timing uncertainties.  This refers to the control design problem in which we aim to design a control law that maximizes the temporal robustness of a system. The control design problem has not been addressed before and is subject to  numerical challenges as the temporal robustness semantics are not differentiable, so that gradient-based optimization methods fail to be applicable. To address this challenge, we presented a Mixed-Integer Linear Programming (MILP) encoding in our previous work  \cite{rodionova2021time} where we describe how to control a dynamical system in order to maximize its temporal robustness.

In this paper, we introduce and analyze the notions of synchronous and asynchronous temporal robustness. While the temporal robustness as presented in \cite{Donze10STLRob} is similar to what we define as asynchronous temporal robustness, we provide a slightly modified definition and a detailed analysis of the synchronous and asynchronous temporal robustness. In particular, we show that these notions are sound in the sense that a positive (negative) robustness implies satisfaction (violation) of the specification in hand. Furthermore, we analyze the meaning of temporal robustness in terms of permissible time shifts in the predicates of the specification and in the underlying signal. For both notions we present MILP encodings, following ideas of our initial work in \cite{rodionova2021time}, that solves the control design problem in which we aim to maximize the synchronous and asynchronous temporal robustness.

\subsection{Related Work}

Time-critical systems have been studied in the literature dealing with real-time systems \cite{laplante2004real,Jane_liu}. The design of task scheduling algorithms for real-time systems has been  extensively studied, see e.g.,  \cite{sha2004real} for an overview. Scheduling algorithms aim at finding an execution order for a set of tasks with corresponding deadlines. Solutions consist, for instance, of periodic scheduling  \cite{serafini1989mathematical} and of event triggered scheduling \cite{tabuada2007event}. While these scheduling algorithms are useful in control design, no consideration has been given to the  temporal robustness of the system. Hence we view these works  as complementary.

Another direction in the study of real-time systems are  timed automata as a modeling formalism \cite{alur1994theory,bengtsson2003timed}. Timed automata allow automatic verification using model checking tools such as UPPAAL \cite{behrmann2004tutorial}.  Robustness of timed automata was investigated in \cite{gupta1997robust} and \cite{bendik2021timed}, while control of timed automata was considered in  \cite{maler1995synthesis} and \cite{asarin1998controller}. A connection between the aforementioned scheduling and timed automata was made in \cite{fersman2006schedulability}. 

Interestingly, timed automata have been connected to temporal logics. Particularly, it has been shown that every Metric Interval Temporal Logic (MITL) formula can be translated into a language equivalent timed automaton \cite{alur1996benefits}. This means that the satisfiability of an MITL formula can be reduced to an emptiness checking problem of a timed automaton. Signal Temporal Logic (STL) \cite{maler2004monitoring} is in spirit similar to MITL, but additionally allows to consider predicates instead of only propositions. It has been shown that STL formulas can be translated into a language equivalent timed signal transducer in \cite{lindemann2019efficient}. 

Various notions of robustness have been presented for temporal logics. Spatial robustness of MITL specifications over deterministic signals was considered in \cite{Donze10STLRob,fainekos2009robustness}. Spatial robustness particularly allows to quantify permissible uncertainty of the signal for each point in time, e.g., caused by additive disturbances. Other notions of spatial robustness are the arithmetic-geometric integral mean robustness~\cite{mehdipour2019average} and the smooth cumulative robustness \cite{haghighi2019control} as well as notions that are tailored for use in reinforcement learning applications  \cite{varnai2020robustness}. Spatial robustness for stochastic systems was considered in \cite{bartocci2013robustness,bartocci2015system} as well as in \cite{lindemann2021stl} when considering the risk of violating a specification. Control of dynamical systems  under  STL specifications has first been considered in \cite{raman2014model} by means of an MILP encoding that allows to maximize spatial robustness. Other optimization-based methods that also follow the idea of maximizing spatial robustness have been proposed in \cite{mehdipour2019average,pant2018fly,gilpin2020smooth}. Another direction has been to design transient feedback control laws that maximize the spatial robustness of fragments of STL specification \cite{lindemann2018control} and \cite{charitidou2021barrier}. 

Analysis and control design for STL specifications focuses mainly on spatial robustness, while less attention has been  on temporal robustness of real-time systems. The authors in \cite{akazaki2015time} proposed averaged STL that captures a form of temporal robustness by averaging over time intervals. Other forms of temporal robustness include the notion of system conformance   to quantify closeness of systems in terms of spatial and temporal closeness of system trajectories \cite{abbas2014formal,gazda2020logical,deshmukh2015quantifying}. Conformance, however, only allows to reason about temporal robustness with respect to synchronous time shifts of a signal and not with respect to synchronous and asynchronous time shifts  as considered in this work. As remarked before, temporal robustness for STL specifications was initially presented without further analysis in \cite{Donze10STLRob}. Temporally-robust control has been considered in \cite{baras_runtime} for special fragments of STL, and in~\cite{rodionova2021time} for general fragments of STL using MILP encodings. In an orthogonal direction that is worth mentioning, the authors in \cite{penedo2020language} and \cite{kamale2021automata} consider the problem of finding temporal relaxations for time window temporal logic specifications \cite{vasile2017time}.

\subsection{Contributions and Paper Outline}

	When dealing with time-critical system, a natural objective is to analyze robustness to various forms of timing uncertainties and to design a control system to maximize this robustness.
We consider the temporal robust interpretation of STL formulas over continuous and discrete-time signals. 
Our goal is to establish a comprehensive theoretical framework for  temporal robustness of STL specifications which has not been studied before, especially from a control design point of view. We make the following contributions.
\begin{itemize}
	\item  We define synchronous and asynchronous temporal robustness to  quantify the robustness with respect to synchronous and asynchronous time shifts in the predicates of the underlying signal temporal logic specification.
	\item We present various desirable properties of the presented robustness notions and show that the synchronous temporal robustness upper bounds the asynchronous temporal robustness in its absolute value. Moreover, we show under which conditions these two robustness notions are equivalent.
	\item We then study the control design problem in which we aim to design a control law that maximizes the temporal robustness of a dynamical system. To solve this problem, we present Mixed-Integer Linear Programming (MILP) encodings following ideas of our initial work in \cite{rodionova2021time}. We provide correctness guarantees and a complexity analysis of the encodings. 
	\item All theoretical results are highlighted in three case studies. Particularly, we show how the proposed MILP encodings can be used to perform  temporally-robust control of multi-agent systems.
\end{itemize}

The remainder of the paper is organized as follows.
Section \ref{sec:prelim} provides background on STL. In Section~\ref{sec:timerob}, the synchronous and asynchronous temporal robustness is introduced. Soundness properties of temporal robustness are presented in Section~\ref{sec:prop_sound} while the properties with respect to time shifts are presented in Section~\ref{sec:robustness_shifts_time}. In Section~\ref{sec:control} we present MILP encodings to solve the temporally-robust control problem.
Extensive simulations and case studies are presented in Section~\ref{sec:experiments}. Finally, we summarize with conclusions in Section~\ref{sec:conclusions}.

\section{Background on Signal Temporal Logic (STL)}
\label{sec:prelim}

Let $\Re$ and $\mathbb{Z}$ be the set of real numbers and integers, respectively. Let $\Re^n$ be the $n$-dimensional real vector space. The supremum operator is written as $\sqcup$ and the infimum operator is written as $\sqcap$. We define the set $\Be\defeq\{\top,\bot\}$, where $\top$ and $\bot$ are the Boolean constants \textit{true} and \textit{false}, respectively.
We also define the sign function as 
\begin{equation*}
	\sign(x)\defeq \begin{cases}
		\po,&\text{if }x\geq0\\
		-1 &\text{if }x<0.
	\end{cases}
\end{equation*}

%We use $t+I = \{t+\tau\,|\, \tau\in I\}$ and $t +_{\TDom}I = (t + I) \cap \TDom$ to define operations on the timing constraints $I$ of the temporal operators for any $t\in\TDom$.
%We use use the following notation to define operations on the timing constraints $I$ of the temporal operators for any $t\in\TDom$:
%\begin{align}
%	&t+I = \{t+\tau\,|\, \tau\in I\} %&&\text{and}&t +_{\TDom}I = (t + I) \cap \TDom 
%\end{align}

In general, we say that a signal $\sstraj$ is a map $\sstraj:\TDom \to X$ where $\TDom$ is a time domain and where the state space $X$ is a metric space. 
In this work, we particularly consider the cases of real-valued continuous-time signals which naturally includes the case of discrete-time signals. In other words,  the state space is $X\subseteq\Re^n$ and the time domain is either $\Te \defeq \Re$ or $\Te \defeq \mathbb{Z}$.\footnote{For convenience, we assume  that the time domain $\TDom$  is unbounded in both directions. This assumption is made without loss
of generality and in order to avoid technicalities.}
For any set $\Te$ we denote by $\overline{\Te} = \Te\cup\{\pm \infty\}$, for instance,
$\CoRe\defeq\Re\cup \{\pm \infty\}$ and $\CoZe\defeq\Ze\cup \{\pm \infty\}$ are the extended real numbers and integers, respectively. Finally, we  denote the set of all signals $\sstraj: \TDom \rightarrow X$ as the \textit{signal space} $\SigSpace$.
%When we consider the bounded-time trajectory, $\TDom=[0, H]\subset \Re_{\geq 0}$ for some time horizon $H\in\Re$. In case of the unbounded-time trajectories $\TDom=\Re$. 

\subsection{Signal Temporal Logic}
For a signal $\sstraj:\TDom \to X$, let the signal state at time step $t$ be $x_t\in X$. 
Let $\mu:X\to\mathbb{R}$ be a real-valued function and let $p:X\to \mathbb{B}$ be a \textit{predicate} defined via $\mu$ as $p(x) \defeq \mu(x) \geq 0$. 
Thus, a predicate $\mu$ defines a set in which $p$ holds true, namely $p$ defines the set $\{x\in X\such \mu(x)\geq 0\}$ in which $p$ is true.
Let $AP \defeq \{p_1,\ldots,p_L\}$ be a set of $L$ predicates defined via the set of predicate functions $M\defeq\{\mu_1,\ldots,\mu_L\}$. Let $I$ be a time interval of the form $[a,b]$, $[a,b)$, $(a,b]$ or $(a,b)$ where $a,b\in\TDom$, $0 \leq a \leq b$. 
For any $t\in\TDom$ and a time interval $I\subseteq\TDom$, we define the set
$t+I \defeq \{t+\tau\,|\, \tau\in I\}$.

%For $t\in\TDom$, the time interval $[t+a, t+b]$ is denoted as $t+I$.
%We use $t+I = \{t+\tau\,|\, \tau\in I\}$ and $t +_{\TDom}I = (t + I) \cap \TDom$ to define 
%operations on the timing constraints $I$ of the temporal operators for any $t\in\TDom$.
%We use use the following notation to define operations on the timing constraints $I$ of the temporal operators for any $t\in\TDom$:
%\begin{align}
%	&t+I = \{t+\tau\,|\, \tau\in I\} &&\text{and}&t +_{\TDom}I = (t + I) \cap \TDom 
%\end{align}

The syntax of Signal Temporal Logic (STL) is defined as follows~\cite{MalerN2004STL}:
\begin{equation*}
	\formula :\defeq \top\ |\ p\ |\ \neg \formula \ |\ \formula_1 \land \formula_2 \ |\ \formula_1 \until_I \formula_2
\end{equation*}
where 
$p\in AP$ is a predicate, $\neg$ and $\land$ are the Boolean negation and conjunction, respectively, and $\until_I$ is the Until temporal operator over an interval $I$. 
The disjunction ($\lor$) and implication ($\implies$) are defined as usual. 
Additionally, the temporal operators Eventually ($\eventually$) and Always ($\always$) can be defined as $\eventually_I\varphi \defeq \top\until_I\varphi$ and $\always_I\varphi \defeq \neg\eventually_I\neg\varphi$.
%Informally, $\formula_1 \until_I \formula_2$ means that $\formula_2$ must hold at some point in $I$, and \textit{until} then, $\formula_1$ must hold without interruption.

Formally, the \textit{semantics} of an STL formula $\formula$ define when a signal $\sstraj$ satisfies  $\formula$ at time point $t$.
	%, denoted as $(\sstraj,t)\models \varphi$
%	When satisfaction does not hold it is denoted as $(\sstraj,t)\not\models \varphi$.
We use the characteristic function $\chi_\varphi(\sstraj,t)$, as defined in the sequel, to indicate when a formula is satisfied.
\begin{definition}[Characteristic function \cite{Donze10STLRob}]
	\label{def:char_func}
	The characteristic function $\chi_{\varphi}(\sstraj,t): \SigSpace \times \TDom\to \{\pm 1\}$ 
	of an  STL formula $\varphi$ relative to
	a signal $\sstraj$ at time $t$ is defined recursively as:
	\begin{equation}
		\label{eq:boolean_sat_char}
		\begin{aligned}
%			\chi_\top(\sstraj,t) &=1
%			\\
			\chi_p(\sstraj, t) &\defeq 
			\sign(\mu(x_t))
			%p(t)
			\\
			\chi_{\neg \formula}(\sstraj,t) &\defeq -\chi_\formula(\sstraj,t)
			\\
			\chi_{\formula_1 \land \formula_2}(\sstraj,t) &\defeq \chi_{\formula_1}(\sstraj,t) \ \sqcap\ \chi_{\formula_2}(\sstraj,t)
			\\
			\chi_{\formula_1 \until_I \formula_2}(\sstraj,t) &\defeq 
			\bigsqcup_{t'\in t+ I} \left(
			\chi_{\formula_2}(\sstraj,t') \ \sqcap \
			\bigsqcap_{t'' \in [t,t')} \chi_{\formula_1}(\sstraj,t'')
			\right)
		\end{aligned}
		%	\vspace{5pt}
	\end{equation}
\end{definition}

%\begin{definition}
%	For a given predicate $p\in AP$ and a system trajectory $\sstraj\in\SigSpace$, we 
%	call a sequence $\bm{\chi}_{p}(\sstraj)\in\{\pm 1\}^\TDom$ a \textbf{predicate trajectory}.
%%	Then $\mathbf{p}(\mathbf{x})=
%%	\begin{bmatrix}
%%		\mathbf{p}_1(\mathbf{x})\\
%%		\vdots\\
%%		\mathbf{p}_L(\mathbf{x})
%%	\end{bmatrix}\in \Be^{L \times (H+1)}$, where $\Be=\{\pm 1\}$.
%\end{definition}

When $\chi_\varphi(\sstraj,t)=1$, it holds that the signal $\sstraj$ satisfies the formula  $\formula$ at time $t$, while $\chi_\varphi(\sstraj,t)=-1$ indicates  that  $\sstraj$ does not satisfy  $\formula$ at time $t$.
%Trajectory $\sstraj$ satisfies $\varphi$ if $(\sstraj,0)\models\varphi$, we denote it as $\sstraj\models\varphi$.

While these semantics show \textit{whether or not} a signal
$\sstraj$
satisfies a given specification $\varphi$ at time $t$, there have been various notions of robust STL semantics. These robust semantics measure \textit{how robustly} the signal $\sstraj$ satisfies  the formula $\formula$ at time $t$.
One widely used notion analyzes how much spatial perturbation a signal can tolerate without changing the satisfaction of the formula. Such \emph{spatial robustness} was presented in \cite{FainekosP09tcs} and indicates the robustness of satisfaction  with respect to point-wise changes in the value of the signal $\sstraj$ at time $t$, i.e., changes in $x_t$.
However, in this paper we are interested in an orthogonal direction by considering temporal robustness. 
For a lot of systems, one should not only robustly satisfy the spatial requirements but also robustly satisfy the temporal requirements. In this paper, we hence focus on timing perturbations, which we define in terms of the time shifts in the predicates $p_k$ of the formula $\formula$.
%\textcolor{blue}{However, in this paper we focus on robustness with respect to time shifts in the predicates of the formula $\formula$. A notion of temporal robustness, which we refer to as \emph{asynchronous temporal robustness}, was presented in \cite{Donze10STLRob}. }

\section{Temporal Robustness}
\label{sec:timerob}

In this section, we introduce synchronous and asynchronous temporal robustness to measure how robustly a signal $\sstraj$ satisfies a formula $\formula$ at time $t$ with respect to time shifts. In particular, these notions quantify the robustness with respect to synchronous and asynchronous time shifts in the predicates of the formula $\formula$. While one can argue that the asynchronous temporal robustness is a more general notion of temporal robustness, we show that the synchronous temporal robustness is  easier to calculate, which is a useful property as it induces less computational complexity during control design. We also show that the synchronous temporal robustness is an upper bound to the asynchronous temporal robustness. We remark that the idea of asynchronous temporal robustness was initially presented in  \cite{Donze10STLRob}. We provide a slightly modified definition and  complement the definition with formal guarantees  in Sections \ref{sec:prop_sound} and \ref{sec:robustness_shifts_time} and show how it can be used for control design in Section \ref{sec:control}. 

\subsection{Synchronous Temporal Robustness}
\label{sec:time_rob_sync}

\begin{figure}[t!]
	\centering
	\includegraphics[width=\textwidth]{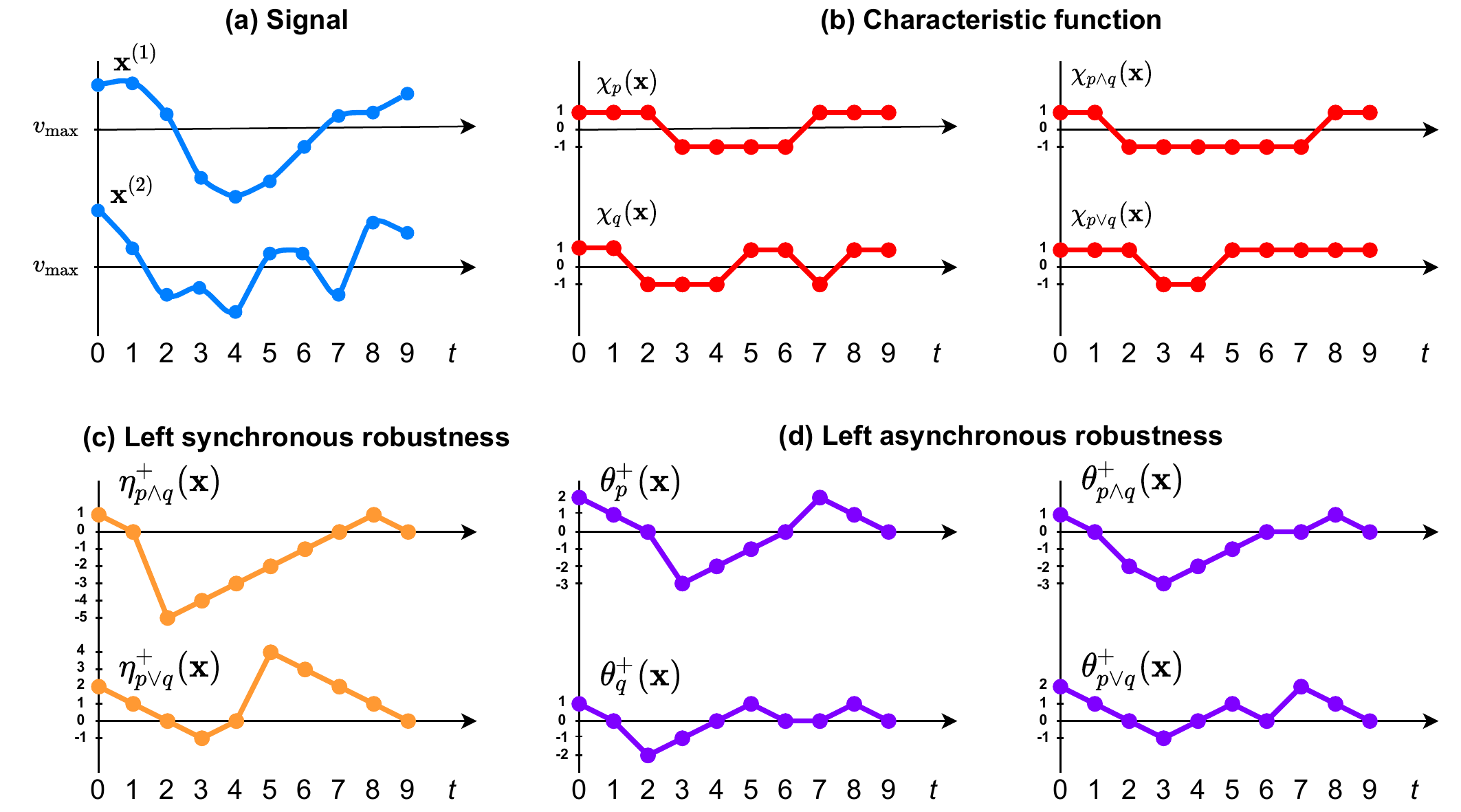}
	\caption{Evolution of the (a) signal, (b) characteristic function, (c) \robplus{} synchronous temporal robustness and (d) \robplus{} asynchronous temporal robustness from Example~\ref{ex:running}.}
	\label{fig:ex1}
\end{figure}

We first define the notion of \robplus{} and \robminus{} synchronous temporal robustness. The idea behind synchronous temporal robustness is to quantify the maximal amount of time by which we can shift the characteristic function $ \chi_\varphi(\sstraj, t)$ of $\formula$  to the \robplus{} and \robminus{}, respectively, without changing the value of $ \chi_\varphi(\sstraj, t)$.

\begin{definition}[Synchronous Temporal Robustness]
	\label{def:time_rob_all}
	The \robplus{} and \robminus{} synchronous temporal robustness $\eta^{\pm}_\varphi(\sstraj,\! t):\SigSpace\! \times\! \TDom\! \to \!\CoTe$ of an STL formula $\varphi$ with respect to a signal $\sstraj$ at time $t$ are defined as\footnote{When $\TDom=\mathbb{Z}$ we implicitly assume that the intervals $[t,t+\tau]$ and $[t-\tau,t]$ encode $[t,t+\tau]\cap \Ze$ and $[t-\tau,t]\cap \Ze$, which results in discrete intervals.}:
	
	\begin{align}
		\label{eq:time_rob_all_p}
		\etap_\varphi(\sstraj,t) &\defeq \chi_\varphi(\sstraj, t)\cdot
		\sup\{\tau\geq0 \ :\ \forall t'\in[t,t+\tau],\  \chi_\varphi(\sstraj,t')=\chi_\varphi(\sstraj,t)\}\\
		\label{eq:time_rob_all_m}
		\etam_\varphi(\sstraj,t) &\defeq \chi_\varphi(\sstraj, t)\cdot
		\sup\{\tau\geq0 \ :\ \forall t'\in[t-\tau,t],\ \chi_\varphi(\sstraj,t')=\chi_\varphi(\sstraj,t)\}.
	\end{align}
\end{definition}
For brevity, we often use the combined notation as follows:
\begin{equation*}
\eta^{\pm}_\varphi(\sstraj,t)\defeq \chi_\varphi(\sstraj, t)\cdot
\sup\{\tau\geq0 \ :\ \forall t'\in t\pm[0,\tau],\ \chi_\varphi(\sstraj,t')=\chi_\varphi(\sstraj,t)\}.
\end{equation*}

\begin{exmp}%[Running example]
		\label{ex:running}
		Take a look at  Fig.~\ref{fig:ex1}. 
		The signal $\sstraj\defeq[\sstraj^{(1)},\, \sstraj^{(2)}]$ presented in Fig.\ref{fig:ex1}(a) is finite and discrete-time, its state at each time step $t=0,\ldots,9$ is $x_t\defeq [x_t^{(1)},\,x_t^{(2)}]\in\Re^2$.
		Let the two predicates be $p\defeq x^{(1)}_t\geq v_{\max}$ and $q\defeq x^{(2)}_t\geq v_{\max}$ for some given $v_{\max}$. The characteristic functions for the predicates $p$ and $q$ and the STL formulas $\varphi_1\defeq p\wedge q$ and $\varphi_2\defeq p\vee q$ 
		are shown in Fig.~\ref{fig:ex1}(b). The evolution of the \robplus{} synchronous temporal robustness $\etap_\varphi(\sstraj,t)$, $t=0,\ldots,9$ for the STL formulas $\varphi_1$ and $\varphi_2$ is presented in Fig.~\ref{fig:ex1}(c). 
		For instance, consider the time point $t=0$ and the formula $\varphi_2$. Since $\chi_{\varphi_2}(\sstraj,0)=\chi_{\varphi_2}(\sstraj,1)=\chi_{\varphi_2}(\sstraj,2)=\po$ and $\chi_{\varphi_2}(\sstraj,3)=-1$, then by Definition~\ref{def:time_rob_all} it holds that $\etap_{\varphi_2}(\sstraj,0)=2$. On the other hand, for $\varphi_1$ and $t=6$, we have that $\chi_{\varphi_1}(\sstraj,6)=\chi_{\varphi_1}(\sstraj,7)=-1\not = \po = \chi_{\varphi_1}(\sstraj,8)$ so that $\etap_{\varphi_1}(\sstraj,6)=-1$. 
	\end{exmp}

Some remarks are in place. Let $\mathbf{s}$ be a signal that is equivalent to $\sstraj$ but shifted by $\tau$ time units to the left, i.e., $s_t\defeq x_{t+\tau}$ for all $t\in \TDom$. Then it holds that $\chi_\varphi(\mathbf{s}, t)=\chi_\varphi(\sstraj, t+\tau)$.
Similarly, $\chi_\varphi(\mathbf{s}, t)=\chi_\varphi(\sstraj, t-\tau)$  where $\mathbf{s}$ is equivalent to $\sstraj$ shifted by $\tau$ to the right, i.e., $s_t\defeq x_{t-\tau}$ for all $t\in \TDom$.\footnote{To see this, note that  the characteristic function $\chi_\varphi(\mathbf{s}, t)$  is built recursively from the characteristic function ${\chi}_{p}(\mathbf{s},t)$ of each predicate $p\in AP$ in $\varphi$, see Definition~\ref{def:char_func}. For the shifted signal $\mathbf{s}$, each ${\chi}_{p}(\mathbf{s},t)$  is simultaneously shifted to the left or right compared to $\chi_\varphi(\sstraj, t)$.
%Note also that predicates evaluated over $\mathbf{s}$ and $\sstraj$, i.e., $\mu(s_t)$ and $\mu(x_t)$, are then only  shifted versions of another. 
Consequently, the shifted version $\chi_\varphi(\mathbf{s}, t)$  can be understood as $\chi_\varphi(\sstraj, t\pm\tau)$  as $\mathbf{s}$ shifts \emph{all} predicates $p\in AP$ that appear in $\formula$  \emph{synchronously} by $\tau$.} This indicates, and is later formally shown in Section \ref{sec:robustness_shifts_time},  that the \robplus{} and \robminus{} synchronous temporal robustness quantify the maximal amount of time by which we can synchronously shift the signal $\sstraj$ (or alternatively each predicate in $\formula$) to the left and right, respectively, without changing the value of $ \chi_\varphi(\sstraj, t)$. 

\subsection{Asynchronous Temporal Robustness}
\label{sec:time_rob_async}

While the synchronous temporal robustness quantifies the amount by which we can shift all predicates synchronously in time, the \robplus{} and \robminus{} asynchronous temporal robustness, which is inspired by \cite{Donze10STLRob} and defined next, quantifies the maximal amount of time by which we can \textit{asynchronously} shift \emph{individual} predicates in $\formula$ to the left and right, respectively. 

\begin{definition}[Asynchronous Temporal Robustness (inspired by \cite{Donze10STLRob})]
	%\label{def:time_rob}
	\label{def:time_rob_rec}
	The \robplus{} and \robminus{} asynchronous temporal robustness $\theta^{\pm}_\varphi(\sstraj,t) :\SigSpace \times \TDom\to \CoTe$ of an STL formula $\varphi$ with respect to a signal $\sstraj$ at time $t$ are defined recursively as:
%	Furthermore, we fixed the typos that were made in \cite{Donze10STLRob} related to the inductive definition of $\theta^{\pm}_\varphi$.
	\begin{align}
		\label{eq:rec_pp}
		\thetap_p(\sstraj,t)\defeq \chi_p(\sstraj, t)\cdot
		\sup\{\tau\geq0 \ :\ \forall t'\in[t,t+\tau],\ \chi_p(\sstraj,t')=\chi_p(\sstraj,t)\} 
		%= \chi_p(\sstraj, t)\cdot \tau^+(\sstraj,t)
		\\
		\label{eq:rec_pm}
		\thetam_p(\sstraj,t) \defeq \chi_p(\sstraj, t)\cdot
		\sup\{\tau\geq0\ :\ \forall t'\in[t-\tau,t],\ \chi_p(\sstraj,t')=\chi_p(\sstraj,t)\} 
		%=  \chi_p(\sstraj, t)\cdot \tau^-(\sstraj,t)
	\end{align}
	and then applying to each $\theta^{\pm}_p(\sstraj,t)$ the
	recursive rules of the operators similarly to Definition.~\ref{def:char_func}, which leads to:
	\begin{align}
%		{\color{blue} \theta^{\pm}_\top(\sstraj, t)=+\infty}
%		\\
		\label{eq:t_neg}
		\theta^{\pm}_{\neg \formula}(\sstraj,t) &\defeq -\theta^{\pm}_\formula(\sstraj,t)
		\\
		\label{eq:t_and}
		\theta^{\pm}_{\formula_1 \land \formula_2}(\sstraj,t) &\defeq\theta^{\pm}_{\formula_1}(\sstraj,t) \ \sqcap\ \theta^{\pm}_{\formula_2}(\sstraj,t)
		\\
		\label{eq:t_until}
		\theta^{\pm}_{\formula_1 \until_I \formula_2}(\sstraj,t) &\defeq 
		\bigsqcup_{t'\in t+I} \left(
		\theta^{\pm}_{\formula_2}(\sstraj,t') \ \sqcap \
		\bigsqcap_{t'' \in [t,t')} \theta^{\pm}_{\formula_1}(\sstraj,t'')
		\right)
	\end{align}
	%	(It is natural to define $\thetam_\top(\sstraj,t) =\thetap_\top(\sstraj,t) = +\infty$)
\end{definition}

\begin{rem}
There are two subtle difference between Definition \ref{def:time_rob_rec} and temporal robustness in  \cite[Definition 4]{Donze10STLRob}. First, we use the supremum operator instead of the maximum operator in our definition. The benefit is that the supremum always exists. 
		%due to the interpretation of the supremum over the extended real numbers. 
		Second, compared with \cite[Definition 4]{Donze10STLRob}, we  use  the characteristic function $\chi_p(\sstraj, t)$ of $p$ instead of the characteristic function $\chi_\formula(\sstraj, t)$ of $\formula$ in the definitions of $\thetap_p(\sstraj,t)$ and $\thetam_p(\sstraj,t)$ in \eqref{eq:rec_pp} and \eqref{eq:rec_pm}, respectively.
\end{rem}

\begin{exmp}[continues=ex:running]
Take a look again at the signal $\sstraj$ shown in Fig.~\ref{fig:ex1}.
	The evolution of the \robplus{} asynchronous temporal robustness $\thetap_p(\sstraj,t)$ , $\thetap_q(\sstraj,t)$ , $\thetap_{\varphi_1}(\sstraj,t)$ and $\thetap_{\varphi_2}(\sstraj,t)$ is presented in Fig.~\ref{fig:ex1}(d).
In order to estimate the \robplus{} asynchronous temporal robustness for $\varphi_1=p\wedge q$ at time point $t$ over the given signal $\sstraj$, one  must first obtain the $\thetap_p(\sstraj, t)$ and $\thetap_q(\sstraj, t)$ values and then apply the conjunction rule \eqref{eq:t_and}. 
For instance,  
$\thetap_{\varphi_1}(\sstraj,0) \defeq \thetap_{p}(\sstraj,0)\,\sqcap\, \thetap_{q}(\sstraj,0) = 1$. Analogously, one could obtain that for $\varphi_2=p\vee q$,  the left asynchronous temporal robustness $\thetap_{\varphi_2}(\sstraj,0) \defeq \thetap_{p}(\sstraj,0)\,\sqcup\, \thetap_{q}(\sstraj,0) = 2$.
\end{exmp}

Note that though the asynchronous temporal robustness $\theta^\pm_\varphi(\sstraj, t)$ is defined in a recursive manner, the synchronous temporal robustness $\eta^\pm_\varphi(\sstraj, t)$ does not follow the recursive rules. 
%From the Definition~\ref{def:time_rob_all} one can see that $\eta^{\pm}_{\neg\varphi}(\sstraj,t) = -\eta^{\pm}_\varphi(\sstraj,t)$ but the rest of the recursive rules similar to Definition~\ref{def:time_rob_rec} do not hold. For example it does not hold that $\eta^{\pm}_{\varphi_1\wedge\varphi_2}(\sstraj,t)=
%\eta^{\pm}_{\varphi_1}(\sstraj,t) \sqcap \eta^{\pm}_{\varphi_2}(\sstraj,t)$. A counterexample can be formulated using Example~\ref{ex:running} and Figure~\ref{fig:ex1}. For instance, one can see on Fig.~\ref{fig:ex1}(d) that $\etap_{p\wedge q}(\sstraj, 2)=-5$ but $\etap_p(\sstraj,2) \sqcap \etap_{q}(\sstraj,2) = 0 \sqcap -2 = -2$. 

\section{Soundness of Temporal Robustness for Continuous-Time Signals}
\label{sec:prop_sound}
In this section, we provide  soundness results that state the relationship between the synchronous and asynchronous temporal robustness  and the Boolean semantics of STL.    Section~\ref{sec:time_rob_sync_prop} presents the main results for synchronous temporal robustness, while Section~\ref{sec:time_rob_async_prop} present analogous results for the asynchronous temporal robustness. While we provide our main results for continuous-time signals, we remark how the results simplify for discrete-time signals.
The proofs of our technical results are provided in Appendix \ref{appndx:soundness}.

\subsection{Properties of Synchronous Temporal Robustness}
\label{sec:time_rob_sync_prop}
The next theorem follows directly from  Definition~\ref{def:time_rob_all} and is fundamental to correctness of our solution to the control synthesis problem defined in Section~\ref{sec:control}.
\begin{theorem}[Soundness]
	\label{thm:sat_eta}
	For an STL formula $\varphi$, signal $\sstraj:\TDom \rightarrow X$ and some time $t\in\TDom$, the following results hold: 
	\begin{enumerate}
		\item\label{ti1} $\eta^{\pm}_\varphi(\sstraj,t) > 0 \quad\, \Longrightarrow\ \chi_\varphi(\sstraj,t)= +1$
		\item\label{ti2} $\eta^{\pm}_\varphi(\sstraj,t) < 0 \quad\, \Longrightarrow\ \chi_\varphi(\sstraj,t)= -1$
		\item\label{ti3} $\chi_\varphi(\sstraj,t)= +1 \  \Longrightarrow\ 
		\eta^{\pm}_\varphi(\sstraj,t) \geq 0$
		\item\label{ti4} $\chi_\varphi(\sstraj,t)= -1 \  \Longrightarrow\ 
		\eta^{\pm}_\varphi(\sstraj,t) \leq 0$ 
	\end{enumerate}
\end{theorem}

To illustrate the previous result, let us again consider Example \ref{ex:running}.

\begin{exmp}[continues=ex:running]
	Take again a look at Fig.~\ref{fig:ex1} and consider the formula $\varphi_2 = p\vee q$. From Fig.~\ref{fig:ex1}(c) one can see that $\etap_{\varphi_2}(\sstraj, 0)=2>0$ so that, due to Theorem~\ref{thm:sat_eta}, $\chi_\varphi(\sstraj,0)= \po$ has to hold, which can indeed be verified in Fig.~\ref{fig:ex1}(b). Note that the equivalence $\eta^{\pm}_\varphi(\sstraj,t) \geq 0 \Longleftrightarrow \chi_\varphi(\sstraj,t)= \po$ does not hold.\footnote{The equivalence $\eta^{\pm}_\varphi(\sstraj,t) \leq 0 \Longleftrightarrow \chi_\varphi(\sstraj,t)= -1$ does not hold either, as the same reasoning applies.} In other words, when $\eta^{\pm}_\varphi(\sstraj,t) =0$, we can not determine if the formula is satisfied or violated. For instance, consider the time points $t=2$ and $t=4$ in Fig.~\ref{fig:ex1} and note that $\etap_{\varphi_2}(\sstraj, 2)=\etap_{\varphi_2}(\sstraj, 4)=0$, but that $\chi_{\varphi_2}(\sstraj, 2)=\po$ and $\chi_{\varphi_2}(\sstraj, 4)=-1$.
\end{exmp}

The next result follows directly from Theorem~\ref{thm:sat_eta} and represents the connection between the \robplus{} and the \robminus{} synchronous temporal robustness. 
\begin{cor}
	\label{cor:eta_lr}
	For an STL formula $\varphi$, signal $\sstraj:\TDom \rightarrow X$ and some time $t\in\TDom$, the following properties hold: 
	\begin{enumerate}
		\item $\eta^{\pm}_\varphi(\sstraj,t)> 0\quad\Longrightarrow\quad \eta^{\mp}_\varphi(\sstraj,t)\geq 0$.
		\item $\eta^{\pm}_\varphi(\sstraj,t)< 0\quad\Longrightarrow\quad \eta^{\mp}_\varphi(\sstraj,t)\leq 0$.
	\end{enumerate}
\end{cor}

In this paper, we are interested in time shifts. As a first step, we next present a  result towards understanding the connection between the synchronous temporal robustness $\eta^{\pm}_\varphi(\sstraj,t)$ and  values of $\tstep$ for which $\chi_{\varphi}(\sstraj,t)$ and $\chi_{\varphi}(\sstraj,t+\tstep)$ are the same. This result  establishes an equivalence between the \robplus{} (\robminus{}) synchronous temporal robustness and the maximum time in the future (past) without changing the satisfaction of the formula.  While this gives a first interpretation of temporal robustness, we analyze temporal robustness in terms of synchronous and asynchronous time shifts in the signal $\sstraj$ itself in detail in Section \ref{sec:robustness_shifts_time}.  
\begin{theorem}
	\label{thm:chi_sync_stronger} 
	For an STL formula $\varphi$, signal $\sstraj:\Re \rightarrow X$, time $t\in\Re$ and some value $\trv\in\CoRe_{\geq 0}$, the following result holds:	
	\begin{align*}
		|\eta^{\pm}_\varphi(\sstraj, t)|=\trv\quad\Longleftrightarrow\quad
		&\forall t'\in t\pm [0, \trv),\ \chi_{\varphi}(\sstraj,t')=\chi_\varphi(\sstraj,t) \qquad\text{and}\\
		&
		\text{if } \trv<\infty, \text{ then } \forall\epsilon>0,\ \exists \tstep\in[\trv, \trv+\epsilon),\ \chi_{\varphi}(\sstraj,t \pm \tstep)\not=\chi_\varphi(\sstraj,t).
	\end{align*}
\end{theorem}
The interpretation of Theorem \ref{thm:chi_sync_stronger}  is as follows. Consider the case of  finite \robplus{} synchronous temporal robustness, i.e., $\etap_\varphi(\sstraj, t)=\trv< \infty$. 
Then for all times $t+[0, \trv)$, the formula satisfaction  is the same, i.e., for all $t'\in t + [0, \trv)$ we have that $\chi_\varphi(\sstraj, t')=\chi_\varphi(\sstraj, t)$.
Note that the interval $[0, \trv)$ is right-open and that we can not in general guarantee that $\chi_\varphi(\sstraj, t+\trv)=\chi_\varphi(\sstraj, t)$. We can, however, guarantee that in  close proximity of $t+\trv$ (quantified by $\epsilon$ in Theorem \ref{thm:chi_sync_stronger})  the satisfaction $\chi_\varphi(\sstraj, t'')$ must change.
%Note that if $i>0$, the above formulation can be simplified because we know that $\chi_\varphi(\sstraj, t)=+1$ or $\chi_\varphi(\sstraj, t)=-1$, respectively, due to Theorem~\ref{thm:sat_eta}. 

Note that the right-open interval $[0, \trv)$ and the existence of $\epsilon$ in Theorem~\ref{thm:chi_sync_stronger} appears due to the $\sup$ operator in  Definition~\ref{def:time_rob_rec} For discrete-time signals,  the interval $[0, \trv)$ becomes closed and  $\epsilon$ disappears. Particularly, for the special case of a discrete-time signal $\sstraj:\Ze \rightarrow X$ and if $\eta^{\pm}_\varphi(\sstraj, t)$ is finite, we remark that  Theorem \ref{thm:chi_sync_stronger}  can instead be stated as:
	\begin{equation}
		\label{eq:chi_eta_disc}
		\begin{aligned}
			|\eta^{\pm}_\varphi(\sstraj, t)|=\trv\quad\Longleftrightarrow\quad
			&\forall t'\in t\pm [0, \trv],\ \chi_{\varphi}(\sstraj,t')=\chi_\varphi(\sstraj,t) \qquad\text{and}\\
			&
			\chi_{\varphi}(\sstraj,t \pm (\trv+1))\not=\chi_\varphi(\sstraj,t).
		\end{aligned}
		\end{equation}

The next result states that the synchronous temporal robustness $\eta^{\pm}_\varphi(\sstraj, t)$ is a piece-wise linear function with segments either increasing with slope 1 or decreasing with slope $-1$, depending on $\chi_\varphi(\sstraj, t)$ and on the \robplus{} or \robminus{} synchronous temporal robustness.
\begin{theorem}
	\label{thm:eta_decrease} 
	For an STL formula $\varphi$, signal $\sstraj:\Re \rightarrow X$, time $t\in\Re$ and some value $\trv\in\CoRe_{\geq 0}$, the following result holds:
	\begin{equation*}
		\label{eq:eta_decrease}
		|\eta^{\pm}_\varphi(\sstraj, t)|= \trv\quad\Longrightarrow\quad \forall \tstep\in [0, \trv),\ |\eta^{\pm}_{\varphi}(\sstraj,t \pm \tstep)| = \trv -\tstep.	
	\end{equation*}
\end{theorem}

	For the special case of a discrete-time signal $\sstraj:\Ze \rightarrow X$ and if $\eta^{\pm}_\varphi(\sstraj, t)$ is finite, we remark that Theorem \ref{thm:eta_decrease}  can instead be stated as:
\begin{equation}
	\label{eq:eta_decrease_disc}
	|\eta^{\pm}_\varphi(\sstraj, t)|= \trv\quad\Longrightarrow\quad \forall \tstep\in [0, \trv],\ |\eta^{\pm}_{\varphi}(\sstraj,t \pm \tstep)| = \trv -\tstep.	
\end{equation}
%\begin{figure}[t!]
%	\centering
%	\includegraphics[width=\textwidth]{figures/f1.jpg}
%	\caption{Trajectory, characteristic function, left and right synchronous temporal robustness}
%	\label{f1}
%\end{figure}

The above result can be interpreted in the sense that the absolute value of the synchronous temporal robustness decreases proportionally with the amount of time shift $\tstep$. For instance, Fig.~\ref{fig:ex1}(c) shows how the function $\etap_{p\wedge q}(\sstraj, t)$ consists of the three linear segments: 
%$1-t$ if $t\in\{0,1\}$, $t-7$ if $t\in\{2,\ldots,7\}$ and $9-t$ if $t\in\{8,9\}$.
\begin{equation*}
	\etap_{p\wedge q}(\sstraj, t) = \begin{cases}
		1-t,&\text{if}\ t\in\{0,1\}\\
		t-7,&\text{if}\ t\in\{2,\ldots,7\}\\
		9-t,&\text{if}\ t\in\{8,9\}
	\end{cases}
\end{equation*}

\subsection{Properties of Asynchronous Temporal Robustness}
\label{sec:time_rob_async_prop}

In this section, we provide similar soundness results for the asynchronous temporal robustness. Theorem~\ref{thm:sat_theta} and Corollary~\ref{cor:theta_lr}, presented below, resemble Theorem \ref{thm:sat_eta} and Corollary \ref{cor:eta_lr} for  the synchronous temporal robustness  as one would expect. However, Theorems \ref{thm:chi_async}  and  \ref{thm:theta_bound}, also presented below, do not directly resemble the previous Theorems \ref{thm:chi_sync_stronger}  and \ref{thm:eta_decrease}  due to the recursive nature of the definition for the asynchronous temporal robustness. 
For instance, in contrast to Theorem \ref{thm:chi_sync_stronger}, Theorem \ref{thm:chi_async} only states a sufficient condition. In Theorem \ref{thm:eta_decrease}, compared to  Theorem~\ref{thm:eta_decrease}, we can only provide a lower bound  instead of an equality.
%\textcolor{red}{I don't like the "weaker" term. Can we use something else, or give more intuition on what we mean by weaker?}

\begin{theorem}[Soundness \cite{rodionova2021time}]
	\label{thm:sat_theta}
	For an STL formula $\varphi$, signal $\sstraj:\TDom \rightarrow X$ and some time $t\in\TDom$, the following results hold: 
	\begin{enumerate}
		\item\label{i1} $\theta^{\pm}_\varphi(\sstraj,t) > 0 \quad\, \Longrightarrow\ \chi_\varphi(\sstraj,t)= +1$
		\item\label{i2} $\theta^{\pm}_\varphi(\sstraj,t) < 0 \quad\, \Longrightarrow\ \chi_\varphi(\sstraj,t)= -1$
		\item\label{i3} $\chi_\varphi(\sstraj,t)= +1 \  \Longrightarrow\ 
		\theta^{\pm}_\varphi(\sstraj,t) \geq 0$
		\item\label{i4} $\chi_\varphi(\sstraj,t)= -1 \  \Longrightarrow\ 
		\theta^{\pm}_\varphi(\sstraj,t) \leq 0$ 
	\end{enumerate}
\end{theorem}
Note that, again, the equivalence $\theta^{\pm}_\varphi(\sstraj,t) \geq 0 \Longleftrightarrow \chi_\varphi(\sstraj,t)= +1$ does not hold\footnote{Equivalence $\theta^{\pm}_\varphi(\sstraj,t) \leq 0 \Longleftrightarrow \chi_\varphi(\sstraj,t)= -1$ does not hold either, same reasoning applied.}. In other words, when $\theta^{\pm}_\varphi(\sstraj,t) =0$ we can not determine if the formula is satisfied or violated. The next result is a straightforward corollary from Theorem \ref{thm:sat_theta}.

\begin{cor}
	\label{cor:theta_lr}
	For an STL formula $\varphi$, signal $\sstraj:\TDom \rightarrow X$ and some time $t\in\TDom$, the following results represent the connections between the \robplus{} and the \robminus{} asynchronous temporal robustness: 
	\begin{enumerate}
		\item $\theta^{\pm}_\varphi(\sstraj,t)> 0\quad\Longrightarrow\quad \theta^{\mp}_\varphi(\sstraj,t)\geq 0$.
		\item $\theta^{\pm}_\varphi(\sstraj,t)< 0\quad\Longrightarrow\quad \theta^{\mp}_\varphi(\sstraj,t)\leq 0$.
	\end{enumerate}
\end{cor}

Similarly to the previous section and Theorem \ref{thm:chi_sync_stronger} that was stated for synchronous temporal robustness, we now analyze the connection between the asynchronous temporal robustness $\theta^{\pm}_\varphi(\sstraj,t)$ and  values of $\tstep$ for which $\chi_{\varphi}(\sstraj,t)$ and $\chi_{\varphi}(\sstraj,t+\tstep)$ are the same. 

\begin{theorem}
	\label{thm:chi_async} 
	For an STL formula $\varphi$, signal $\sstraj:\Re \rightarrow X$, time $t\in\Re$ and some value $\trv\in\CoRe_{\geq 0}$, the following result holds:
	\begin{equation*}
		|\theta^{\pm}_\varphi(\sstraj,t)| = \trv 
		\quad \Longrightarrow\quad 
		\forall t'\in t \pm \,[0,\trv),\ \chi_\varphi(\sstraj,t')=\chi_\varphi(\sstraj,t)
	\end{equation*}
\end{theorem}
Note that Theorem~\ref{thm:chi_async} only provides a sufficient condition unlike Theorem~\ref{thm:chi_sync_stronger}. In other words, due to the recursive definition of the asynchronous temporal robustness,  one can not guarantee the existence of the time shift that would change the formula satisfaction as in the second line of Theorem~\ref{thm:chi_sync_stronger}.
For instance, in Fig.~\ref{fig:ex1} one could get that $\thetap_{p\vee q}(\sstraj,5)=1$ but $\chi_{p\vee q}(\sstraj,5)=...=\chi_{p\vee q}(\sstraj,9)=\po$. 	For the special case of a discrete-time signal $\sstraj:\Ze \rightarrow X$ and if $\theta^{\pm}_\varphi(\sstraj, t)$ is finite, we remark that Theorem \ref{thm:chi_async}  can instead be stated as:
	\begin{equation}
		\label{eq:chi_theta_disc}
	|\theta^{\pm}_\varphi(\sstraj, t)|=\trv\quad\Longrightarrow\quad
\forall t'\in t \pm \,[0,\trv],\ \chi_\varphi(\sstraj,t')=\chi_\varphi(\sstraj,t).
	\end{equation}

We next show how the asynchronous temporal robustness changes with time. 
\begin{theorem}
	\label{thm:theta_bound}
	For an STL formula $\varphi$, signal $\sstraj:\Re \rightarrow X$, time $t\in\Re$ and some value $\trv\in\CoRe_{\geq 0}$, the following result holds:
	\begin{equation*}
		\label{eq:theta_decrease}
		|\theta^{\pm}_\varphi(\sstraj, t)|= \trv\quad\Longrightarrow\quad \forall \tstep\in [0, \trv),\ |\theta^{\pm}_{\varphi}(\sstraj,t \pm \tstep)| \geq \trv -\tstep.	
	\end{equation*}
\end{theorem}
Note that, compared to  Theorem~\ref{thm:eta_decrease}, no equality can be stated on the right side of the implication operator. One can not guarantee the steady increase or decrease of the asynchronous temporal robustness  in time and obtain an exact value of the shifted temporal robustness. In fact, one can only provide a lower bound on its value. For the special case of a discrete-time signal $\sstraj:\Ze \rightarrow X$ and if $\theta^{\pm}_\varphi(\sstraj, t)$ is finite, we remark that Theorem \ref{thm:theta_bound}  can instead be stated as:
	\begin{equation}
	\label{eq:theta_decrease_disc}
	|\theta^{\pm}_\varphi(\sstraj, t)|= \trv\quad\Longrightarrow\quad \forall \tstep\in [0, \trv],\ |\theta^{\pm}_{\varphi}(\sstraj,t \pm \tstep)| \geq \trv -\tstep.	
	\end{equation}

\subsection{The Relationship between Synchronous and Asynchronous Temporal Robustness}
\label{sec:prop}

Until  now we have not yet established the precise relationship between the two notions of temporal robustness and how  they relate to each other. The properties specified in the previous Section~\ref{sec:prop_sound} suggest some similarities and differences between these temporal robustness notions.
For instance, it is easy to see from Definitions~\ref{def:time_rob_all} and \ref{def:time_rob_rec} that for the simple case of the specification $\varphi$ being a predicate $p$ the two notions are equal to each other.
\begin{cor}
	\label{rem:eq}
	$\theta^{\pm}_p(\sstraj,t)=\eta^{\pm}_p(\sstraj,t)$ for any $t\in\TDom$ and any $\sstraj:\TDom \rightarrow X$.
\end{cor}

The following theorem shows that the asynchronous temporal robustness is upper bounded by the synchronous temporal robustness value. 
\begin{theorem}
	\label{thm:theta_bounded}
	Given an STL formula $\varphi$ and a signal $\sstraj:\TDom \rightarrow X$, then for any $t\in\TDom$, the following inequality holds: 	
	\begin{equation*}		
		|\theta^{\pm}_\varphi(\sstraj,t) | \leq |\eta^{\pm}_\varphi(\sstraj,t)|
	\end{equation*}
	
\end{theorem}

Essentially, Theorem~\ref{thm:theta_bounded} states that the asynchronous temporal robustness is upper bounded by the synchronous temporal robustness. Together with the soundness Theorems \ref{thm:sat_eta} and \ref{thm:sat_theta}, the following holds
\begin{align*}
\chi_{\formula}(\sstraj,t)=+1 \;\;\; \implies \;\;\; 0\leq \theta^{\pm}_\varphi(\sstraj,t) \leq \eta^{\pm}_\varphi(\sstraj,t),\\
\chi_{\formula}(\sstraj,t)=-1 \;\;\; \implies \;\;\; \eta^{\pm}_\varphi(\sstraj,t)\leq \theta^{\pm}_\varphi(\sstraj,t) \leq 0.
\end{align*}

As Example~\ref{ex:running} suggests, the above inequalities are often strict. At this point, one may ask when equality holds. In fact there are fragments of STL for which  equality indeed holds. These STL fragments include  formulas in Negation Normal Form \cite{fainekos2009robustness}, i.e., negations only occur in front of predicates and the only other allowed operators are either the conjunction and always operators, denoted by $\varphi\in\stlnnf(\wedge,\always_I)$, or the disjunction and eventually operators, denoted by $\varphi\in\stlnnf(\vee,\eventually_I)$. For the former STL fragment, the following holds.
\begin{lemma}
	\label{lemma:p1_eq}
	Consider a formula $\varphi\in\stlnnf(\wedge,\always_I)$ and a signal $\sstraj:\TDom \rightarrow X$, then for any $t\in\TDom$ such that $\chi_\varphi(\sstraj,t)=\po$ it follows that $\eta^{\pm}_\varphi(\sstraj,t)=\theta^{\pm}_\varphi(\sstraj,t)$.
\end{lemma}
For $\varphi\in\stlnnf(\vee,\eventually_I)$, we can obtain a similar result as follows.
\begin{lemma}
	\label{lemma:m1_eq}
	Consider a formula $\varphi\in\stlnnf(\vee,\eventually_I)$ and a signal $\sstraj:\TDom \rightarrow X$, then for any $t\in\TDom$ such that $\chi_\varphi(\sstraj,t)=-1$ it follows that $\eta^{\pm}_\varphi(\sstraj,t)=\theta^{\pm}_\varphi(\sstraj,t)$.
\end{lemma}

%\begin{lemma}
%	\label{aeq}
%	Consider a signal $\sstraj:\TDom \rightarrow X$. The following results hold:
%	\begin{enumerate}
%		\item\label{lemma:p1_eq} For a formula $\varphi\in\stlnnf(\wedge,\always_I)$ and any $t\in\TDom$ such that $\chi_\varphi(\sstraj,t)=+1$ it follows that $\eta^{\pm}_\varphi(\sstraj,t)=\theta^{\pm}_\varphi(\sstraj,t)$.
%		\item\label{lemma:m1_eq} For a formula $\varphi\in\stlnnf(\vee,\eventually_I)$ and any $t\in\TDom$ such that $\chi_\varphi(\sstraj,t)=-1$ it follows that $\eta^{\pm}_\varphi(\sstraj,t)=\theta^{\pm}_\varphi(\sstraj,t)$.
%	\end{enumerate}
%\end{lemma}

%\section{Properties of Time Robustness with Respect to Trajectory Shifts}
\section{Robustness of Continuous-Time Signals with Respect to Time Shifts}
\label{sec:robustness_shifts_time}
So far, we have analyzed the connection between synchronous and asynchronous temporal robustness $\eta^{\pm}_\varphi(\sstraj, t)$ and $\theta^{\pm}_\varphi(\sstraj, t)$ and the satisfaction $\chi_\varphi(\sstraj,t)$ of an STL formula $\varphi$. In this section, we quantify the permissible synchronous and asynchronous time shifts in the signal $\sstraj$ that do not lead to a change in the satisfaction of the formula. By Definition~\ref{def:char_func}, the satisfaction $\chi_\varphi(\sstraj,t)$ of a formula $\varphi$ is recursively defined through the characteristic functions $\chi_{p_k}(\sstraj,t)$ of each predicate $p_k\in AP$ contained in $\varphi$. Therefore, 
from a satisfaction point of view, one could think about the permissible time shifts in terms of shifting the predicates via the characteristic function $\chi_{p_k}(\sstraj,t)$ in time, either synchronously (all $p_k$ together) or asynchronously (each $p_k$ independently).

\subsection{Synchronous Temporal Robustness}

%Now, let $\sstraj\in\SigSpace$ again be a signal and $p\in AP$ be a predicate of interest. For convenience, we call $\bm\chi_{p}(\sstraj):\SigSpace \to \{\pm 1\}^{\TDom}$ the  \textit{predicate signal} and define it to be the  map from the signal $\sstraj$ to the set of characteristic function evaluations $\chi_p(\sstraj,t)$ for all times $t\in \TDom$. 
We  start with the properties of the synchronous temporal robustness and, therefore, first define the notion of synchronous $\tstep$-early  and synchronous $\tstep$-late signals. 

\begin{definition}[Synchronously shifted early and late signals]
	\label{def:set_sync}
	Let $\sstraj\in\SigSpace$ be a signal, $AP$ be a given set of predicates and $\tstep\in\Te_{\geq 0}$. Then
	
	\begin{itemize}
		\item the signal $\sstraj^{\leftarrow \tstep}$ is called a \textbf{synchronous $\tstep$-early signal} 
		if  
		\begin{equation}
			\label{eq:synch_shift}
			\forall p_k\in AP,\ \forall t\in\TDom, \quad \chi_{p_k}^{}(\sstraj^{\leftarrow \tstep}, t)=
			\chi_{p_k}(\sstraj, t+\tstep)
		\end{equation}
		\item 	the signal $\sstraj^{\rightarrow \tstep}$ is called a \textbf{synchronous $\tstep$-late signal} if 
		\begin{equation}
			\forall p_k\in AP,\ \forall t\in\TDom,\quad \chi_{p_k}(\sstraj^{\rightarrow \tstep}, t)=
			\chi_{p_k}(\sstraj, t-\tstep)
		\end{equation}
	\end{itemize}

	\end{definition}

\begin{figure}[t!]
	\centering
	\includegraphics[width=\textwidth]{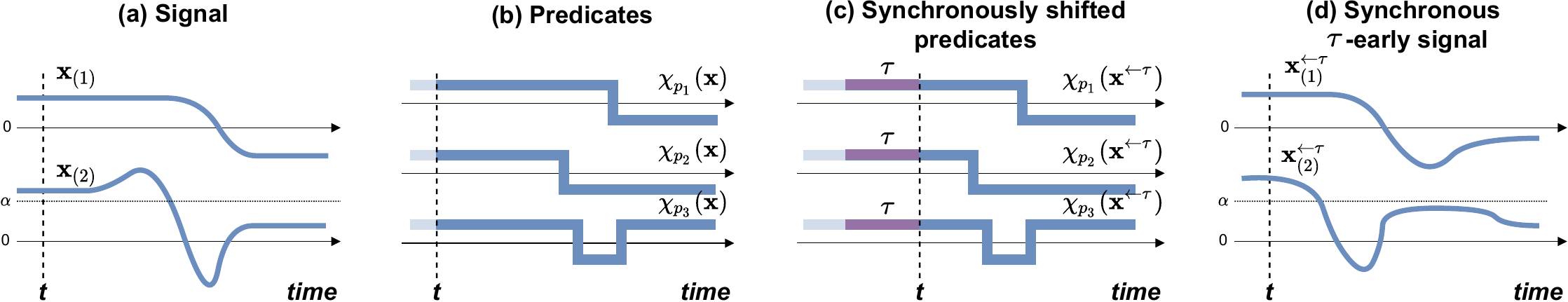}
	\caption{Synchronous $\tstep$-early signal. 
		(a) Signal $\sstraj\defeq(\sstraj\coord{1},\, \sstraj\coord{2})$; 
		(b) Evolution of the predicates 
		$p_1\defeq \sstraj\coord{1} \geq 0$, $p_2 \defeq \sstraj\coord{2}-\alpha \geq 0$ for given $\alpha>0$ and $p_3 \defeq \sstraj\coord{2} \geq 0$ over signal $\sstraj$ through their characteristic function
		$\chi_{p_k}(\sstraj, t)$, $\forall t\in\TDom$, where $k\in\{1,2,3\}$;
		(c) Evolution of predicates 
		$p_1$, $p_2$ and $p_3$ over signal $\sstraj^{\leftarrow \tstep}$ through their characteristic functions
		$\chi_{p_k}(\sstraj^{\leftarrow \tstep}, t)$ which are equal to the characteristic functions $\chi_{p_k}(\sstraj, t)$ synchronously shifted to the left in time by $\tstep$ for all $k\in\{1,2,3\}$ and $\forall t\in\TDom$;
		(d) Synchronous $\tstep$-early signal $\sstraj^{\leftarrow \tstep}$.}
	\label{fig:early_traj}
\end{figure}

Fig.~\ref{fig:early_traj} presents an example to illustrate the notion of a synchronous $\tstep$-early signal. Fig.~\ref{fig:early_traj}(a) depicts an initial signal $\sstraj\defeq(\sstraj\coord{1},\, \sstraj\coord{2})$. Assume that the formula of interest $\varphi$ is built upon three predicates, i.e., $AP\defeq \{p_1,p_2,p_3\}$, 
where $p_1\defeq \sstraj\coord{1} \geq 0$, $p_2 \defeq \sstraj\coord{2}-\alpha \geq 0$ and $p_3 \defeq \sstraj\coord{2} \geq 0$, for $\alpha>0$ as shown in Fig.~\ref{fig:early_traj}(a) and (d).
 Fig.~\ref{fig:early_traj}(b) shows the evolution of these three predicates $p_1$, $p_2$ and $p_3$ through their characteristic function
$\chi_{p_k}(\sstraj, t)$ for all $t\in\TDom$ where $k\in\{1,2,3\}$.
	By synchronously shifting all three characteristic functions $\chi_{p_k}(\sstraj, t)$, $\forall t\in\TDom$ to the left by $\tstep$ we obtain $\chi_{p_k}(\sstraj, t+\tstep)$, $\forall t\in\TDom$, which are depicted in Fig.~\ref{fig:early_traj}(c). Fig. \ref{fig:early_traj}(d) then shows a signal $\sstraj^{\leftarrow \tstep}$ that is a synchronous $\tstep$-early signal
	due to \eqref{eq:synch_shift} as the three synchronously shifted predicates in Fig.~\ref{fig:early_traj}(c) correspond to the predicates of  $\sstraj^{\leftarrow \tstep}$.

\begin{rem}
	\label{rem:shift_sync}
	If $\mathbf{s}$ is defined as $s_t\defeq x_{t \pm \tstep}$, i.e., we shift the entire signal $\sstraj$ by $\tstep$, then it holds that $\forall p_k\in AP$ and $\forall t\in\TDom$ we have that $\chi_{p_k}^{}(\mathbf{s}, t)=
	\chi_{p_k}(\sstraj, t\pm \tstep)$, 
	i.e., $\mathbf{s}=\sstraj^{\leftrightarrows \tstep}$ is a synchronous $\tstep$-early/late signal.
\end{rem}
%But  $\QA^{\leftarrow h}(\sstraj)$ is not limited to only $\sstraj^{\leftarrow h}$. In fact, it might contain infinite number of trajectories, since they can vary in shape in $X$. As long as the equality \eqref{eq:QA} holds on the APT level, such trajectories belong to $\QA^{\leftarrow h}(\sstraj)$. 

 The next result  follows by Definitions \ref{def:char_func} and \ref{def:set_sync}.
\begin{cor}
	\label{cor:sync_h}
	For an STL formula $\varphi$ built upon the predicate set $AP$,
	some $\tstep\in\TDom_{\geq 0}$ and signal $\sstraj:\TDom \rightarrow X$ it holds that $\forall t\in\TDom$,
	\begin{equation*}
		\chi_\varphi(\sstraj^{\leftrightarrows \tstep}, t)=\chi_\varphi(\sstraj, t\pm \tstep).
	\end{equation*} 

%	i.e. if trajectory $\mathbf{s}$ is a synchronously shifted trajectory $\sstraj$ to the left (right) by $\tstep$, then the overall formula trajectory  $\bm{\chi}_\varphi(\mathbf{s})$ is going to be a shifted to the left (right) by $\tstep$ formula trajectory $\bm{\chi}_\varphi(\sstraj)$. 
%
%	If all predicate trajectories are synchronously shifted by $\tstep$ then the overall formula trajectory will also be shifted by $\tstep$. 
	\end{cor}

\begin{figure}[t!]
	\centering
		\includegraphics[width=.9\textwidth]{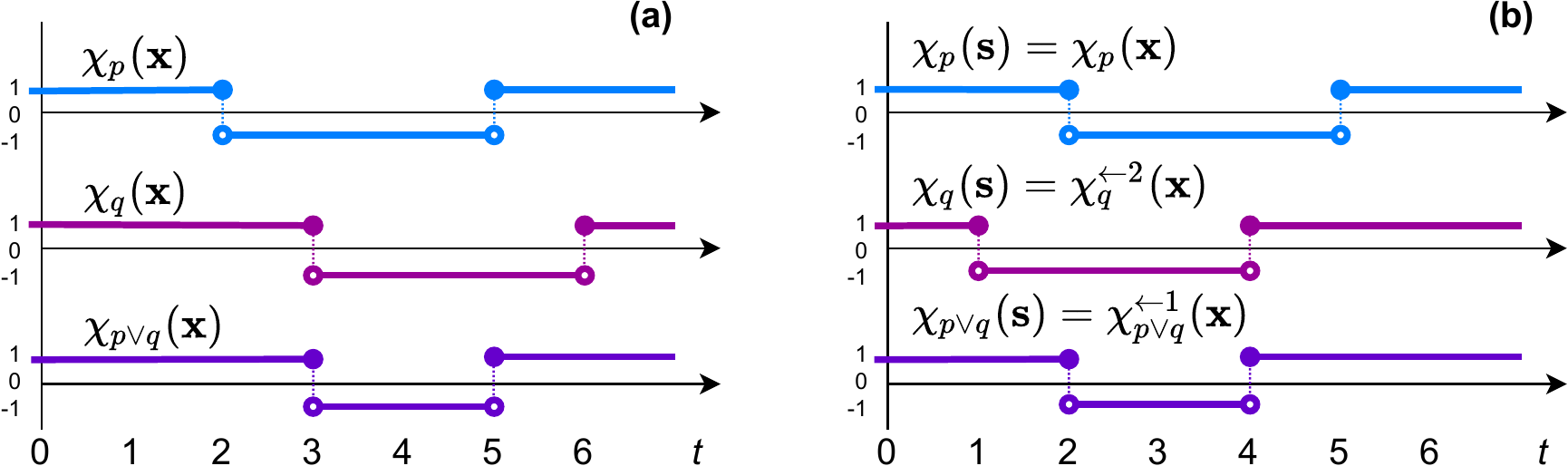}
	\caption{Predicates $p$ and $q$ and formula $p\vee q$ satisfaction over (a) signal $\sstraj$ and (b) signal $\mathbf{s}$. One can see that $\forall t\in\TDom$, $\chi_{p\vee q}(\mathbf{s}, t)=\chi_{p\vee q}(\sstraj, t+1)$ but $\mathbf{s}\not=\sstraj^{\leftarrow 1}$.}
	\label{fig:phi_shift}
\end{figure}

Note that  Corollary~\ref{cor:sync_h} is only a sufficient but not a necessary condition in the sense that there may exist signals $\mathbf{s}\not=\sstraj^{\leftrightarrows \tstep}$ for which $\chi_\varphi(\mathbf{s}, t)=\chi_\varphi(\sstraj, t \pm \tstep)$. For example, Fig.\ref{fig:phi_shift}(a) shows the characteristic function  for the predicates $p$ and $q$ and the formula $p\vee q$ over a signal $\sstraj$. Fig.\ref{fig:phi_shift}(b) shows the same functions but over another signal $\mathbf{s}$. One can see that $\forall t\in\TDom$, $\chi_{p\vee q}(\mathbf{s}, t)=\chi_{p\vee q}(\sstraj, t+1)$ but $\mathbf{s}\not=\sstraj^{\leftarrow 1}$. This is so because $\chi_{p}(\mathbf{s}, t)=\chi_{p}(\sstraj, t)$ and $\chi_{q}(\mathbf{s}, t)=\chi_{q}(\sstraj, t+2)$, therefore, ~\eqref{eq:synch_shift} is not satisfied.

We are now ready to state the main result that establishes a connection between the synchronous temporal robustness $\eta^\pm_\varphi(\sstraj, t)$ and the permissible time shifts $\tstep$ via $\sstraj^{\leftrightarrows {\tstep}}$. We note, that the proofs of all technical results presented in this section are provided in Appendix \ref{appndx:shifts}.

%\begin{theorem}
%	\label{thm:shift_sync}
%	Let $\varphi$ be an STL formula built upon the predicate set $AP$ and $\sstraj:\TDom \rightarrow X$ be a trajectory. For any time $t\in\TDom$ and $h\in\Re_{\geq 0}$, it holds that:
%	\begin{equation}
%		\label{eq:shift_sync0}
%		h < |\eta^{\pm}_\varphi(\sstraj,t)| \quad\Longrightarrow\quad 
%		\chi_\varphi(\sstraj^{\leftrightarrows h},t )=\chi_\varphi(\sstraj,t).	
%	\end{equation}
%%\begin{equation}
%%	\label{eq:shift_sync0}
%%	|\eta^{\pm}_\varphi(\sstraj,t)|=i \quad\Longrightarrow\quad
%%	\forall h\in[0, i),\quad 
%%	\chi_\varphi(\sstraj^{\leftrightarrows h},t )=\chi_\varphi(\sstraj,t).	
%%\end{equation}
%\end{theorem}

	\begin{theorem}
		\label{thm:shift_sync_stronger}
		Let $\varphi$ be an STL formula built upon the predicate set $AP$ and $\sstraj:\Re \rightarrow X$ be a signal. For any time $t\in\Re$ and $\trv\in\CoRe_{\geq 0}$, it holds that:
	\begin{align*}
		|\eta^\pm_\varphi(\sstraj, t)|=\trv\quad\Longleftrightarrow\quad
		&\forall \tstep\in[0,\trv),\ 
		\chi_\varphi(\sstraj^{\leftrightarrows \tstep}, t )=\chi_\varphi(\sstraj, t ) \qquad\text{and}\\
		&\text{if } \trv<\infty, \text{ then }
		\forall\epsilon>0,\ \exists \tstep\in[\trv, \trv+\epsilon),\ 
		\chi_\varphi(\sstraj^{\leftrightarrows \tstep}, t )\not=\chi_\varphi(\sstraj, t ).
	\end{align*}
\end{theorem}
	
Note particularly that the previous result closely resembles Theorem \ref{thm:chi_sync_stronger}. For the special case of a discrete-time signal $\sstraj:\Ze \rightarrow X$ and if $\eta^{\pm}_\varphi(\sstraj, t)$ is finite, we remark that Theorem \ref{thm:shift_sync_stronger} can instead be stated as:
\begin{equation}
	\label{eq:shift_eta_disc}
	\begin{aligned}
		|\eta^\pm_\varphi(\sstraj, t)|=\trv\quad\Longleftrightarrow\quad
		&\forall \tstep\in[0,\trv],\ 
		\chi_\varphi(\sstraj^{\leftrightarrows \tstep}, t )=\chi_\varphi(\sstraj, t ) \qquad\text{and}\\
		&
		\chi_\varphi(\sstraj^{\leftrightarrows \trv+1}, t )\not=\chi_\varphi(\sstraj, t ).
	\end{aligned}
\end{equation}

%\newpage
\subsection{Asynchronous Temporal Robustness}

Let us now continue by analyzing the asynchronous temporal robustness and first define the notion of asynchronous $\bar{\tstep}$-early  and asynchronous $\bar{\tstep}$-late signals.
\begin{definition}[Asynchronously shifted signal]
	\label{def:set_async}
	Let $\sstraj\in\SigSpace$ be a signal, $AP\defeq\{p_1,\ldots,p_L\}$ be a given set of predicates and 
	$\tstep_1,\ldots,\tstep_L\in\TDom_{\geq 0}$. Denote $\bar{\tstep}\defeq(\tstep_1,\ldots,\tstep_L)$. Then	
	\begin{itemize}
		\item the signal $\sstraj^{\leftarrow \bar{\tstep}}$ is called an \textbf{asynchronous $\bm{\bar{\tstep}}$-early signal} 
		if 
		\begin{equation}
			\label{eq:asynch_shift}
			\forall p_k\in AP,\ \forall t\in\TDom,\quad \chi_{p_k}^{}(\sstraj^{\leftarrow \bar{\tstep}}, t)=
			\chi_{p_k}(\sstraj, t+\tstep_k)
		\end{equation}
		\item 	the signal $\sstraj^{\rightarrow \bar{\tstep}}$ is called an \textbf{asynchronous $\bm{\bar{\tstep}}$-late signal} if,
		\begin{equation}
			\forall p_k\in AP,\ \forall t\in\TDom \quad \chi_{p_k}^{}(\sstraj^{\rightarrow \bar{\tstep}}, t)=
			\chi_{p_k}(\sstraj, t-\tstep_k)
		\end{equation}
	\end{itemize}
\end{definition}

\begin{figure}[t!]
		\includegraphics[width=.97\textwidth]{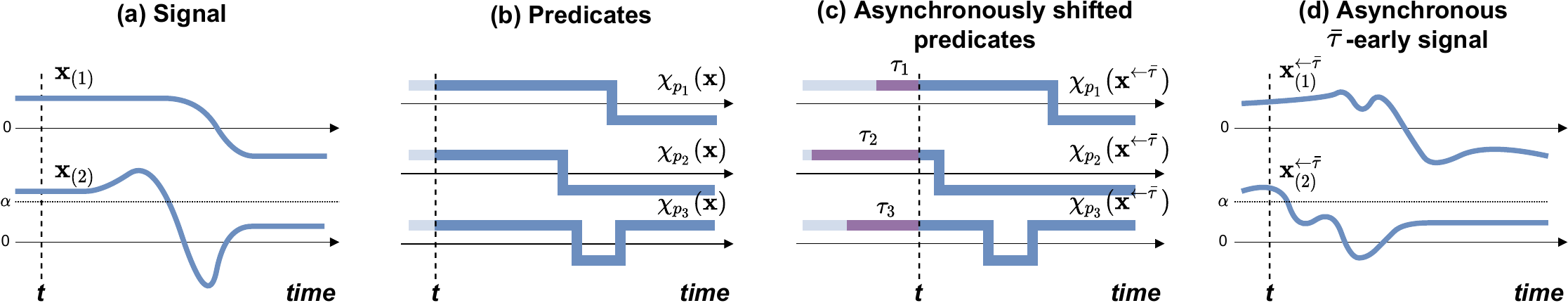}
		\caption{Asynchronous $\bar{\tstep}$-early signal. 
			(a) Signal $\sstraj\defeq(\sstraj\coord{1},\, \sstraj\coord{2})$; 
				(b) Evolution of the predicates 
				$p_1\defeq \sstraj\coord{1} \geq 0$, $p_2 \defeq \sstraj\coord{2}-\alpha \geq 0$ for given $\alpha>0$ and $p_3 \defeq \sstraj\coord{2} \geq 0$ over signal $\sstraj$ through their characteristic function
				$\chi_{p_k}(\sstraj, t)$, $\forall t\in\TDom$, where $k\in\{1,2,3\}$;
			(c) Evolution of predicates 
			$p_1$, $p_2$ and $p_3$ over the signal $\sstraj^{\leftarrow \bar{\tstep}}$, $\bar{\tstep}\defeq(\tstep_1, \tstep_2, \tstep_3)$ through their characteristic functions
			$\chi_{p_k}(\sstraj^{\leftarrow \bar{\tstep}}, t)$ which are equal to the characteristic functions $\chi_{p_k}(\sstraj, t)$ asynchronously shifted in time by $\tstep_k$ for each $k\in\{1,2,3\}$ and $\forall t\in\TDom$;
			(d) Asynchronous $\bar{\tstep}$-early signal $\sstraj^{\leftarrow \bar{\tstep}}$.}
		\label{fig:asynch_traj}
	\end{figure}

Fig.~\ref{fig:asynch_traj} presents an example to illustrate the notion of an asynchronous $\bar{\tstep}$-early signal. Similar to Fig.~\ref{fig:early_traj}, Fig.~\ref{fig:asynch_traj}(a)-(b) show the evolution of a signal  $\sstraj$ and the same three predicates of interest over this signal, respectively. In Fig.~\ref{fig:asynch_traj}(c) one can see each predicate $p_k$ being asynchronously shifted to the left by an individual $\tstep_k$, where $k\in\{1,2,3\}$. Fig.~\ref{fig:asynch_traj}(d) then shows a signal $\sstraj^{\leftarrow \bar{\tstep}}$ that is an asynchronous $\bar{\tstep}$-early  signal due to \eqref{eq:asynch_shift} as the three asynchronously shifted predicates in Fig.~\ref{fig:asynch_traj}(c)  correspond to the predicates of $\sstraj^{\leftarrow \bar{\tstep}}$. 
	
%Note that $\sstraj^{\leftrightarrows \bar{\tstep}}=\sstraj^{\leftrightarrows h}$ if $\tstep_1=\ldots=\tstep_L=\tstep$.
Note particularly that the above Definition~\ref{def:set_async} allows to shift the characteristic function $\chi_{p_k}(\sstraj, t+\tstep_k)$ of each predicate $p_k$ individually (and hence asynchronously) via $\tstep_k$. This is in contrast to Definition \ref{def:set_sync} where we shift the characteristic function $\chi_{p_k}(\sstraj, t+\tstep)$ of each predicate $p_k$ by the same amount (and hence synchronously) via $\tstep$. We are now ready to state the main result that establishes a connection between the asynchronous temporal robustness $\theta^\pm_\varphi(\sstraj, t)$ and the permissible time shifts $\bar{\tstep}$ in the predicates via $\sstraj^{\leftrightarrows \bar{\tstep}}$.

\begin{theorem}
\label{thm:shift_async}
Let $\varphi$ be an STL formula built upon the predicate set $AP\defeq\{p_1,\ldots,p_L\}$ and $\sstraj:\Re \rightarrow X$ be a signal. For any time $t\in\Re$ and $\trv\in\CoRe_{\geq 0}$, it holds that:
%\begin{equation}
%	\max(\tstep_1,\ldots,\tstep_L) < |\theta^{\pm}_\varphi(\sstraj,t)| \quad\Longrightarrow\quad 
%	\chi_\varphi(\sstraj^{\leftrightarrows \bar{\tstep}},t )=\chi_\varphi(\sstraj,t).	
%\end{equation}
	\begin{equation*}
	|\theta^{\pm}_{\varphi}(\sstraj, t)|=\trv  \quad\Longrightarrow\quad
	\forall \tstep_1,\ldots,\tstep_L\in[0, \trv),\quad 
	\chi_\varphi(\sstraj^{\leftrightarrows  \bar{\tstep}},t )=\chi_\varphi(\sstraj, t)
\end{equation*}
where $\bar{\tstep}\defeq(\tstep_1,\ldots,\tstep_L)$.
\end{theorem}
	Note that Theorem~\ref{thm:shift_async} only provides a sufficient condition unlike Theorem~\ref{thm:shift_sync_stronger}. In other words, due to the recursive definition of the asynchronous temporal robustness,  one can not guarantee the existence of the time shift that would change the formula satisfaction as in the second line of Theorem~\ref{thm:shift_sync_stronger}. For the special case of a discrete-time signal $\sstraj:\Ze \rightarrow X$ and if $\theta^{\pm}_\varphi(\sstraj, t)$ is finite, we remark that Theorem \ref{thm:shift_async} can instead be stated as:
		\begin{equation}
			\label{eq:shift_theta_disc}
		|\theta^{\pm}_{\varphi}(\sstraj, t)|=\trv  \quad\Longrightarrow\quad
		\forall \tstep_1,\ldots,\tstep_L\in[0, \trv],\quad 
		\chi_\varphi(\sstraj^{\leftrightarrows  \bar{\tstep}},t )=\chi_\varphi(\sstraj, t).
		\end{equation}

The definitions of asynchronous $\bar{\tstep}$-early  and asynchronous $\bar{\tstep}$-late signals together with Theorem \ref{thm:shift_async} quantify the permissible time shift $\bar{\tstep}$ in terms of how much each predicate $p_k$ in $\varphi$ can be shifted via the characteristic function $\chi_{p_k}(\sstraj,t\pm \tstep_k )$. Let us next analyze under which conditions these time shifts directly correlate with time shifts in the underlying signal $\sstraj$.	Consider a signal $\sstraj:\TDom \rightarrow X $ with its state $x_t\defeq (x\coordup{1}_{t}\ldots,x\coordup{n}_{t})\in\Re^n$. We will use $\kappa$ to denote  time shifts in the elements of $x_t$, in contrast to $\tstep$ for the time shifts in the predicates.
	First, note that if all elements $x\coordup{i}_{i}$ of the signal $\sstraj$ are shifted in time by the same amount $\kappa$, i.e.,
	$s\coordup{i}_t\defeq x\coordup{i}_{t \pm \kappa}$, $\forall i\in\{1,\ldots,n\}$, then 
	due to Remark~\ref{rem:shift_sync} we have that $\mathbf{s}$ is an asynchronous $\bar{\tstep}$-early/late signal\footnote{The signal $\mathbf{s}$ is a synchronous $\tstep$-early/late signal with $\tstep=\kappa$.}, i.e., $\mathbf{s}=\sstraj^{\leftrightarrows \bar{\tstep}}$ for $\bar{\tstep}=(\kappa,\ldots,\kappa)$. 
	In contrast, if time shifts are different across different elements of the state, the following may apply.

	\begin{rem}
		\label{rem:shift_async}
		
		If the signal $\mathbf{s}$ is defined as 
		$s_t\defeq (x\coordup{1}_{t \pm \kappa_1},\ldots,x\coordup{n}_{t \pm \kappa_n})$, i.e., each element $x_t^{(i)}$  of $x_t$ is shifted individually by $\kappa_i$, then there might not exist any $\bar{\tstep}\defeq (\tstep_1,\ldots,\tstep_L)$ such that $\mathbf{s}=\sstraj^{\leftrightarrows \bar{\tstep}}$. In other words, the signal $\mathbf{s}$ might not be an asynchronous $\bar{\tstep}$-early/late signal.
	\end{rem}

Next we consider a specific example that will help us understand under what  conditions asynchronously shifting the elements of the state in time leads to asynchronous early or late signals.

\begin{exmp}
	Consider three drones 
	which are, for simplicity, assumed to only move in vertical direction. 
	The state of each drone $d$ is $x\coordup{d} \defeq (z\coordup{d}, v\coordup{d})\in \Re^2$ and comprises of altitude and velocity. We define the signal state as
	$x \defeq (x\coordup{1}, x\coordup{2}, x\coordup{3})
	\in \Re^{6}$. Assume that we allow time shifts only across the full state of each drone $x\coordup{d}$, i.e., different drones can be shifted in time differently, but the  altitude and velocity of each drone $d$ is shifted in time by the same amount $\kappa_d$. Formally, such shifted signal $\mathbf{s}$ can be defined as following:
	\begin{equation*}
	s_t \defeq (s\coordup{1}_t,\, s\coordup{2}_t,\,	s\coordup{3}_t)=
	(x\coordup{1}_{t+\kappa_1},\, x\coordup{2}_{t+\kappa_2},\, x\coordup{3}_{t+\kappa_3}).	
	\end{equation*}
	Also assume that predicates are defined over separate drones only. In this case, any predicate $p_k\in AP \defeq \{p_1,\ldots,p_L\}$ that is defined over a drone $d$ can be represented as $\mu_k(x_t)\defeq \nu_k(x\coordup{d}_t)$, where $\nu_k$ is some real-valued function.
	Therefore, the following holds for the predicate $p_k$:
	\begin{align*}
	\chi_{p_k}(\mathbf{s}, t)\overset{\text{def}~\ref{def:char_func}}{\defeq}
	\sign \mu_k(s_t)=
	&\sign \nu_k(s\coordup{d}_t) \\= 
	&\sign \nu_k(x\coordup{d}_{t+\kappa_i}) =\sign\mu_k(x_{t+\kappa_d}) = 
	\chi_{p_k}(\sstraj, t+\kappa_d).
	\end{align*}

	Consequently, it holds that $\mathbf{s}= \sstraj^{\leftarrow \bar{\tstep}}$ where $\bar{\tstep}=(\tstep_1,\ldots, \tstep_L)$ and each $\tstep_k\in\{\kappa_1, \kappa_2, \kappa_3\}$.
	By applying Theorem~\ref{thm:shift_async}, we can conclude that  $\forall \kappa_1,\kappa_2,\kappa_3 \in [0, \trv)$ it holds that
	$\chi_\varphi(\mathbf{s},t )=\chi_\varphi(\sstraj,t)$ if $|\theta^+_{\varphi}(\sstraj, t)|=\trv$. This means that, 
	as long as within each drone the elements of the state are shifted in time dy the same amount, different drones can be shifted in time by the different amounts up to $|\theta^+_{\varphi}(\sstraj, t)|$, while formula satisfaction will not change. 
\end{exmp}
\newcommand{\cluster}{\zeta}
We now formally state the result that establishes when asynchronously shifting the elements of the state in time leads to asynchronous early or late signals. Let us therefore consider that the signal $\sstraj$ is defined as $x_t\defeq(\cluster\coordup{1}_t,\ldots,\cluster\coordup{c}_t)$ where $c$ the number of groups in which $x_t$ is clustered. For $d\in\{1,\ldots,c\}$, let us define the state as $\cluster\coordup{d}_t\defeq(x\coordup{d_1}_t,\ldots,x\coordup{d_m}_t)$ for dimensions $d_1,\hdots,d_m$.

\begin{lemma}
	\label{lemma:shift_async}
	Assume that every predicate  $p_k\in AP\defeq\{p_1,\ldots,p_L\}$ in $\varphi$ is defined over only a single $\cluster\coordup{d}$ for some $d$ i.e., $\exists d\in\{1,\ldots,c\}$ and a real-valued function $\nu_k$, such that 
	$p_k
	%\defeq\mu_k(x_t)\geq 0
	\defeq\nu_k(\cluster\coordup{d}_t)\geq 0$. 
	If $\mathbf{s}$ is defined as 
	$s_t\defeq(\cluster\coordup{1}_{t \pm \kappa_1}, \ldots, \cluster\coordup{c}_{t\pm \kappa_c})$, i.e. each $\cluster\coordup{d}$ is shifted by a different amount $\kappa_d$, then $\mathbf{s}=\sstraj^{\leftrightarrows \bar{\tstep}}$, i.e., 
	$\mathbf{s}$ is an asynchronous $\bar{\tstep}$ early/late signal,	
	where $\bar{\tstep} \defeq (\tstep_1,\ldots,\tstep_L)$ and each $\tstep_k\in\{\kappa_1,\ldots,\kappa_c\}$. 
\end{lemma}

\section{Temporally-Robust STL Control Synthesis}
\label{sec:control}

As mentioned before, in time-critical systems one is  not only interested in satisfying an STL formula, but also in satisfying the formula \textit{robustly} in terms of \textit{temporal robustness}. In this section, our goal is to design control laws that maximize the temporal robustness of a dynamical system. We refer to this problem as the \textit{temporally-robust control synthesis} problem and consider in the remainder discrete-time dynamical systems.
 
\subsection{Problem Formulation}
\label{sec:problem_form}

Consider a discrete-time, linear control system
\begin{equation}
	\label{eq:system}
	x_{t+1} \defeq A x_t + Bu_t, \;\;\; x_0\in X_0
\end{equation}
	where $x_t\in X\subseteq \mathbb{R}^n$ and $u_t\in U\subseteq \mathbb{R}^m$ are the state and control input of a linear dynamical system, respectively, where $X$ and $U$ are the workspace and  the set of permissible control inputs.
	We assume that sets $X$ and $U$ can be represented as a set of MILP constraints, e.g., when $X$ and $U$ are polytopes. Let the system have an initial condition $x_0$ from the set $X_0$. Let $A$ and $B$ be the system matrices of appropriate dimensions. Before defining the temporally-robust control synthesis problem, we make two assumptions on the STL formula $\varphi$. First, we assume that $\varphi$ is built from  linear predicate function, i.e., $\mu(x_t)$ is a linear function of the state. Second, we assume that $\varphi$ is  bounded-time  with formula length $\textit{len}(\varphi)$.  For the definition of the formula length, we refer the reader to \cite{raman2014model}.

\newcommand{\objs}{\vartheta}
\begin{prob}[Temporally-Robust  Control Synthesis]
	\label{prob:synthesis}
	Given an STL specification $\varphi$, a temporal robustness of interest $\objs \in\{\thetap_\varphi , \thetam_\varphi , \etap_\varphi ,\etam_\varphi \}$, a
	time horizon $H\ge\text{len}(\varphi)$, a discrete-time  control system as in \eqref{eq:system}, and a desired lower bound $\objs^*>0$ on the temporal robustness $\objs$, solve
	\begin{equation}
		\label{eq:control_problem}
		\begin{aligned}
			\inpSig^*=
			&\underset{\inpSig:=(u_0,\hdots,u_{H-1})}{\argmax}
			\quad \objs(\sstraj,0) \\
			\text{s.t.}
			&\quad x_{t+1} \defeq A x_t + B u_t,\quad t= 0,\ldots,H-1\\
			& \quad u_t\in U,\quad t=0,\ldots,H-1\\
			&\quad x_t\in X,\quad t=0,\ldots,H\\
			&\quad \objs(\sstraj,0)\geq \objs^*>0
		\end{aligned}
	\end{equation} 

\end{prob}
Problem \ref{prob:synthesis} aims at finding an optimal control sequence $\inpSig^*$ such that the corresponding signal $\sstraj$ not only respects the dynamics and input-state constraints presented by $X$ and $U$, but also robustly satisfies the given specification $\varphi$. Note that the last constraint in \eqref{eq:control_problem} particularly implies that the formula $\varphi$ is satisfied because we require the temporal robustness value to be strictly positive, see Theorems~\ref{thm:sat_eta} and \ref{thm:sat_theta}. Moreover, one can also control the desired lower bound $ \objs^*$ on the temporal robustness that must be achieved by the system. 

Solving Problem \ref{prob:synthesis} poses several challenges as both the synchronous as well as the asynchronous temporal robustness are neither continuous nor
smooth functions. Particularly  the counting of time shifts, as expressed by the $\sup$ operator in the Definitions \ref{def:time_rob_all} and \ref{def:time_rob_rec}, does not allow for the use of the variety of existing methods for control under STL specifications, e.g., gradient-based solutions and smooth approximations as in \cite{pant2017smooth} and \cite{AbbasF_NonlinearDescent13}. This motivates the use of Mixed-Integer Linear Programming (MILP) to explicitly encode and solve Problem~\ref{prob:synthesis}. In the remainder, we present an MILP encoding for the \robplus{} temporal robustness and we remark that the \robminus{} temporal robustness can be encoded analogously with only minor modifications, and is hence omitted. 

\begin{rem}
	We note that  
	Problem \ref{prob:synthesis} can be encoded as an MILP for an even  broader class of dynamical systems than  in \eqref{eq:system}. As long as the system dynamics can be expressed as a set of MILP constraints, the overall encoding will lead to an MILP formulation. Such systems include, for instance,  hybrid systems such as mixed logical dynamical systems \cite{bemporad1999control}, 
	piecewise affine systems \cite{sontag1981nonlinear}, 
	linear complementarity systems \cite{van1998complementarity}, and max-min-plus-scaling systems \cite{de2000model}.
	\end{rem}
%https://citeseerx.ist.psu.edu/viewdoc/download?doi=10.1.1.384.1977&rep=rep1&type=pdf

\subsection{MILP Encoding of Synchronous Temporal Robustness}
\label{sec:milp_eta}

For the case of a finite left synchronous temporal robustness 	$\etap_\varphi(\sstraj,t)$, e.g., when considering a finite time horizon $H$ as in Problem \ref{prob:synthesis}, the $\sup$ operator in Definition \ref{def:time_rob_all} can be replaced by a $\max$ operator, i.e., we have that
	\begin{equation}
		\label{eq:milp_eta}
	\etap_\varphi(\sstraj,t) \defeq \chi_\varphi(\sstraj, t)\cdot
	\max\{\tau\geq0 \ :\ \forall t'\in[t,t+\tau],\  \chi_\varphi(\sstraj,t')=\chi_\varphi(\sstraj,t)\}.\\
\end{equation}
Note that the calculation of $\etap_\varphi(\sstraj,t)$ is based on the
characteristic function  $\chi_\varphi(\sstraj, t)$.  Therefore, we start our MILP encoding of the synchronous temporal robustness $\etap_\varphi(\sstraj,t) $ with the  encoding of the STL formula $\varphi$ followed by the encoding of the $\max$ operator in \eqref{eq:milp_eta}. 

\paragraph{\textbf{Boolean encoding of STL constraints}}
Consider a binary variable $z^\varphi_t\in\{0,1\}$ that corresponds to the satisfaction of the  formula $\varphi$ by the signal $\sstraj$ at time point $t$, i.e. we let 
\begin{equation}
	\label{eq:milp_zphi}
	z_t^\varphi\defeq\begin{cases}
		1&\text{if}\ \chi_\varphi(\sstraj,t)=1\\
		0&\text{if}\ \chi_\varphi(\sstraj,t)=-1.\\
	\end{cases}
\end{equation}
 In \cite{raman2014model}, a recursive MILP encoding of the variable $z_t^\varphi$ has been presented such that exactly the above relationship holds. We use this set of MILP constraints in the remainder to obtain an MILP encoding for the \robplus{} synchronous temporal robustness $\etap_\varphi(\sstraj,t)$.

%Therefore, it is straightforward that the characteristic function $\chi_\varphi(\sstraj,t)\in\{\pm 1\}$ can be encoded as
%\begin{equation}
%	\label{eq:milp_char}
%	\chi_\varphi(\sstraj, t)=2z^\varphi_t -1.
%\end{equation}
%and the $\max$ operator that is responsible for counting the maximum number of time steps to the right for which the satisfaction of the $\varphi$ is uniform. 

\paragraph{\textbf{Encoding of the \robplus{} synchronous temporal robustness}}
Using \eqref{eq:milp_zphi}, the definition of the \robplus{} temporal robustness $\etap_\varphi(\sstraj,t)$ in \eqref{eq:milp_eta}  can be written in terms of $z_t^\varphi$ as following:
\begin{equation}
	\label{eq:milp_eta_cases}
	\etap_\varphi(\sstraj,t)\defeq\begin{cases}
		\quad \max\{\tau\geq0 \ :\ \forall t'\in[t,t+\tau],\  z_{t'}^\varphi=1\}&\text{if}\ z_t^\varphi=1\\
		-\max\{\tau\geq0 \ :\ \forall t'\in[t,t+\tau],\  z_{t'}^\varphi=0\}&\text{if}\ z_t^\varphi=0
	\end{cases}
\end{equation}
In other words, if $z_t^\varphi=1$ one has to count the maximum number of sequential time points $t'>t$ in the future for which $z_{t'}^\varphi=1$.
If $z_t^\varphi=0$, one has to count the maximum number of sequential time points $t'>t$ in the future  for which $z_{t'}^\varphi=0$, and then multiply this  number with $-1$. To implement this idea, we first construct the counter variable $\cophi_t\in\Ze_{\geq 0}$ that counts sequential $z_{t'}^\varphi=1$ for $t'\geq t$.
We also construct the second counter variable $\czphi_t\in\Ze_{\geq 0}$ that counts the number of sequential $z_{t'}^\varphi=0$ for $t'\geq t$ and then multiplies the counted value by $-1$. 
Since we count steps into the future from $t$, we do this recursively and backwards in time as follows:
\begin{align}
	\label{eq:milp_c1}
	&\cophi_t \defeq (\cophi_{t+1} +1) \cdot z^\varphi_t, \qquad\ \,\qquad \cophi_{H+1}\defeq0, \\
	\label{eq:milp_c0}
	&\czphi_t \defeq (\czphi_{t+1} -1) \cdot (1-z^\varphi_t), \qquad \czphi_{H+1}\defeq0.
\end{align}
Note that temporal robustness is defined by $z^\varphi_{t'}$ for which $t'>t$ rather than $t'\geq t$. Therefore, the previously defined counters $\cophi_t$, $\czphi_t$ must be adjusted by the value of 1 as follows:
%the corresponding value $z^\varphi_t$ must be excluded:
 \begin{align}
 	\label{eq:milp_ctp1}
 	&\acophi_{t}\defeq \cophi_t - z^\varphi_t,\\
 	\label{eq:milp_ctp0}
 	&\aczphi_{t} \defeq \czphi_t + (1-z^\varphi_t).
 \end{align}
Since the two cases in \eqref{eq:milp_eta_cases} are disjoint or, in other words, $z_t^\varphi$ is either $1$ or $0$, the overall \robplus{} synchronous temporal robustness $\etap_\varphi(\sstraj, t)$ can be implemented as a sum of the two adjusted counters:
\begin{equation}
	\label{eq:milp_eta_final}
	\etap_\varphi(\sstraj,t) \defeq \acophi_t + \aczphi_t
	%= c^1_{t} + c^0_t + (1- 2z^\varphi_t).
\end{equation}  

\begin{rem}
The constraints \eqref{eq:milp_c1} and \eqref{eq:milp_c0} contain a product of an integer variable with a Boolean variable. Such product can be expressed as a set of MILP constraints following Lemma~\ref{lemma:product}.  
\end{rem}

%%%%%%%%%%%%%%%%%%%%%%%%%%%%%%%%%%%%%%%%%%%%%%%%%%%%%%%%%%%%%%%%%%%%%%
\begin{algorithm}[t!]
	\SetAlgoLined
	%\Notation{HOW?}
	\KwIn{Specification $\varphi$, signal $\sstraj$.}
	\KwOut{The \robplus{} synchronous temporal robustness $\bm{\eta}^+_{\varphi}(\sstraj)$, the set of MILP constraints $\mathcal{H}$.}
	
	{1: Let $z^\varphi_t\in\Be$ be constrained by \eqref{eq:milp_zphi}, where  $t=0,\ldots,H$.}
	
	{2: Let $\cophi_t, \czphi_t\in\Ze$  be constrained by \eqref{eq:milp_c1} and \eqref{eq:milp_c0}, where $t=0,\ldots,H+1$.}
	
	{3: Let $\acophi_t, \aczphi_t\in\Ze$  be constrained by \eqref{eq:milp_ctp1} and \eqref{eq:milp_ctp0}, where $t=0,\ldots,H$.}
	
	{4: $\etap_{\varphi}(\sstraj, t)\overset{\eqref{eq:milp_eta_final}}{\defeq}\acophi_t + \aczphi_t$, where $t=0,\ldots,H$.}
	
	{5: $\mathcal{H}$ consists of the MILP constraints \eqref{eq:milp_zphi}, \eqref{eq:milp_c1},  \eqref{eq:milp_c0}, \eqref{eq:milp_ctp1}, \eqref{eq:milp_ctp0} and \eqref{eq:milp_eta_final}.}
	
	\caption{The function $(\bm{\eta}^+_{\varphi}(\sstraj),\,\mathcal{H})\defeq\texttt{MILP\_SYNC}(\varphi,\sstraj)$}
	\label{alg:eta}
	%	\vspace{-7pt}
\end{algorithm}
%%%%%%%%%%%%%%%%%%%%%%%%%%%%%%%%%%%%%%%%%%%%%%%%%%%%%%%%%%%%%%%%%%%%%%

\begin{table}[b!]
	\caption{\small Estimation of $\etap_\varphi(\sstraj,t)$ for $\varphi\defeq p\wedge q$ from Example~\ref{ex:running} following Alg.~\ref{alg:eta}.}
	\renewcommand{\arraystretch}{1.1}
	\setlength{\tabcolsep}{3pt}
	\centering
	\begin{tabular}{|l|c|c|c|c|c|c|c|c|c|c|c|c|}
		\hline
		$t$   & 0 & 1  & 2  & 3 & 4 & 5 & 6 & 7 & 8 & 9& 10\\ \hline
		\hline
		$\chi_{\varphi}(\sstraj,t)$ & 1 & 1 & -1 & -1 & -1 & -1 & -1 & -1 & 1& 1&\\ \hline
		$z_t^{\varphi}$, see \eqref{eq:milp_zphi} & 1 & 1 & 0 & 0 & 0 & 0 & 0 & 0 & 1& 1&\\ \hline
		$\cophi_t\defeq (\cophi_{t+1} +1) \cdot z^\varphi_t$, $\quad \cophi_{H+1}\defeq0$ & 2 & 1 & 0 & 0 & 0 & 0 & 0 & 0 & 2& 1&0\\ \hline
		$\czphi_t \defeq (\czphi_{t+1} -1) \cdot (1-z^\varphi_t)$, $\quad \czphi_{H+1}\defeq0$ & 0 & 0 & -6 & -5 & -4 & -3 & -2 & -1 & 0& 0&0\\ \hline
		$\acophi_{t}\defeq \cophi_t - z^\varphi_t$ & 1 & 0 & 0 & 0 & 0 & 0 & 0 & 0 & 1& 0&\\ \hline
		$\aczphi_{t} \defeq \czphi_t + (1-z^\varphi_t)$& 0 & 0 & -5 & -4 & -3 & -2 & -1 & 0 & 0& 0&\\ \hline\hline
		$\etap_\varphi(\sstraj,t) \defeq \acophi_t + \aczphi_t$ & 1   & 0  & -5  & -4 &-3 & -2  &  -1 & 0  &  1 &0&\\ \hline
	\end{tabular}
	%\vspace{0pt}
	\label{tab:exmp_eta}
	%\vspace{-6mm}
\end{table}

The MILP encoding  \eqref{eq:milp_zphi}-\eqref{eq:milp_eta_final}  of the \robplus{} synchronous temporal robustness $\etap_\varphi(\sstraj,t)$
is formally summarized in Alg.~\ref{alg:eta}.  
It outputs a set of MILP constraints $\mathcal{H}$ and $\etap_{\varphi}(\sstraj, t)$ for all time steps $t=0,\ldots,H$, denoted as $\bm{\eta}^+_{\varphi}(\sstraj)$.
We summarize the main properties of the above MILP encoding in the following Proposition \ref{thm:milp_sync_encoding}, while the proof is available in Appendix~\ref{appndx:proof_sec_control}.
\begin{prop}
	\label{thm:milp_sync_encoding}
	Let $\sstraj$ be a signal satisfying the recursion of the discrete-time dynamical system \eqref{eq:system} and let $\varphi$ be an STL specification that is built upon linear predicates. Then:
	\begin{enumerate}
		\item  The \robplus{} synchronous temporal robustness sequence $\bm{\eta}^+_\varphi(\sstraj)\defeq(\etap_\varphi(\sstraj,0),\ldots,\etap_\varphi(\sstraj,H))$, with $\etap_\varphi(\sstraj,t)$ as in \eqref{eq:milp_eta} 
		is equivalent to the output $(\bm{\eta}^+_{\varphi}(\sstraj),\,\mathcal{H})\defeq\texttt{MILP\_SYNC}(\varphi,\sstraj)$ of Alg.~\ref{alg:eta}.
		\item The MILP encoding of $\bm{\eta}^+_{\varphi}(\sstraj)$ in Alg.~\ref{alg:eta} is a function of $O(H\cdot|AP|)$ binary and $O(H\cdot|\varphi|)$ continuous variables, where $H$ is the time horizon, $|\varphi|$ is the number of operators in the formula $\varphi$, and $|AP|$ is the number of used predicates.  
	\end{enumerate}
\end{prop}

\begin{exmp}[continues=ex:running]
	Take a look again at the signal $\sstraj$ and an STL formula $\varphi=p\wedge q$ evaluation shown in Fig.~\ref{fig:ex1}.
	In Table~\ref{tab:exmp_eta} we consider a step-by-step estimation of the \robplus{} synchronous temporal robustness $\bm{\eta}^+_{\varphi}(\sstraj)$ following the MILP encoding procedure described above in Section~\ref{sec:milp_eta} and formally defined in Alg.~\ref{alg:eta}. 
	From Figure~\ref{fig:ex1}(b) one can see that $\bm{\chi}_{p\wedge q}(\sstraj)=(1, 1, -1, -1, -1, -1, -1, -1, 1, 1)$. Then following Alg.~\ref{alg:eta}, one can see that $\bm{\eta}^+_\varphi(\sstraj)$ sequence is indeed equal to $(1, 0, -5, -4, -3, -2, -1, 0, 1, 0)$ shown in Fig.~\ref{fig:ex1}(c). Therefore, the MILP encoding procedure leads to the same result as its estimation by the definition.
\end{exmp}

Finally, using the MILP encoding of the \robplus{} synchronous temporal robustness $\etap_\varphi(\sstraj,t)$ presented in Alg.~\ref{alg:eta}, we summarize the overall MILP encoding of  Problem~\ref{prob:synthesis} in case of $\objs= \etap_{\varphi}$ in Alg.~\ref{alg:milp_eta}. 
 The function $\texttt{SYSTEM\_CONSTRAINTS}(x_0,\inpSig)$ defines linear constraints on the decision variable $u_t$ according to \eqref{eq:system} and such that $x_t\in X$. 
 The below Proposition~\ref{thm:milp_sync_problem} states the correctness of our encoding and follows directly from the definition of Problem~\ref{prob:synthesis}.
\begin{prop}
	\label{thm:milp_sync_problem}
The temporally-robust STL control synthesis problem as defined in Problem~\ref{prob:synthesis} with synchronous \robplus{} temporal robustness, i.e., $\objs= \etap_{\varphi}$, is equivalent to an MILP as described in Alg.~\ref{alg:milp_eta}.
\end{prop}

We again remark that the case of synchronous \robminus{} temporal robustness  $\objs = \etam_\varphi$ can be handled almost identically.
%%%%%%%%%%%%%%%%%%%%%%%%%%%%%%%%%%%%%%%%%%%%%%%%%%%%%%%%%%%%%%%%%%%%%%
\begin{algorithm}[t!]
	\SetAlgoLined
	\KwIn{Specification $\varphi$, initial state $x_0$, time horizon $H$, minimum synchronous temporal robustness $\eta^*$}
	\KwOut{Control input sequence $\inpSig^* = (u^*_0,\ldots, u^*_{H-1})$}
	
	Set $\inpSig=(u_0,\ldots,u_{H-1})$ to be the decision variable

	$\,\inpSig^*=
	\underset{\inpSig}{\argmax}\quad \eta^{+}_\varphi(\sstraj,0)$
	%\vspace{-1pt}
	
	$\qquad\quad\ 
	\text{s.t.}\quad\ \ \,\eta^{+}_\varphi(\sstraj,0)\geq \eta^*>0$
	
	$\qquad\qquad\qquad\ \,\sstraj\defeq\texttt{SYSTEM\_CONSTRAINTS}(x_0,\inpSig)$
	
	$\qquad\qquad\qquad\ \,(\bm{\eta}^+_{\varphi}(\sstraj),\,\mathcal{H})\defeq\texttt{MILP\_SYNC}(\varphi,\sstraj)$ // Sec.~\ref{sec:milp_eta}, Alg.~\ref{alg:eta}
	
	$\qquad\qquad\qquad\ \,\mathcal{H},\ u_t\in U$ for $t=0,\hdots,H-1$
	
	\caption{MILP encoding of Problem~\ref{prob:synthesis} with the \robplus{} synchronous temporal robustness objective $\objs\defeq \etap_{\varphi}$}
	\label{alg:milp_eta}
\end{algorithm}
%%%%%%%%%%%%%%%%%%%%%%%%%%%%%%%%%%%%%%%%%%%%%%%%%%%%%%%%%%%%%%%%%%%%%%

%\begin{table}[t!]
%	\renewcommand{\arraystretch}{1.1}
%	\setlength{\tabcolsep}{3pt}
%	\centering
%	\begin{tabular}{|l|c|c|c|c|c|c|c|c|c|c|c|}
%		\hline
%		$t$   & 0 & 1  & 2  & 3 & 4 & 5 & 6 & 7 & 8 & 9\\ \hline
%		\hline
%		$\chi_p(\sstraj,t)$ & 1 & 1 & 1 & 0 & 0 & 0 & 0 & 1 & 1&1 \\ \hline
%		$\chi_q(\sstraj,t)$ & 1 & 1 & 0 & 0 & 0 & 1 & 1 & 0 & 1&1 \\ \hline \hline
%		$\chi_{p\wedge q}(\sstraj,t)$ & 1 & 1 & 0 & 0 & 0 & 0 & 0 & 0 & 1& 1\\ \hline
%		$\chi_{p\vee q}(\sstraj,t)$ & 1 & 1 & 1& 0 & 0 & 1 & 1 & 1 &1 &1 \\ \hline
%\hline		
%		$\thetap_p(\sstraj, t)$    & 2 & 1  &  0 & -3  &  -2 &  -1 &  0   & 2 &1&0 \\ \hline
%		$\thetap_q(\sstraj, t)$   & 1&  0  &  -2  & -1  & 0  & 1  & 0  &  0  &1&0\\ \hline\hline
%		$\thetap_{p\wedge q}(\sstraj, t)$ & 1   & 0  & -2  & -3  & -2  &  -1 & 0  & 0  &1&0\\ \hline
%		$\thetap_{p\vee q}(\sstraj, t)$ & 2   & 1  & 0  & -1  & 0  &  1 & 0  & 2  &1&0\\
%		\hline\hline
%		$\etap_{p\wedge q}(\sstraj, t)$ & 1   & 0  & -5  & -4 &-3 & -2  &  -1 & 0  &  1 &0\\ \hline
%		$\etap_{p\vee q}(\sstraj, t)$ & 2   & 1  & 0  & -1  & 0  &  4  & 3 & 2  &1&0\\
%		\hline
%	\end{tabular}
%	%\vspace{0pt}
%	\caption{\small Estimation of $\thetap_p\!(\sstraj,t)$ from Example~\ref{ex:running} following Alg.~\ref{alg:milp_theta}.}
%	\label{tab:exmp_theta}
%	%\vspace{-6mm}
%\end{table}

\subsection{MILP Encoding of Asynchronous Temporal Robustness}
\label{sec:milp_async}

Recall that the \robplus{} asynchronous temporal robustness $\thetap_\varphi(\sstraj,t)$ from Definition~\ref{def:time_rob_rec}
is defined recursively on the structure of $\varphi$.
In contrast to the MILP encoding for the synchronous temporal robustness $\etap_p(\sstraj,t)$, we hence encode the \robplus{} asynchronous temporal robustness $\thetap_p(\sstraj,t)$ for every predicate $p$ containing in $\varphi$ first, and then encode the recursive rules \eqref{eq:t_neg}-\eqref{eq:t_until} that are applied to every $\theta^{+}_p(\sstraj,t)$.

%Recall from Section~\ref{sec:disc} that the definition of the right asynchronous temporal robustness of predicates $p$ for discrete-time signals is as follows
%\begin{equation}
%	\label{eq:milp_theta}
%	\thetap_p(\sstraj,t) \defeq \chi_p(\sstraj, t)\cdot
%	\max\{\tau\geq0 \ :\ \forall t'\in[t,t+\tau],\  \chi_p(\sstraj,t')=\chi_p(\sstraj,t)\}.\\
%\end{equation}
\paragraph{\textbf{Encoding of STL predicates}}
Note that due to Corollary~\ref{rem:eq}, $\theta^{\pm}_p(\sstraj,t)=\eta^{\pm}_p(\sstraj,t)$ for any $t\in\TDom$. Therefore, the MILP encoding of $\thetap_p(\sstraj)$ is defined by the function $\texttt{MILP\_SYNC}(p,\sstraj)$ described by Alg.~\ref{alg:eta}.
Therefore, we can already summarize the encoding of STL predicates using MILP in the following corollary. 
\begin{cor}
	For a signal $\sstraj$ satisfying the recursion of the discrete-time dynamical system \eqref{eq:system} and a linear predicate $p\in AP$, the \robplus{} asynchronous temporal robustness sequence $\bm{\theta}^+_p(\sstraj)\defeq(\thetap_p(\sstraj,0),\ldots,\thetap_p(\sstraj,H))$ with $\thetap_p(\sstraj,t)$ defined by \eqref{eq:rec_pp} 
	is equivalent to the set of MILP constraints produced by the
	function $(\thetap_{p}(\sstraj),\,\mathcal{P})\defeq\texttt{MILP\_SYNC}(p,\sstraj)$ presented in Alg.~\ref{alg:eta}.
	\end{cor}

\paragraph{\textbf{Encoding of STL operators}}
Having encoded the temporal robustness of STL predicates as MILP constraints, the generalization to STL formulas is straight-forward and can use the encoding from \cite{raman2014model}. 
For example, let $\varphi = \wedge_{j=1}^m \varphi_i$ and $\thetap_{\varphi_j}(\sstraj,t) = r_j$. 
By Definition~\ref{def:time_rob_rec} and following \eqref{eq:t_and}, we have that $\thetap_{\varphi}(\sstraj, t)=\min_{j=1,\ldots,m} \thetap_{\varphi_j}(\sstraj, t)$. 
Then $\thetap_{\formula}(\sstraj,t) = r$ if and only if the following MILP constraints hold:
\begin{equation}
	\label{eq:conj}
	\begin{aligned}
		&r_j - M(1-b_j) \leq r \leq r_j, \quad j=1,\ldots m\\
		& \sum_{j=1}^{m} b_j = 1
	\end{aligned}
\end{equation}
where $b_j=\{0,1\}$ are introduced binary variables for $j=1,\ldots,m$ and $M$ is a big-$M$ parameter\footnote{$M$ is a large positive constant that is at least an upper bound on all pair-wise combinations of $r_j$, $M\geq |r_i-r_j|$. For more, see \cite{bemporad1999control}.}. 
%Constraints \eqref{eq:conj} enforce $r=\min r_i$.

%%%%%%%%%%%%%%%%%%%%%%%%%%%%%%%%%%%%%%%%%%%%%%%%%%%%%%%
\begin{algorithm}[t!]
	\SetAlgoLined
	%\Notation{HOW?}
	\KwIn{Specification $\varphi$, signal $\sstraj$}
	\KwOut{The \robplus{} asynchronous temporal robustness $\bm{\theta}^+_{\varphi}(\sstraj)$, set of MILP constraints $\mathcal{Q}$.}
	
	\For{$k=1,\ldots,L$}{
		
		$(\bm{\theta}^+_{p_k}(\sstraj),\mathcal{P}_k)\defeq\texttt{MILP\_SYNC}(p_k,\sstraj)$ // Alg.~\ref{alg:eta}
	}
	
	$(\bm{\theta}^+_{\varphi}(\sstraj),\mathcal{O}) \defeq\texttt{MILP\_OPERATORS}(\varphi,\sstraj,\bm{\theta}^+_{p}(\sstraj))$   %// %Sec.~\ref{sec:milp_op}
	
	$\mathcal{Q} \defeq \mathcal{O} \cup \bigcup_{k=1}^L \mathcal{P}_k$
	
	\caption{The function $(\bm{\theta}^+_{\varphi}(\sstraj),\,\mathcal{Q})\defeq\texttt{MIPL\_ASYN}(\varphi,\sstraj)$}
	\label{alg:milp_theta}
\end{algorithm}
%%%%%%%%%%%%%%%%%%%%%%%%%%%%%%%%%%%%%%%%%%%%%%%%%%%%%

%\paragraph{\textbf{Encoding of the right asynchronous temporal robustness}}

	The complete MILP encoding of the \robplus{} asynchronous temporal robustness $\thetap_\varphi(\sstraj,t)$
	is formally summarized in Alg.~\ref{alg:milp_theta}, which consists of two steps. First, the encoding of the STL predicates, defined formally through the function $\texttt{MILP\_SYNC}(p_k,\sstraj)$ for each predicate $p_k$ within $\varphi$.
	And second, the encoding of STL operators, defined through the function $\texttt{MILP\_OPERATORS}(\varphi,\sstraj,\bm{\theta}^+_{p}(\sstraj))$ that 
	is defined according to \cite{raman2014model} and recursively follows the structure of $\varphi$.
%	takes as an input the formula $\varphi$, signal $\sstraj$ and the MILP encodings of all predicates within $\varphi$. 
	Alg.~\ref{alg:milp_theta} outputs a set of MILP constraints $\mathcal{Q}$ and $\thetap_{\varphi}(\sstraj, t)$ for all time steps $t=0,\ldots,H$, denoted as $\bm{\theta}^+_{\varphi}(\sstraj)$.
We summarize the main properties of the above MILP formulation of the \robplus{} asynchronous temporal robustness in Proposition~\ref{thm:milp_asyn_encoding}, while the proof is available in Appendix~\ref{appndx:proof_sec_control}.
	\begin{prop}
		\label{thm:milp_asyn_encoding}
		Let $\sstraj$ be a signal satisfying the recursion of the discrete-time dynamical system \eqref{eq:system} and let $\varphi$ be an STL specification that is built upon linear predicates. Then:
		\begin{enumerate}
			\item The \robplus{} asynchronous temporal robustness sequence $\bm{\theta}^+_{\varphi}(\sstraj)\defeq(\thetap_\varphi(\sstraj,0),\ldots,\thetap_\varphi(\sstraj,H))$ with $\thetap_\varphi(\sstraj,t)$ as in \eqref{eq:rec_pp} 
			is equivalent to the output $(\thetap_{\varphi}(\sstraj),\,\mathcal{Q})\defeq\texttt{MILP\_ASYN}(\varphi,\sstraj)$ of Alg.~\ref{alg:milp_theta}.
			\item MILP encoding of $\bm{\theta}^+_{\varphi}(\sstraj)$ in Alg.~\ref{alg:milp_theta} is a function of $O(H\cdot(|AP|+|\varphi|))$ binary and continuous variables, where $H$ is the time horizon, $|\varphi|$ is the number of operators in the formula $\varphi$ and $|AP|$ is the number of used predicates.  
		\end{enumerate}
	\end{prop}

	\begin{table}[b!]
	\caption{\small Estimation of $\thetap_{\varphi}(\sstraj,t)$ for $\varphi\defeq p\wedge q$ from Example~\ref{ex:running} following Alg.~\ref{alg:milp_theta}.}
	\renewcommand{\arraystretch}{1.1}
	\setlength{\tabcolsep}{3pt}
	\centering
	\begin{tabular}{|l|c|c|c|c|c|c|c|c|c|c|c|}
		\hline
		$t$   & 0 & 1  & 2  & 3 & 4 & 5 & 6 & 7 & 8 & 9\\ \hline
		\hline
		$\chi_p(\sstraj,t)$ & 1 & 1 & 1 & -1 & -1 & -1 & -1 & 1 & 1&1 \\ \hline
		$\chi_q(\sstraj,t)$ & 1 & 1 & -1 & -1 & -1 & 1 & 1 & -1 & 1&1 \\ \hline 
		\hline		
		$\thetap_p(\sstraj, t)$, see $\texttt{MILP\_SYNC}(p,\sstraj)$ & 2 & 1  &  0 & -3  &  -2 &  -1 &  0   & 2 &1&0 \\ \hline
		$\thetap_q(\sstraj, t)$, see $\texttt{MILP\_SYNC}(q,\sstraj)$  & 1&  0  &  -2  & -1  & 0  & 1  & 0  &  0  &1&0\\ \hline\hline
		$\thetap_{\varphi}(\sstraj, t)\defeq \thetap_{p}(\sstraj, t) \sqcap \thetap_{q}(\sstraj, t)$, see \eqref{eq:conj} & 1   & 0  & -2  & -3  & -2  &  -1 & 0  & 0  &1&0\\ \hline
	\end{tabular}
	%\vspace{0pt}
	\label{tab:exmp_theta}
	%\vspace{-6mm}
\end{table}

		\begin{exmp}[continues=ex:running]
			Take a look again at the signal $\sstraj$ and an STL formula $\varphi=p\wedge q$ evaluation shown in Fig.~\ref{fig:ex1}.
			In Table~\ref{tab:exmp_theta} we consider an estimation of the \robplus{} asynchronous temporal robustness $\bm{\theta}^+_{\varphi}(\sstraj)$ following the MILP encoding procedure described above in Section~\ref{sec:milp_async} and formally defined in Alg.~\ref{alg:milp_theta}. 
			From Figure~\ref{fig:ex1}(b) one can see the values of $\chi_{p}(\sstraj, t)$ and $\chi_{q}(\sstraj, t)$ for all $t=1,\ldots,9$. Then following Alg.~\ref{alg:eta}, we construct $\bm{\theta}^+_p(\sstraj)$ and $\bm{\theta}^+_q(\sstraj)$ sequences shown in Fig.~\ref{fig:ex1}(d). 
			Finally, using \eqref{eq:conj}, we obtain that $\bm{\theta}^+_\varphi(\sstraj)$ sequence is indeed equal to $(1, 0, -2, -3, -2, -1, 0, 0, 1, 0)$ shown in Fig.~\ref{fig:ex1}(d).
			Therefore, the MILP encoding procedure leads to the same result as its estimation by the definition.
		\end{exmp}

	Finally, we define the overall MILP encoding of Problem~\ref{prob:synthesis} in case of $\objs= \thetap_{\varphi}$ in Alg.~\ref{alg:milp_asyn}. 
%	The function $\texttt{SYSTEM\_CONSTRAINTS}(x_0,\inpSig)$ defines linear constraints on the decision variable $u_t$ according to \eqref{eq:system} and such that $x_t\in X$. 
We state correctness of our encoding in the next proposition.

%%%%%%%%%%%%%%%%%%%%%%%%%%%%%%%%%%%%%%%%%%%%%%%%%%%%%%%%%%%%%%%%%%%%%%
\begin{algorithm}[t!]
	\SetAlgoLined
	\KwIn{Specification $\varphi$, initial state $x_0$, time horizon $H$, minimum asynchronous temporal robustness $\theta^*$}
	\KwOut{Control input sequence $\inpSig^* = (u^*_0,\ldots, u^*_{H-1})$}
	
	Set $\inpSig=(u_0,\ldots,u_{H-1})$ to be the decision variable

	$\,\inpSig^*=
	\underset{\inpSig}{\argmax}\quad \theta^{+}_\varphi(\sstraj,0)$
	%\vspace{-1pt}
	
	$\qquad\quad\ 
	\text{s.t.}\quad\ \ \,\theta^{+}_\varphi(\sstraj,0)\geq \theta^*>0$
	
	$\qquad\qquad\qquad\ \,\sstraj\defeq\texttt{SYSTEM\_CONSTRAINTS}(x_0,\inpSig)$
	
	$\qquad\qquad\qquad\ \,(\bm{\theta}^+_{\varphi}(\sstraj),\,\mathcal{Q})\defeq\texttt{MILP\_ASYN}(\varphi,\sstraj)$ // Sec.~\ref{sec:milp_async}, Alg.~\ref{alg:milp_theta}
	
	$\qquad\qquad\qquad\ \,\mathcal{Q},\ u_t\in U$ for $t=0,\hdots,H-1$
	
	\caption{MILP encoding of Problem~\ref{prob:synthesis} with the \robplus{} asynchronous temporal robustness objective $\objs\defeq \thetap_{\varphi}$}
	\label{alg:milp_asyn}
\end{algorithm}

%%%%%%%%%%%%%%%%%%%%%%%%%%%%%%%%%%%%%%%%%%%%%%%%%%%%%%%%%%%%%%%%%%%%%%
\begin{prop}
	\label{thm:milp_asyn_problem}
The temporally-robust control synthesis problem as defined in Problem~\ref{prob:synthesis} with asynchronous \robplus{} temporal robustness, i.e., $\objs = \thetap_\varphi$, is equivalent to an MILP as in Alg.~\ref{alg:milp_asyn}.
\end{prop}

We again remark that the case of asynchronous \robminus{} temporal robustness  $\objs = \thetam_\varphi$ can be handled almost identically.
\section{Experimental Results}
\label{sec:experiments}

In this section, we present three case studies. The first two case studies illustrate the tools presented in Sections \ref{sec:prop_sound}  and \ref{sec:robustness_shifts_time} to analyze temporal robustness. The third case study, on the other hand, illustrates the control design tools proposed in Section \ref{sec:control}. All simulations were performed on a computer with an Intel Core i7-9750H 6-core processor and 16GB RAM, running
Ubuntu 18.04. The MILPs were implemented in MATLAB
using YALMIP~\cite{lofberg2004yalmip} with Gurobi 9.1~\cite{gurobi} as a solver.

\subsection{Sine and Cosine Waves}

\begin{figure}[b!]
	\centering
	\includegraphics[width=0.7\textwidth]{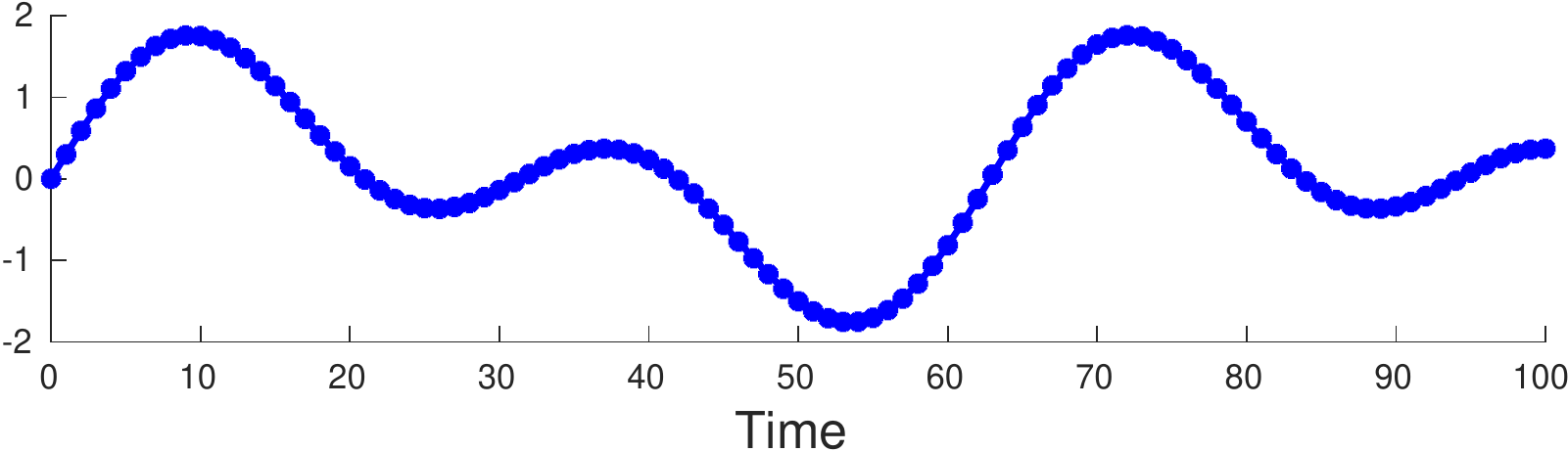}
	\caption{Discrete-time sine wave signal $\sstraj$ defined by \eqref{eq:sin}.}
	\label{fig:sin}
\end{figure}

We first analyze a simple sine wave signal similar to \cite{FainekosP09tcs} to illustrate our theoretical findings.
Assume we are given the following discrete-time signal:
\begin{equation}
	\label{eq:sin}
	x(t) \defeq \sin (a t) + \sin (2a t),\quad t=0,\ldots,100
\end{equation} 
where we set  $a\defeq0.1$, 
see Fig.~\ref{fig:sin} for an illustration. In the remainder, we consider four different STL specifications $\varphi_1$--$\varphi_4$.

\textbf{Specification $\varphi_1$.} First, we would like to verify that the signal $x(t)$ stays above the threshold $-0.2$ within the first $15$ time units. This can be  stated as the
	STL formula
	\begin{equation*}
	\varphi_1\defeq\always_{[0,\, 15]}\, p_1
	\end{equation*}
where $p_1\defeq x\geq -0.2$. Note that $x$ satisfies $\varphi_1$, i.e., $\chi_{\varphi_1}(\sstraj,0)=1$.
	\begin{figure}[t!]
	\centering
	\includegraphics[width=.7\textwidth]{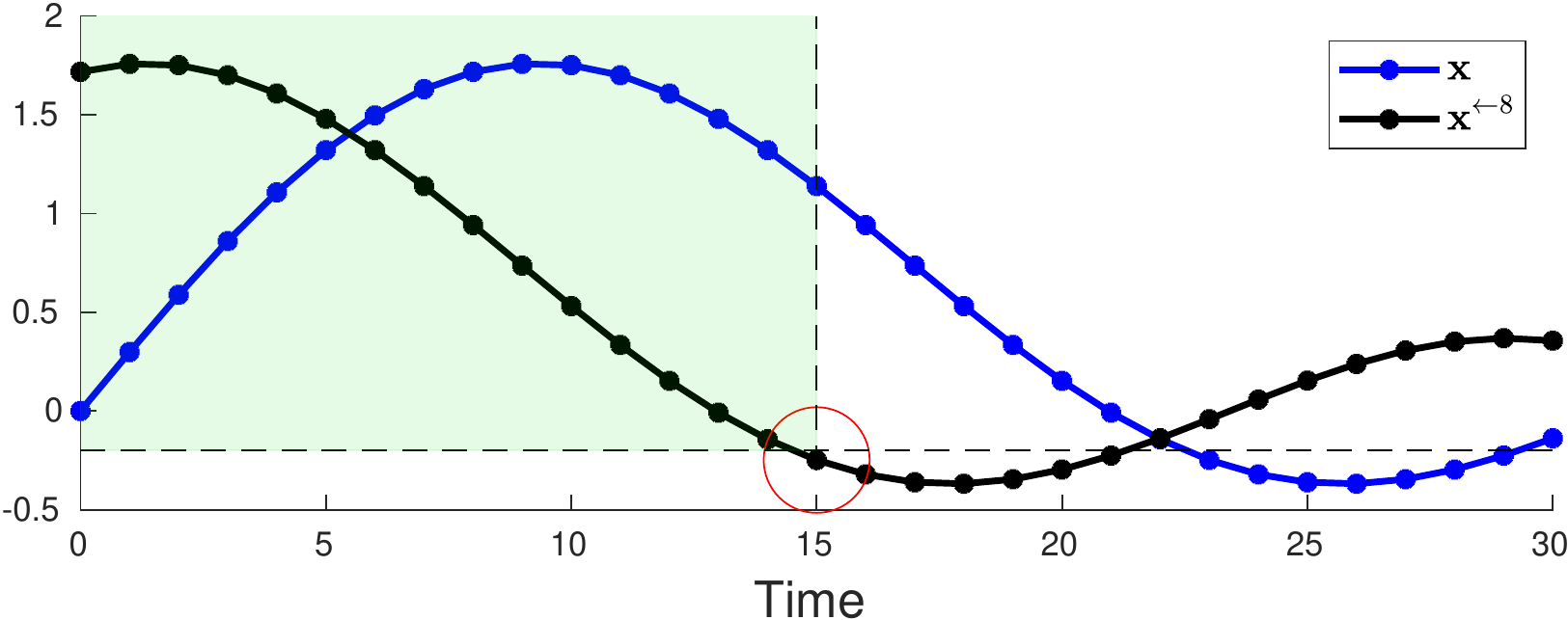}
	\caption{Zoomed in discrete-time sine wave signal $\sstraj$ (blue). The STL formula $\varphi_1$ is satisfied with asynchronous temporal robustness $\thetap_{\varphi_1}(\sstraj,0) = 7$. For the signal $\sstraj$ that is shifted by $8$ time units to the left (black), the STL formula $\varphi_1$ is violated. Violation is depicted in red where $\sstraj^{\leftarrow 8}(15)<-0.2$ so that $\chi_{p_1}(\sstraj^{\leftarrow 8},15)=-1$ and consequently $\chi_{\varphi_1}(\sstraj^{\leftarrow 8},0)=-1$.}
	\label{fig:sin_1}
\end{figure}
The \robplus{} asynchronous temporal robustness  is $\thetap_{\varphi_1}(\sstraj,0) = 7$ where the calculation of $\thetap_{\varphi_1}(\sstraj,0)$ took $0.002$ seconds. We remark that low computation times are obtained throughout this section. Note that since $\varphi_1\in\stlnnf(\wedge,\always_I)$ 
and $\chi_{\varphi_1}(\sstraj,0)=\po$, it holds that  
$\eta^{+}_{\varphi_1}(\sstraj,0)=\theta^{+}_{\varphi_1}(\sstraj,0)=7$ due to Lemma~\ref{lemma:p1_eq}. Consequently, we have that $\chi_{\varphi_1}(\sstraj^{\leftarrow \tstep},0)=1$ for each $\tstep\in\{0,\hdots,7\}$ due to 
%Theorems \ref{thm:shift_sync_stronger} and \ref{thm:shift_async}, in particular, 
\eqref{eq:shift_eta_disc} and \eqref{eq:shift_theta_disc}. In other words, we can shift the signal $x(t)$ by up to $7$ time units without violating the specification $\varphi_1$.
Due to 
%Theorem~\ref{thm:shift_sync_stronger}, 
\eqref{eq:shift_eta_disc} it also holds that $\chi_{\varphi_1}(\sstraj^{\leftarrow 8},0)=-1$ as illustrated in Fig.~\ref{fig:sin_1} for $\sstraj^{\leftarrow 8}$ being depicted in black. This means that once we shift by $8$ time units, we violate the specification $\varphi_1$.

\textbf{Specification $\varphi_2$.}  Let us now increase the timing constraint in $\varphi_1$ from 15 to 30 time units, i.e., consider instead the STL formula 
	$\varphi_2:=\always_{[0,\,30]}\, p_1$. 	
	Then it follows that the specification $\varphi_2$ is violated by $\sstraj$, i.e.,  $\chi_{\varphi_2}(\sstraj,0)=-1$. In fact, the \robplus{} asynchronous temporal robustness is $\theta^{+}_{\varphi_2}(\sstraj,0)=-6$. Unlike in the previous case, note that we now can not apply Lemma~\ref{lemma:p1_eq} anymore since $\chi_{\varphi_2}(\sstraj,0)=-1$ and it holds that $\theta^{+}_{\varphi_2}(\sstraj,0)\neq \eta^{+}_{\varphi_2}(\sstraj,0)$. Particularly, note that the synchronous \robplus{} temporal robustness is now $\eta^{+}_{\varphi_2}(\sstraj,0)=-70$.\footnote{Note here that we limit the time domain by the horizon  $H=100$. If the time domain was unbounded, then one can show that $\eta^{+}_{\varphi_2}(\sstraj,0)=-\infty$.} 

	\begin{figure}[t!]
	\centering
	\includegraphics[width=0.7\textwidth]{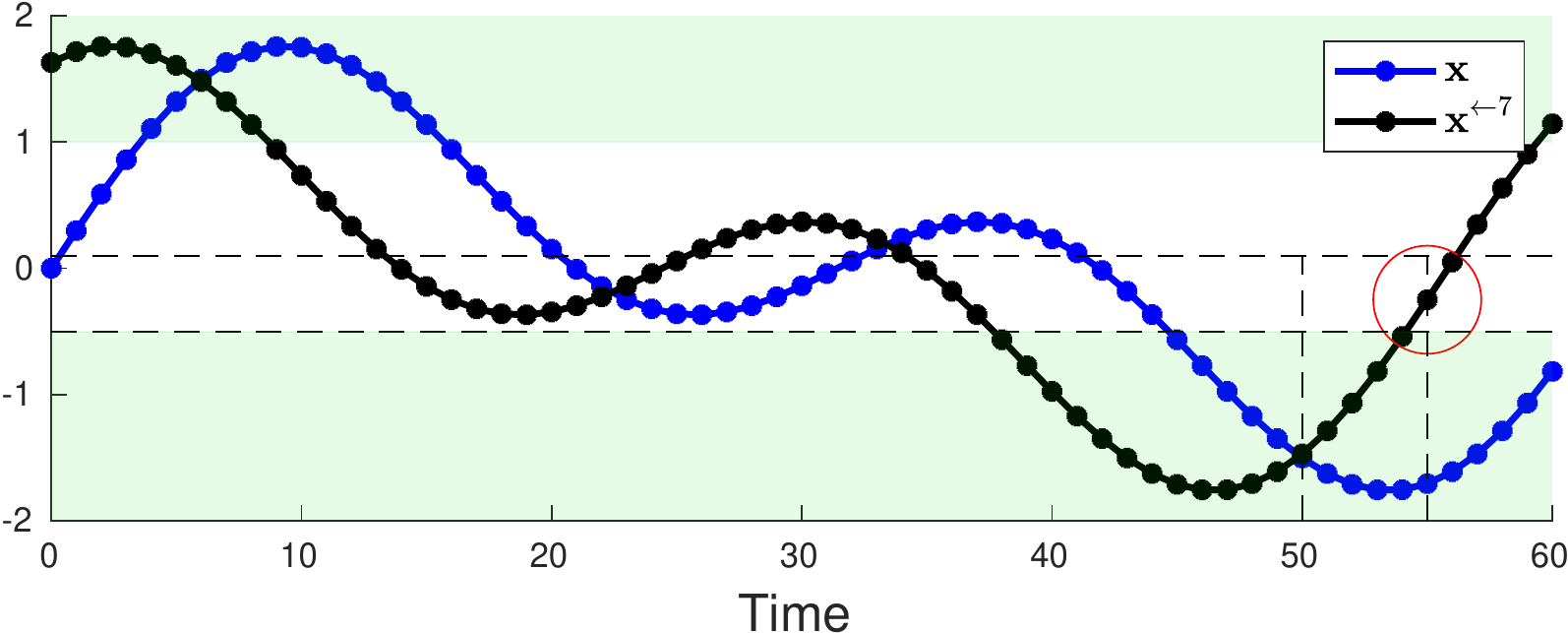}
	\caption{Zoomed in discrete-time sine wave signal $\sstraj$ (blue). The STL formula $\varphi_3$ is satisfied with asynchronous temporal robustness $\thetap_{\varphi_3}(\sstraj,0) = 6$. For the signal $\sstraj$ that is shifted by $7$ time units to the left (black), the STL formula $\varphi_3$ is violated. Violation is depicted in red, $\sstraj^{\leftarrow 7}(55)>-0.5$.}
	\label{fig:sin_2}
\end{figure}
	
\textbf{Specification $\varphi_3$.}  Now, we would like to verify the following property: if in the first 5 time units the value of the signal raises above $0.1$, then it should drop and stay below $-0.5$ from $45$ to $50$ time units. This is expressed by the STL formula
	\begin{equation*}
	\varphi_3 \defeq \always_{[0,\, 5]} \left(p_2 \Longrightarrow \always_{[45,\, 50]}\; p_3\right),	
	\end{equation*}
where $p_2\defeq x\geq 0.1$ and $p_3\defeq x \leq -0.5$.
	The \robplus{} asynchronous temporal robustness for this case is $\thetap_{\varphi_3}(\sstraj,0) = 6$ with a calculation time of $0.005$ seconds.
	Note that the \robplus{} synchronous temporal robustness for this case is also $\etap_{\varphi_3}(\sstraj,0) = 6$ with a computation time of $0.003$ seconds. Consequently, if we shift the signal $\sstraj$ by $7$ time units to the left, i.e. more than the calculated temporal robustness, then the specification $\varphi_3$ is not satisfied anymore, see Fig.~\ref{fig:sin_2}. In fact, 
	one can see that $\sstraj^{\leftarrow 7}(55)>-0.5$. This means that $\chi_{p_3}(\sstraj^{\leftarrow 7},55)=-1$ so that $\chi_{\always_{[45,50]}p_3}(\sstraj^{\leftarrow 7},5)=-1$ and consequently $\chi_{\varphi_3}(\sstraj^{\leftarrow 7},0)=-1$.

%\begin{figure}
%	\centering
%	\begin{subfigure}{.35\textwidth}
%		\centering
%		\includegraphics[width=0.95\textwidth]{figures/sin_1.pdf}
%		\caption{The STL formula $\varphi_1$ is satisfied with asynchronous temporal robustness $\thetap_{\varphi_1}(\sstraj,0) = 0.7$. For the trajectory $\sstraj$ that is shifted by $0.8$ seconds to the left (black), the STL formula $\varphi_1$ is violated. Violation is depicted in red, $\sstraj^{\leftarrow 0.8}(1.5)<-0.2$, thus, $\chi_{p_1}(\sstraj^{\leftarrow 0.8},1.5)=-1$ and therefore, $\chi_{\varphi_1}(\sstraj^{\leftarrow 0.8},0)=-1$.}
%		\label{fig:sub1}
%	\end{subfigure}%
%\quad
%	\begin{subfigure}{.6\textwidth}
%		\centering
%		\includegraphics[width=0.95\textwidth]{figures/sin_2.pdf}
%		\caption{The STL formula $\varphi_1$ is satisfied with asynchronous temporal robustness $\thetap_{\varphi_3}(\sstraj,0) = 0.6$ seconds. For the trajectory $\sstraj$ that is shifted by $0.7$ seconds to the left (black), the STL formula $\varphi_1$ is violated. Violation is depicted in red, $\sstraj^{\leftarrow 0.7}(5.5)>-0.5$.}
%		\label{fig:sub2}
%	\end{subfigure}
%	\caption{Zoomed in discrete-time sine wave trajectory $\sstraj$ (blue). }
%	\label{fig:test}
%\end{figure}

\begin{figure}[t!]
	\centering
	\includegraphics[width=0.7\textwidth]{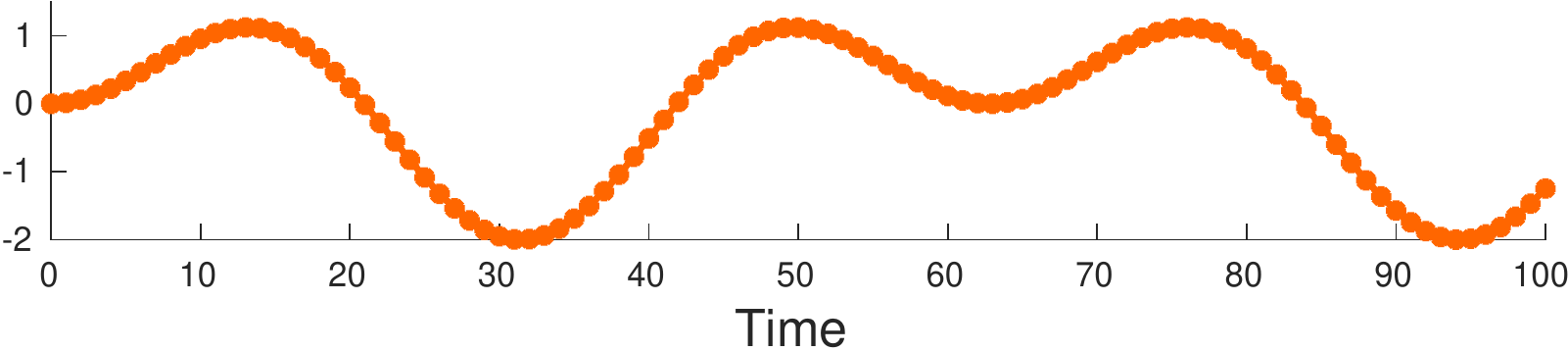}
	\caption{Discrete-time cosine wave signal $\sstraj\coord{2}$ defined by \eqref{eq:cos}.}
	\label{fig:x2}
\end{figure}

\textbf{Specification $\varphi_4$.}  Consider now a specification $\varphi_4$ that depends on a two-dimensional signal. The first component $x\coord{1}(t)$ is  equivalent to $x(t)$ in \eqref{eq:sin}, recall Fig.~\ref{fig:sin}, while the second component $x\coord{2}(t)$ is defined as 
	\begin{equation}
		\label{eq:cos}
		x\coord{2}(t) \defeq \cos (a t) - \cos (2a t),\quad t=0,\ldots,100
	\end{equation} 
	where we set $a\defeq0.1$, see Fig.~\ref{fig:x2} for an illustration. We also redefine $\sstraj\defeq(\sstraj\coord{1},\ \sstraj\coord{2})$.
 We want to verify that during the first ten time units either $x\coord{1}$ is non-positive or $x\coord{2}$ is non-negative which is expressed as
	\begin{equation*}
		\varphi_4 \defeq \always_{[0,\,10]} (p_4 \vee p_5)
	\end{equation*}
where $p_4 \defeq x\coord{1} \leq 0$ and $p_5\defeq x\coord{2} \geq 0$.

\begin{figure}[t!]
	\centering
	\begin{subfigure}{.45\textwidth}
		\centering
		\includegraphics[width=0.95\textwidth]{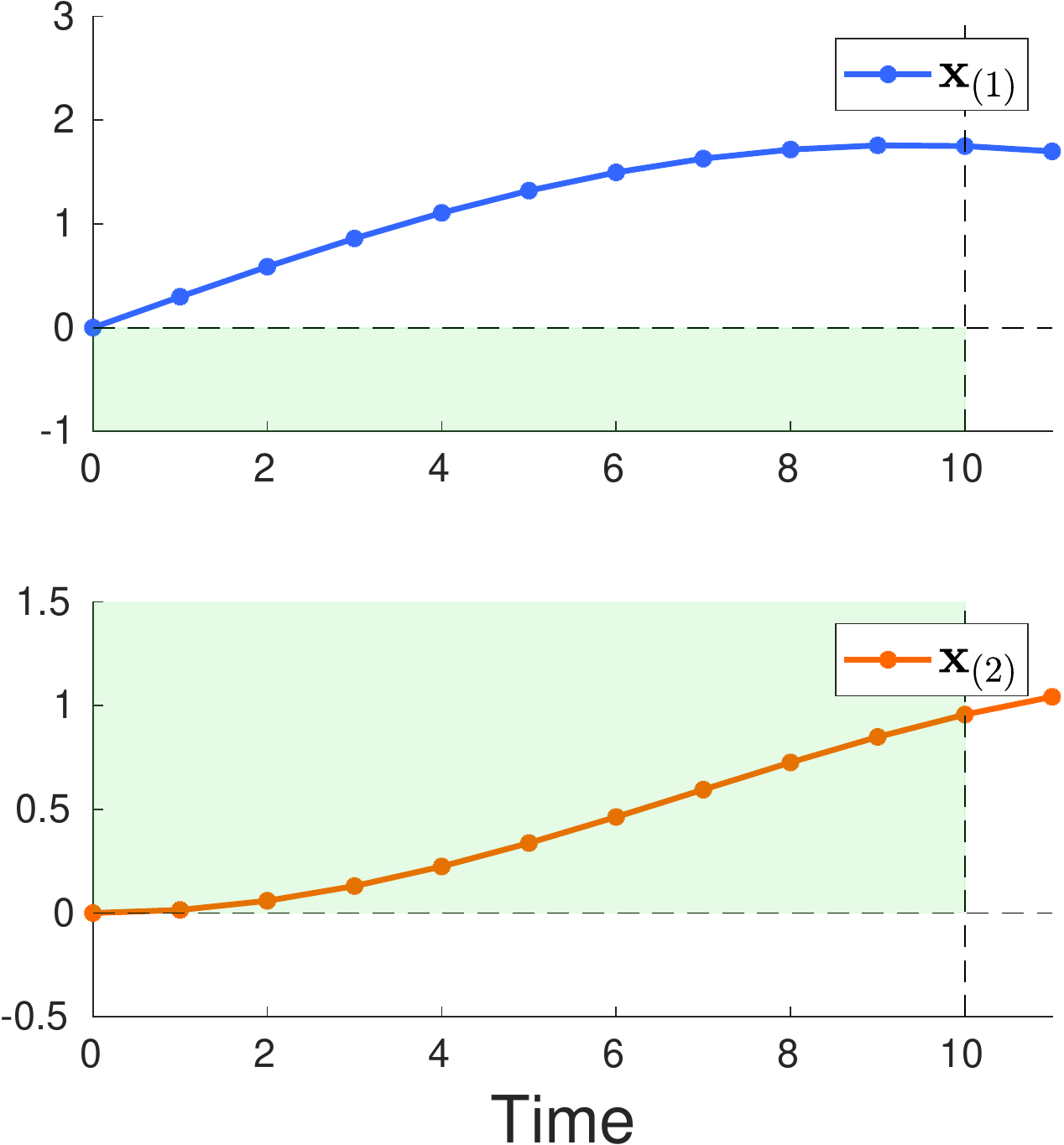}
		\caption{Default signal $\sstraj$. Specification $\varphi_4$ is satisfied.\vspace{10pt}}
		\label{fig:sin_default}
	\end{subfigure}%
	\quad
	\begin{subfigure}{.45\textwidth}
		\centering
		\includegraphics[width=0.95\textwidth]{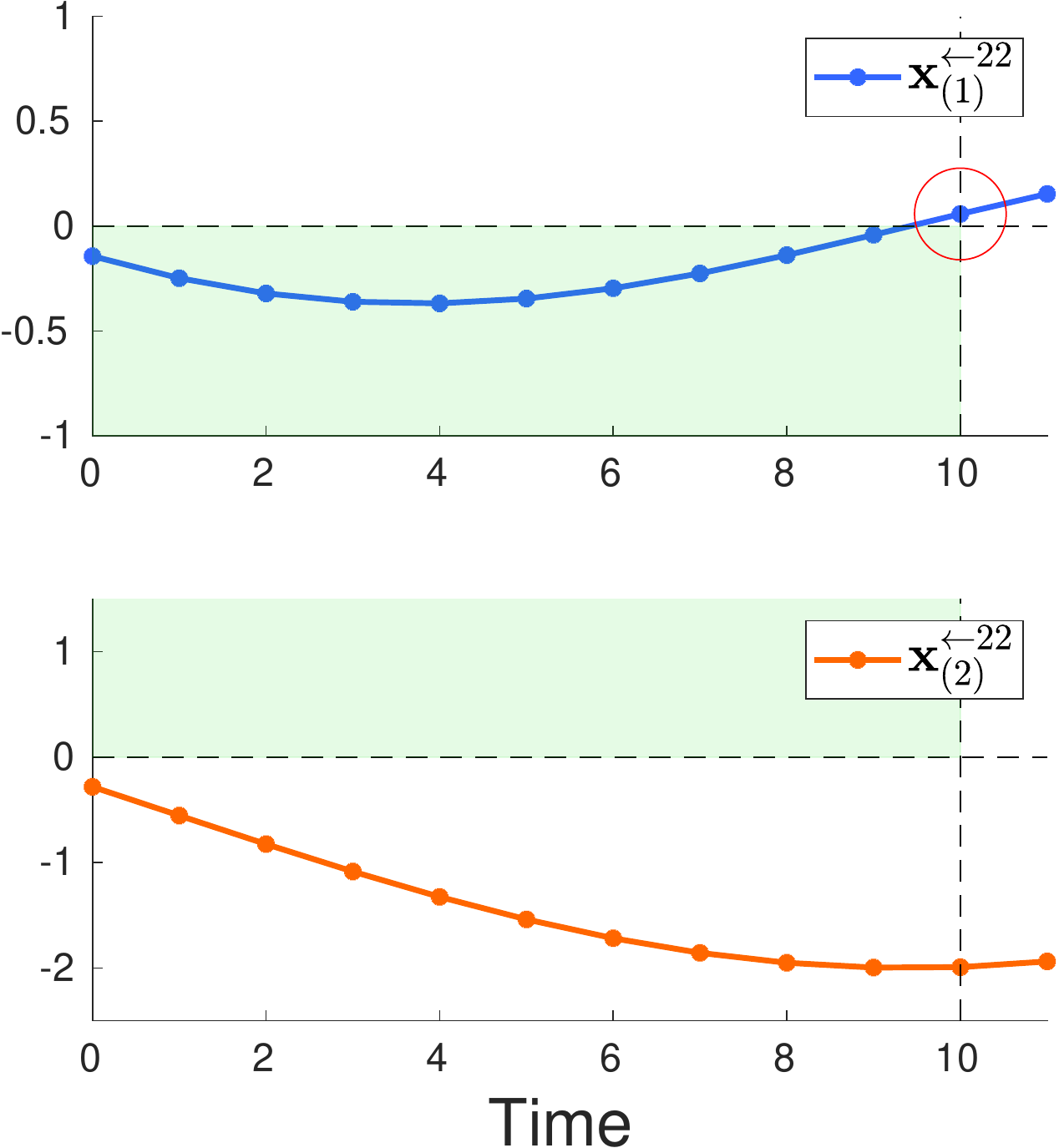}
		\caption{Synchronous early signal $\sstraj^{\leftarrow \tstep}$, where $\tstep=22$. Specification $\varphi_4$ is violated.}
		\label{fig:sin_asyn_22}
	\end{subfigure}
	\caption{Zoomed in analysis of specification $\varphi_4$ satisfaction and violation.}
	\label{fig:phi4_analysis}
\end{figure}

%\begin{figure}[t!]
%	\centering
%	\begin{subfigure}{.45\textwidth}
%		\centering
%		\includegraphics[width=0.95\textwidth]{figures/x2_no_sh.pdf}
%		\caption{Default satisfaction: no shift.}
%		\label{fig:x2_no_sh}
%	\end{subfigure}%
%	\quad
%	\begin{subfigure}{.45\textwidth}
%		\centering
%		\includegraphics[width=0.95\textwidth]{figures/x2_sy_1.pdf}
%		\caption{Synchronous shift within $\etap_{\varphi_4}$ bounds.}
%		\label{fig:x2_sy_1}
%	\end{subfigure}
%	\caption{Zoomed in analysis of specification $\varphi_4$ satisfaction or violation. \textcolor{red} {Change $_1$ to $x\coord{1}$}}
%	\label{fig:x2_analysis1}
%\end{figure}

%\begin{figure}[t!]
%	\centering
%	\begin{subfigure}{.45\textwidth}
%		\centering
%		\includegraphics[width=0.95\textwidth]{figures/x2_as.pdf}
%		\caption{Asynchronous shift outside $\thetap_{\varphi_4}$ bounds.}
%		\label{fig:x2_as}
%	\end{subfigure}%
%	\quad
%	\begin{subfigure}{.45\textwidth}
%		\centering
%		\includegraphics[width=0.95\textwidth]{figures/x2_sy_0.pdf}
%		\caption{Synchronous shift outside $\etap_{\varphi_4}$ bounds.}
%		\label{fig:x2_sy_0}
%	\end{subfigure}
%	\caption{Zoomed in analysis of specification $\varphi_4$ satisfaction or violation. \textcolor{red} {Change $_1$ to $x\coord{1}$}}
%	\label{fig:x2_analysis2}
%\end{figure}

The \robplus{} synchronous temporal robustness for this case is $\etap_{\varphi_4}(\sstraj,0) = 21$, while the \robplus{} asynchronous temporal robustness is $\thetap_{\varphi_4}(\sstraj,0) = 10$. Since both values are positive,  we can conclude that $\varphi_4$ is satisfied due to Theorems~\ref{thm:sat_eta} and \ref{thm:sat_theta}, see Fig.~\ref{fig:phi4_analysis}(a). Due to 
%Theorem~\ref{thm:shift_sync_stronger_disc}, 
\eqref{eq:shift_eta_disc}
for any synchronous early signal $\sstraj^{\leftarrow \tstep}$ where $\tstep \leq |\eta^{+}_{\varphi_4}(\sstraj,0)|=21$ it holds that the formula $\varphi_4$ will still be satisfied. However, the formula $\varphi_4$ will not be satisfied anymore for $\sstraj^{\leftarrow 22}$ which is highlighted in
% For example, in Fig.~\ref{fig:x2_sy_1} we show how $\sstraj^{\leftarrow 11}$ satisfies $\varphi_4$. 
 Fig.~\ref{fig:phi4_analysis}(b). One can indeed see that the specification $\varphi_4$ is  violated by $\sstraj^{\leftarrow 22}$ since $\sstraj\coord{1}^{\leftarrow 22}(10)>0$ and $\sstraj\coord{2}^{\leftarrow 22}(0),\ldots,\sstraj\coord{2}^{\leftarrow 22}(10)<0$ so that
$\chi_{\varphi_4}(\sstraj^{\leftarrow 22},0)=-1$. Analogously, we can reason about asynchronous temporal robustness. Since $\thetap_{\varphi_4}(\sstraj,0) = 10$, then  for any asynchronous early signal $\sstraj^{\leftarrow \bar{\tstep}}$ where $\max(\tstep_{p_4},\tstep_{p_5}) \leq |\theta^{+}_{\varphi_4}(\sstraj,0)|=10$ it holds that the formula $\varphi_4$ will still be satisfied due to 
%Theorem~\ref{thm:shift_async_disc}.
\eqref{eq:shift_theta_disc}.
% In Fig.~\ref{fig:x2_as} we show that with a shift outside of these bounds the specification is violated: in case when $h_1=0$ and $h_2 =11 > |\theta^{+}_{\varphi_4}(\sstraj,0)|$ formula $\varphi_4$ does not hold since $x_2(1)<0$, therefore, $\chi_{p_5}(\sstraj, 10)=-1$ and thus, $\varphi_4$ does not hold at time point 0.  

\subsection{Multi-Agent Coordination}

\begin{figure}[t!]
	\centering
	\begin{subfigure}{.45\textwidth}
		\centering
		\includegraphics[width=0.95\textwidth]{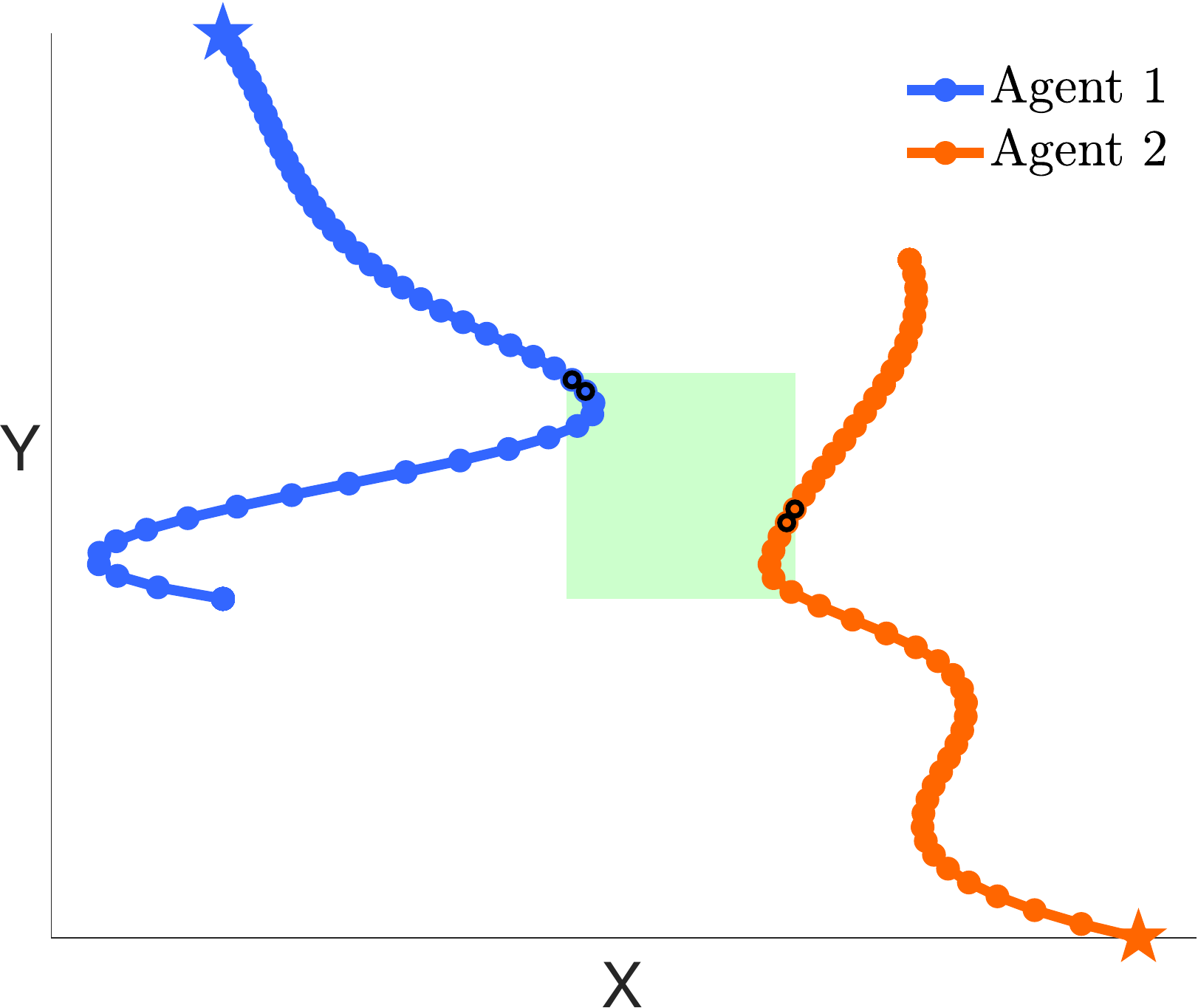}
		\caption{Map and nominal signals.}
		\label{fig:x2_no_sh1}
	\end{subfigure}%
	\quad
	\begin{subfigure}{.45\textwidth}
		\centering
		\includegraphics[width=0.95\textwidth]{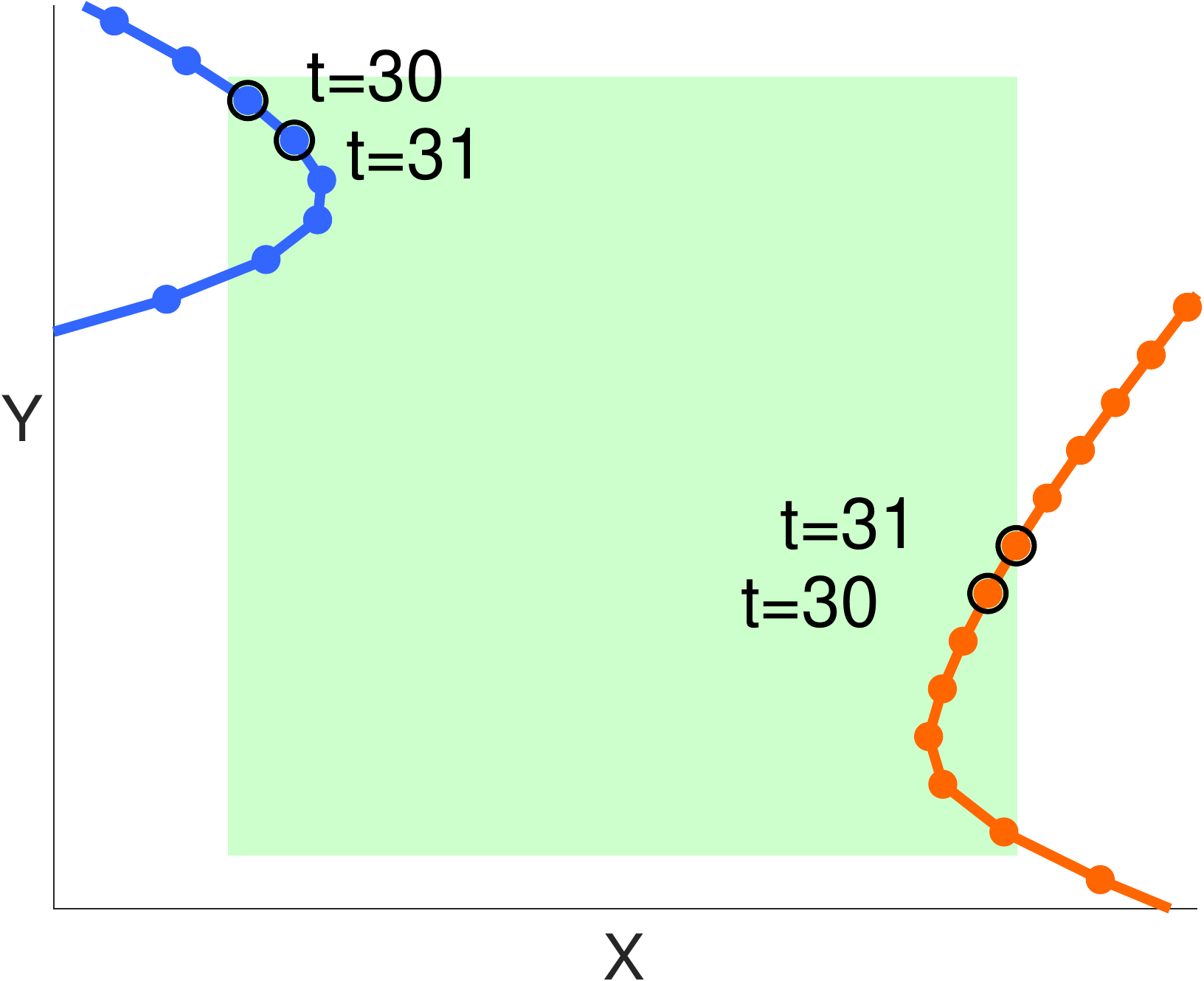}
		\caption{Zoomed in map.}
		\label{fig:x2_sy_12}
	\end{subfigure}
	\caption{Mission workspace for the multi-agent coordination scenario $\varphi_{\text{coord}}$. Initial positions of the agents marked by $\star$, goal set is given as a green rectangle. Specification $\varphi_{\text{coord}}$ is satisfied by the given nominal signals since there are two time instances when both of the agents are within the goal set (black circles).}
	\label{fig:exchange}
\end{figure}

In this example, we consider a multi-agent scenario where two agents  are supposed to coordinate and exchange goods within a designated  goal set and within 50 time units. This scenario is
captured in the following specification:
\begin{equation*}
	\varphi_{\text{coord}} \defeq \eventually_{[0, 50]} \left( x\coordup{1}\in \textit{Goal}\ \wedge\ x\coordup{2} \in \textit{Goal}\right)
\end{equation*}
where $x\coordup{d}\in\Re^2$, $d\in\{1,2\}$ denotes the position of the agent $d$ in a two-dimensional space.
Fig.~\ref{fig:exchange} depicts the goal set $\textit{Goal}$ and the given discrete-time signal $\sstraj\defeq(\sstraj\coordup{1}, \sstraj\coordup{2})$ which has $100$ sampling points. Each $\sstraj\coordup{d}$ denotes the position signal of the agent $d$. 
One can notice that the specification $\varphi_{\text{coord}}$ is satisfied by the nominal signal because there are two time instances ($t=30$ and $t=31$) when both of the agents are within the goal set within the first 50 time units, see Fig.~\ref{fig:exchange}(b).

\begin{figure}[t!]
	\centering
	\begin{subfigure}{.45\textwidth}
		\centering
		\includegraphics[width=0.95\textwidth]{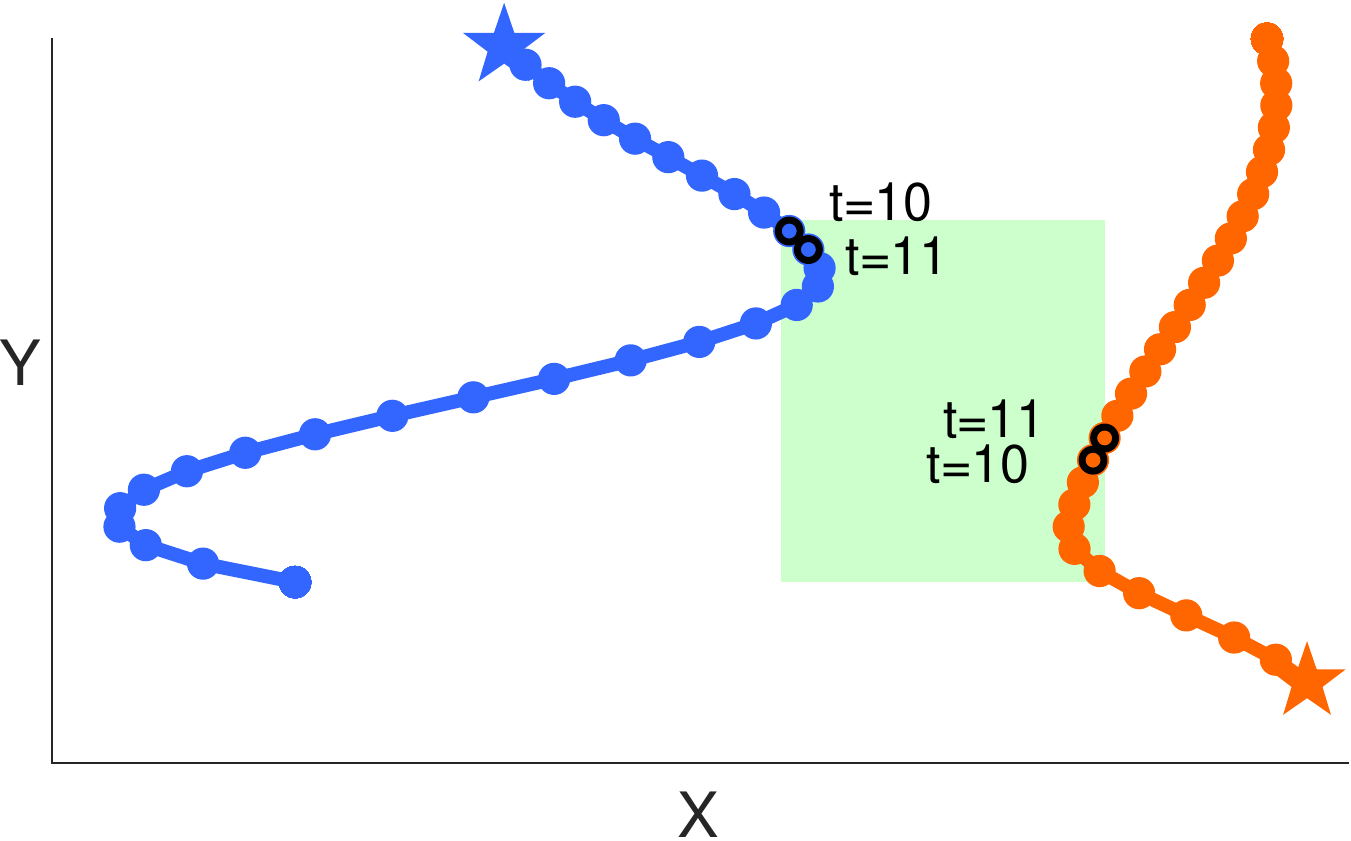}
		\caption{$\mathbf{x}^{\leftarrow \tstep},\ \ \tstep=20$. Specification $\varphi_{\text{coord}}$ is satisfied by $\mathbf{x}^{\leftarrow 20}$ since there are two time instances when both of the agents are within the goal set.}
		\label{fig:h20}
	\end{subfigure}%
	\quad
	\begin{subfigure}{.46\textwidth}
		\centering
		\includegraphics[width=0.95\textwidth]{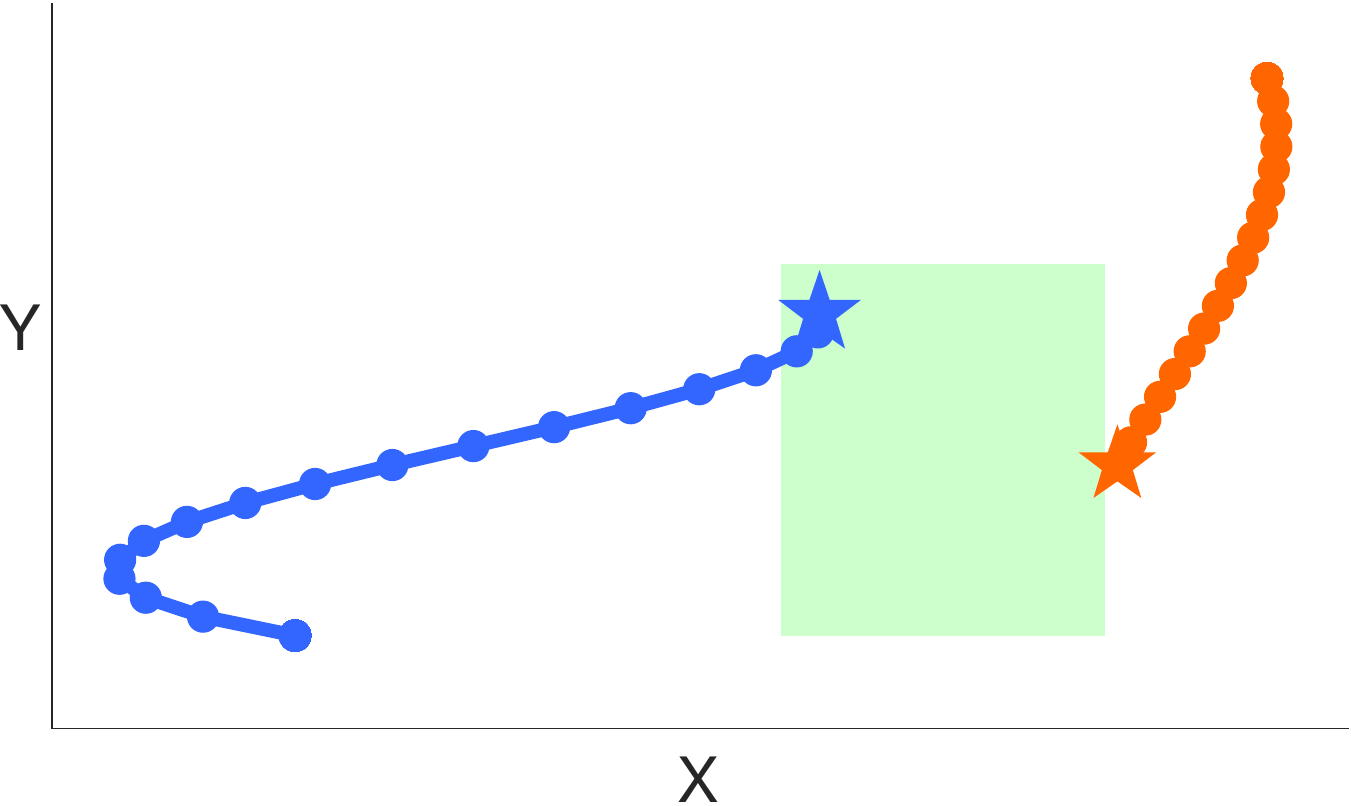}
		\caption{$\mathbf{x}^{\leftarrow \tstep},\ \ \tstep=32$. Specification $\varphi_{\text{coord}}$ is violated by $\mathbf{x}^{\leftarrow 32}$ since agents are never together in the goal set.}
		\label{fig:h32}
	\end{subfigure}
	\caption{Synchronous early signals.}
	\label{fig:h_synch}
\end{figure}

The calculated \robplus{} synchronous temporal robustness for this case is $\etap_{\varphi_{\text{coord}}}(\sstraj,0) = 31$ time units. Due to Theorem~\ref{thm:sat_eta} the specification $\varphi_{\text{coord}}$ is indeed satisfied, i.e., $\chi_{\varphi_{\text{coord}}}(\sstraj, 0)=\po$.
Then due to 
%Theorem~\ref{thm:shift_sync_stronger_disc}
\eqref{eq:shift_eta_disc}, for any synchronous $\tstep$-early signal with $\tstep\leq31$ time units
the specification $\varphi_{\text{coord}}$ will still be satisfied. For the signal $\sstraj^{\leftarrow 32}$, however, the specification $\varphi_{\text{coord}}$ will change its satisfaction to violation, 
i.e.
\begin{equation*}
	\forall \tstep\in[0,31],\   
	\chi_{\varphi_{\text{coord}}}(\sstraj^{\leftarrow \tstep}, 0)=\po\quad  \text{and} \quad
	\chi_{\varphi_{\text{coord}}}(\sstraj^{\leftarrow 32}, 0)=-1.
\end{equation*} 
Fig.~\ref{fig:h_synch} depicts two examples of synchronous early signals. Fig.~\ref{fig:h_synch}(a) shows the signal $\mathbf{x}^{\leftarrow 20}$ for which the specification $\varphi_{\text{coord}}$ is satisfied, and Fig.~\ref{fig:h_synch}(b) shows $\mathbf{x}^{\leftarrow 32}$ for which $\varphi_{\text{coord}}$ is violated.

\begin{figure}[t!]
	\centering
	\begin{subfigure}{.45\textwidth}
		\centering
		\includegraphics[width=0.95\textwidth]{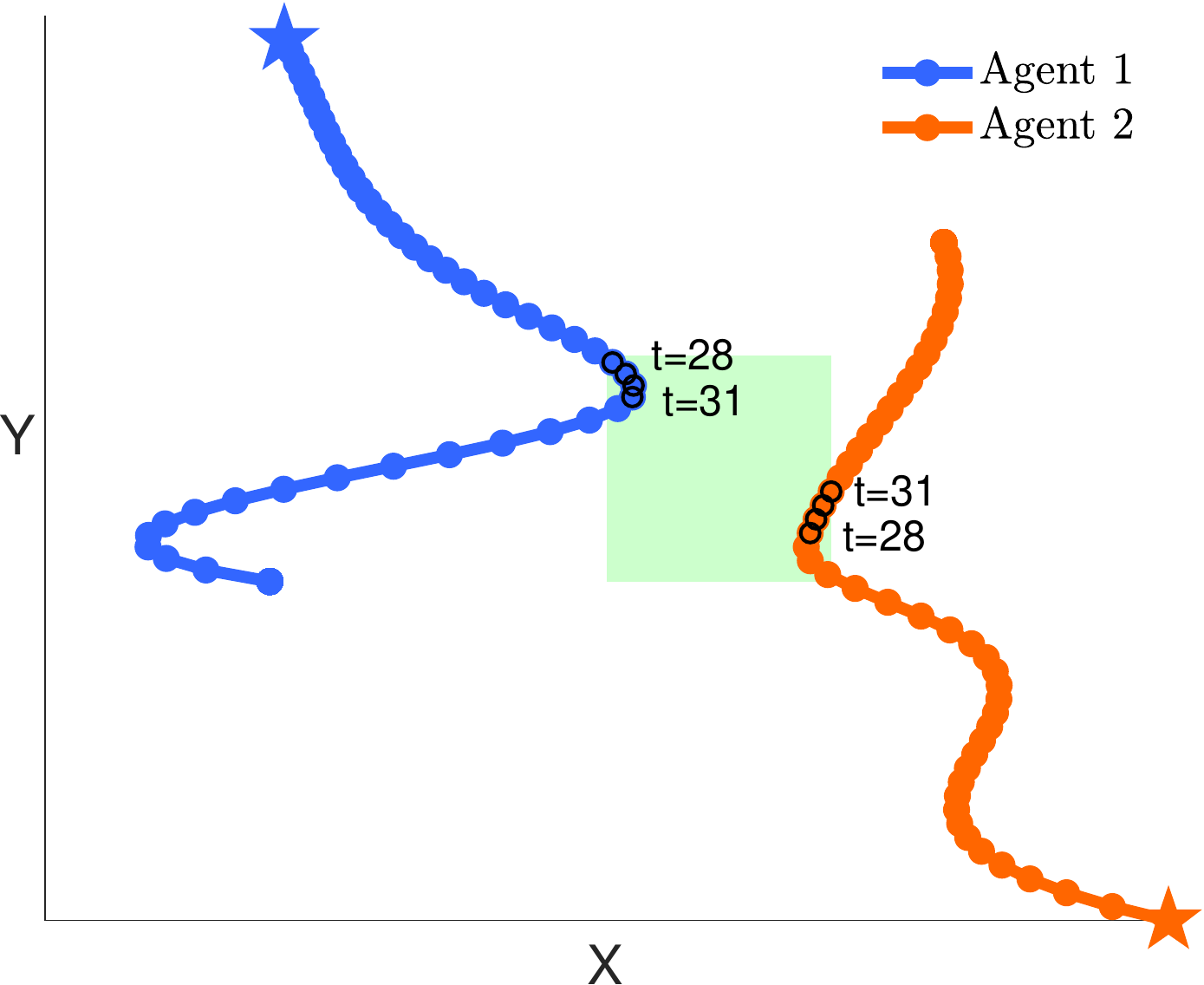}
		\caption{$\mathbf{x}^{\leftarrow \bar{\tstep}},\ \ \bar{\tstep}=(2,0)$.  Specification $\varphi_{\text{coord}}$ is satisfied by $\mathbf{x}^{\leftarrow (2,0)}$ since there are four time instances when both of the agents are within the goal set.}
		\label{fig:h2-0}
	\end{subfigure}%
	\quad
	\begin{subfigure}{.45\textwidth}
		\centering
		\includegraphics[width=0.95\textwidth]{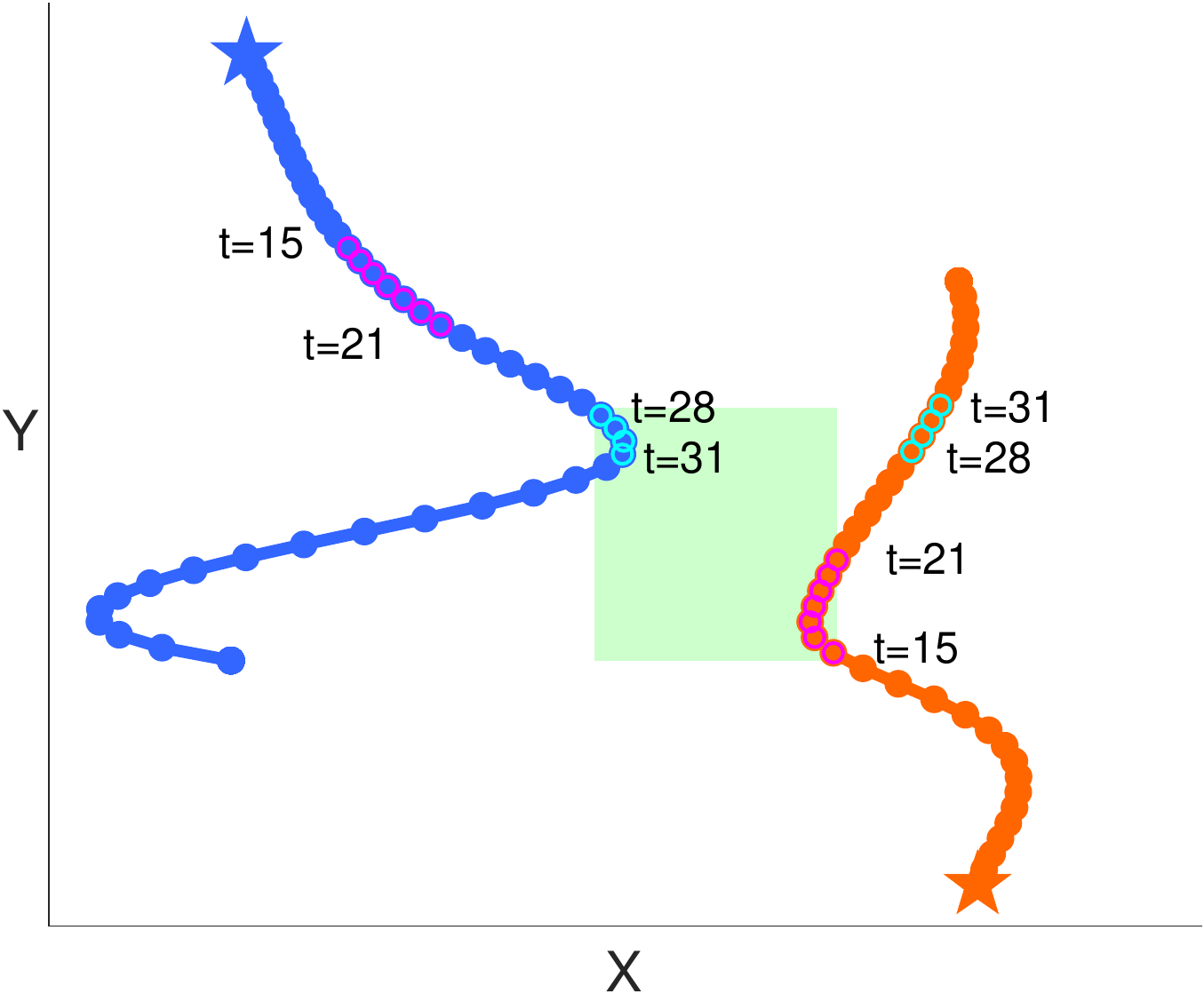}
		\caption{$\mathbf{x}^{\leftarrow \bar{\tstep}},\ \ \bar{\tstep}=(2,10)$. Specification $\varphi_{\text{coord}}$ is violated by $\mathbf{x}^{\leftarrow (2,10)}$ since agents are never together in the goal set.}
		\label{fig:h2-10}
	\end{subfigure}
	\caption{Asynchronous early signals.}
	\label{fig:h_asynch}
\end{figure}

Due to Theorem~\ref{thm:theta_bounded}, one can immediately see that $\thetap_{\varphi_{\text{coord}}}(\sstraj,0) \leq 31$. Now, since $\varphi\not\in\stlnnf(\wedge,\always_I)$ it does not necessarily hold that $\thetap_{\varphi_{\text{coord}}}(\sstraj,0) = 31$. In fact, we find that
$\thetap_{\varphi_{\text{coord}}}(\sstraj,0) = 1$ so that only  asynchronous time shifts by up to $1$ time unit to the left will not violate the specification, i.e. 
$\forall \bar{\tstep}\in\{(0, 0), (0, 1), (1, 0), (1,1)\}$ it holds that $\chi_{\varphi_{\text{coord}}}(\sstraj^{\leftarrow \bar{\tstep}}, 0)=\po$. As discussed before, 
%Theorem \ref{thm:shift_async_disc} 
\eqref{eq:shift_theta_disc} only provides a sufficient condition so that  asynchronous time shifts larger than $\thetap_{\varphi_{\text{coord}}}(\sstraj,0)$ do not necessarily result in a violation of  $\varphi_{\text{coord}}$. Indeed, Fig.~\ref{fig:h_asynch}(a) depicts $\mathbf{x}^{\leftarrow (2, 0)}$ for which the specification $\varphi_{\text{coord}}$ is satisfied since there are four time instances (black circles) for which both agents are inside the goal set. On the other hand, a simple search over the whole space of possible asynchronous time shifts let us find a signal when the specification $\varphi_{\text{coord}}$ becomes violated, see Fig.~\ref{fig:h_asynch}(b). For $\mathbf{x}^{\leftarrow (2, 10)}$, the agents are never together in the goal set: when agent $2$ is inside the goal set (magenta circles), agent $1$ is still on the way to the goal set and the moment agent $1$ reaches the goal (cyan circles), agent $2$ already leaves it.

%\begin{figure}[t!]
%	\centering
%	\begin{subfigure}{.24\textwidth}
	%		\centering
	%		\includegraphics[width=0.95\textwidth]{figures/h20.pdf}
	%		\caption{$\mathbf{x}^{\leftarrow h},\ \ h=20$.}
	%		\label{fig:x2_no_sh1}
	%	\end{subfigure}%
%	\ 
%	\begin{subfigure}{.24\textwidth}
	%		\centering
	%		\includegraphics[width=0.95\textwidth]{figures/h32.pdf}
	%		\caption{$\mathbf{x}^{\leftarrow h},\ \ h=32$.}
	%		\label{fig:x2_sy_12}
	%	\end{subfigure}
%\ 
%\begin{subfigure}{.24\textwidth}
%	\centering
%	\includegraphics[width=0.95\textwidth]{figures/h2-0.pdf}
%	\caption{$\mathbf{x}^{\leftarrow \bar{h}},\ \ \bar{h}=(2,0)$.}
%	\label{fig:x2_no_sh1}
%\end{subfigure}%
%\ 
%\begin{subfigure}{.24\textwidth}
%	\centering
%	\includegraphics[width=0.95\textwidth]{figures/h2-10.pdf}
%	\caption{$\mathbf{x}^{\leftarrow \bar{h}},\ \ \bar{h}=(2,10)$.}
%	\label{fig:x2_sy_12}
%\end{subfigure}
%	\caption{Zoomed in analysis of specification $\varphi_4$ satisfaction or violation.}
%	\label{fig:x2_analysis12}
%\end{figure}

%\input{chapters/test_truck}
%\input{chapters/test_circuit}
\subsection{Multi-Agent Surveillance}

\newcommand{\pos}{pos}
\newcommand{\vel}{vel}

\begin{figure}[t!]
	\centering
	\includegraphics[width=0.9\textwidth]{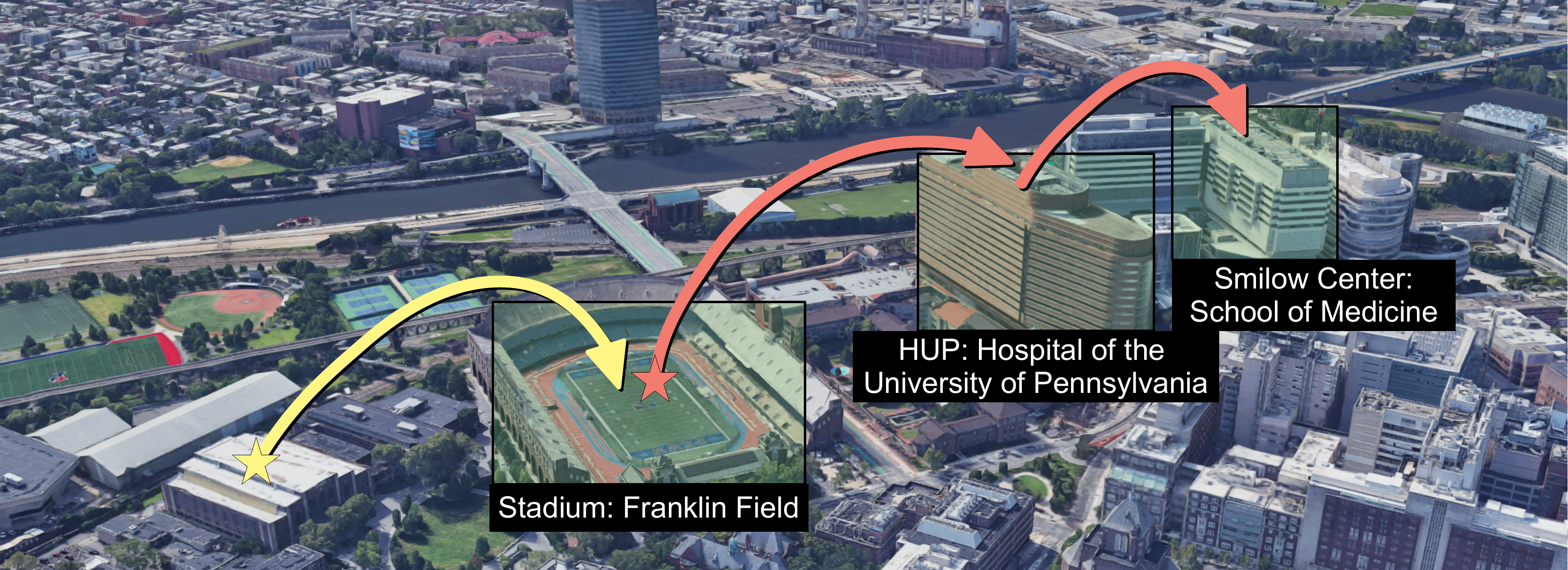}
	\caption{Multi-Agent Surveillance case study map. Initial positions for the two UAVs marked by $\star$.}
	\label{fig:penn_map}
\end{figure}

In this case study we illustrate the control design tools proposed in Section \ref{sec:control}. 
We consider two identical unmanned aerial vehicles (UAVs) and three places of interest on the campus of the University of Pennsylvania, see Fig.~\ref{fig:penn_map} for an overview. In particular, these places of interest are  the Franklin Field (\textit{Stadium}), the Hospital of the University of Pennsylvania (\textit{HUP}), and the School of Medicine Smilow Center (\textit{Smilow}). The UAVs are tasked with a mission where they collaboratively surveil the \textit{Stadium} while one of the UAVs is required to pick-up specific items in the \textit{HUP} and then drop them off at the \textit{Smilow}. 

We assume that the UAVs operate in a two dimensional workspace. 
For each UAV $d\in\{1,2\}$,  
let the state be $x\coordup{d}\defeq(\pos\coordup{d}, \vel\coordup{d})\in\Re^4$ where $\pos\coordup{d}$ and $\vel\coordup{d}$ are the two-dimensional position and velocity, respectively, and let the control input be $u\coordup{d} \in\Re^2$.
We denote the full state of the system
as $x \defeq (x\coordup{1}, x\coordup{2})$ and the stacked control input as $u \defeq (u\coordup{1},u\coordup{2})$.
The
linear state-space representation of the system is driven by discrete-time double integrator
dynamics and is written as follows:
\begin{equation*}
	x_{t+1}\defeq Ax_t+B u_t, \quad ||u_t||_\infty\leq 20,
\end{equation*}
where
$A\defeq I_4 \otimes \begin{bmatrix}
	1&0.1\\0&1
\end{bmatrix}$ and $B\defeq I_4 \otimes\begin{bmatrix}
	0.005\\0.1
\end{bmatrix}$ with $I_n$ being the identity matrix of
dimension $n\times n$ and with $\otimes$ denoting the Kronecker product.
We set the  time horizon to $H\defeq 50$. The
initial positions of the agents are set to $\pos_{0}\coordup{1}\defeq(1, 3)$ and $\pos_{0}\coordup{2}\defeq(8,3)$ (this UAV starts from within the \textit{Stadium}), while the initial velocities are set to zero. 

\begin{figure}[t!]
	\centering
	\includegraphics[width=0.9\textwidth]{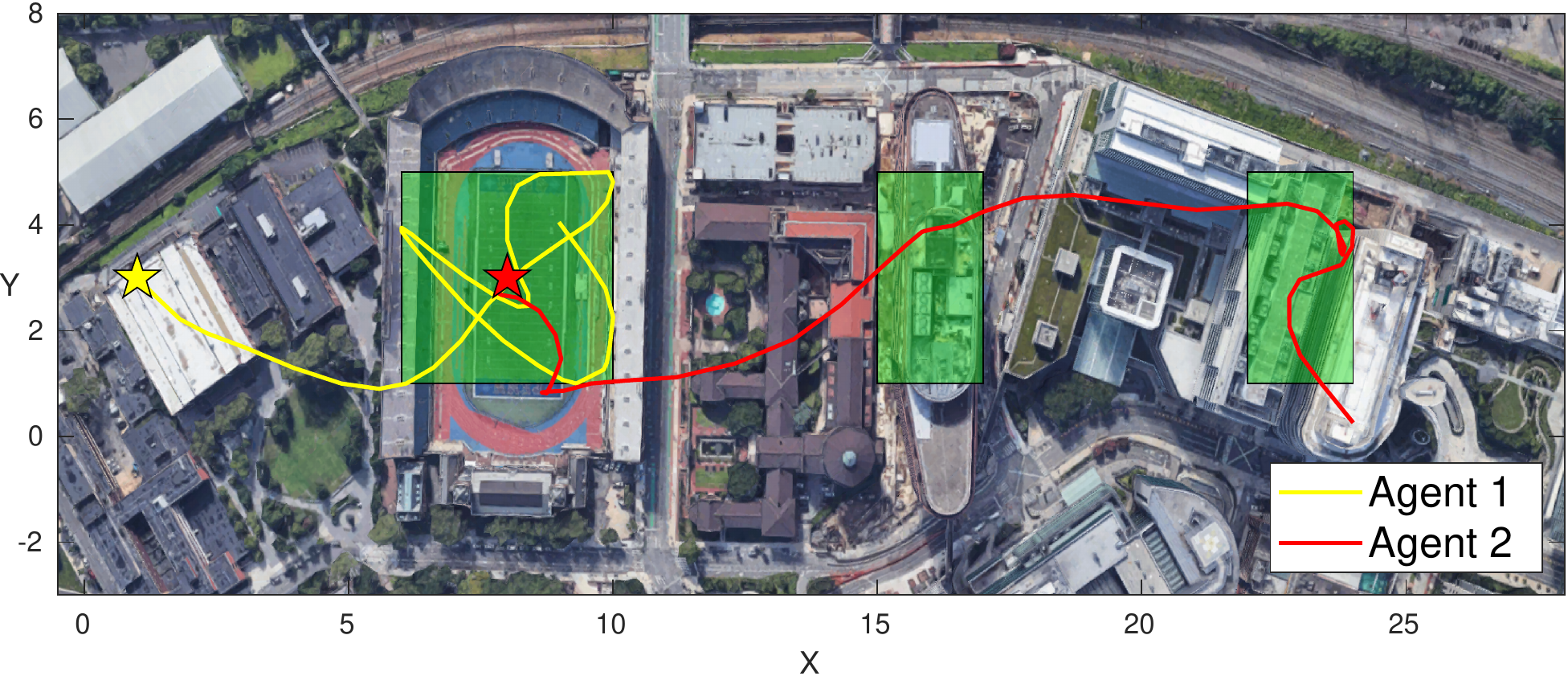}
	\caption{The nominal UAV trajectories under the optimal
		control inputs generated by solving Problem~\ref{prob:synthesis} with the  \robplus{} synchronous temporal robustness as an objective.  The achieved robustness 
		is $\etap_{\varphi_{\text{MAS}}}(\trajsync,0) = 10$ time steps. The initial positions are marked by $\star$.}
	\label{fig:penn_map_eta}
\end{figure}

The overall multi-agent surveillance mission is defined as:
\begin{equation}
\varphi_{\text{MAS}} \defeq \varphi_{\text{surveil}} 
\wedge \varphi_{\text{pick-up}} \wedge
\varphi_{\text{drop-off}}
\end{equation}
where we explain $\varphi_{\text{surveil}} $, $ \varphi_{\text{pick-up}} $, and $\varphi_{\text{drop-off}}$ in the remainder. The surveillance sub-mission $\varphi_{\text{surveil}} $ requires the \textit{Stadium} region
to be surveilled by at least one UAV for the first 35
time steps. Formally,
it is defined as:
\begin{equation*}
	\varphi_{\text{surveil}} \defeq \always_{[0,\, 35]} (\pos\coordup{1} \in \textit{Stadium}\ \vee\ 
	 \pos\coordup{2} \in \textit{Stadium}).
\end{equation*}
where $\textit{Stadium}$ is a rectangle, i.e., $\pos\coordup{d}\in \textit{Stadium}$ is a conjunction of four linear predicates $x\coordup{d} \leq 10$, $x\coordup{d} \geq 6$, $y\coordup{d} \leq 5$, $y\coordup{d} \geq 1$, see Fig.~\ref{fig:penn_map_eta}.

The second submissions $\varphi_{\text{pick-up}}$ and $\varphi_{\text{drop-off}}$ specify that the second UAV must reach \textit{HUP} for a pick-up between 10 to 20 time steps. It is required to stay there for 5 time steps which are needed for the loading before
departing to \textit{Smilow} where it suppose to spend all the time between $30$ to $35$ time steps needed for a drop-off. These two sub-missions are formally defined as follows:
\begin{equation*}
	\varphi_{\text{pick-up}} \defeq \eventually_{[10, 20]}\always_{[0,\, 5]}(\pos\coordup{2} \in \textit{HUP}) 
\end{equation*}
\begin{equation*}
	\varphi_{\text{drop-off}} \defeq \always_{[30,\, 35]} (\pos\coordup{2} \in \textit{Smilow})
\end{equation*}

Now, in order to generate optimal control inputs, we solve Problem \ref{prob:synthesis} for two different objectives: the \robplus{} synchronous temporal robustness according to Section \ref{sec:milp_eta} and the \robplus{} asynchronous temporal robustness according to Section \ref{sec:milp_async}.

First, we solved Problem \ref{prob:synthesis} with the \robplus{} synchronous temporal robustness objective, $\objs\defeq\etap_{\varphi_{\text{MAS}}}$, horizon $H\defeq 50$ and the desired temporal robustness lower bound $\objs^*\defeq1$.
It led to the optimal trajectory $\trajsync$ and the \robplus{} temporal robustness value of $\etap_{\varphi_{\text{MAS}}}(\trajsync,0) = 10$ time units. The visualization of $\trajsync$ is presented in Fig.~\ref{fig:penn_map_eta} and simulation is available at \href{https://tinyurl.com/syncrobust}{https://tinyurl.com/syncrobust}. 
In Table~\ref{tab:experiments}, we show that the solver takes only $0.7$ seconds on average to solve Problem~\ref{prob:synthesis} after the problem has been defined using YALMIP, which takes additional $0.65$ seconds.  
The implementation results in 1165 Boolean and 226 integer variables. 
Since $\etap_{\varphi_{\text{MAS}}}(\trajsync,0)>0$ the produced optimal signal $\trajsync$ leads to a satisfaction of the mission $\varphi_{\text{MAS}}$. Furthermore, any time disturbances that could lead to $\tstep$-early signals by up to $\tstep=10$ time steps would still be tolerated by the system and lead to a satisfaction of $\varphi_{\text{MAS}}$, as analyzed in the two previous case studies. 
\begin{figure}[t!]
	\centering
	\includegraphics[width=0.9\textwidth]{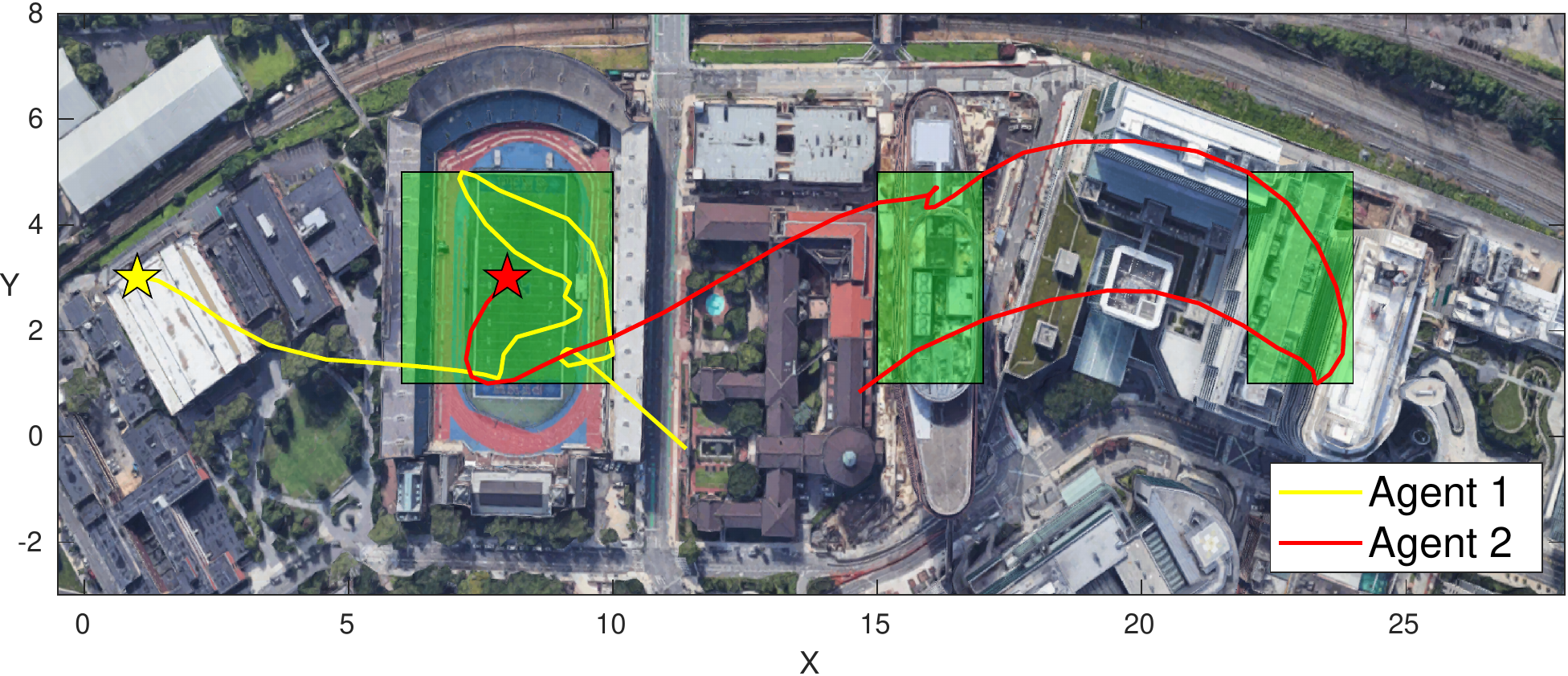}
	\caption{Nominal UAV trajectories under the
		control policies generated by solving Problem~\ref{prob:synthesis} with the \robplus{} asynchronous temporal robustness.  The found robustness value
		is $\thetap_{\varphi_{\text{MAS}}}(\trajasyn,0) = 4$ time steps. Initial positions marked by $\star$.}
	\label{fig:penn_map_theta}
\end{figure}
\begin{figure}[t!]
	\centering
	\includegraphics[width=0.95\textwidth]{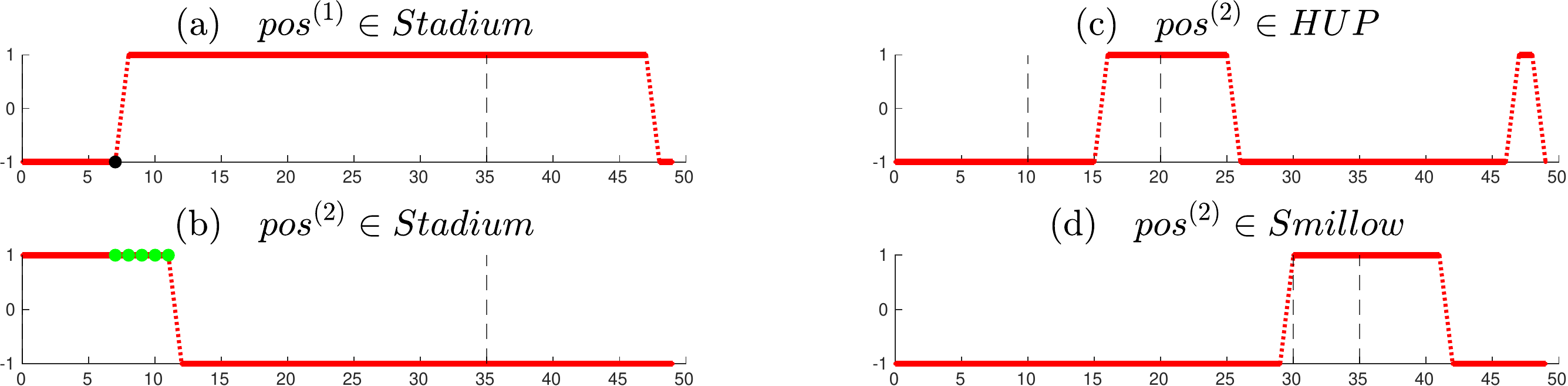}
	\caption{Characteristic function evolution of four predicates consisting in $\varphi_{\text{MAS}}$ over the signal $\trajasyn$.
		%$\chi_{p_k}(\trajasyn, t)$, $\forall t\in [0,H]$
		Found maximum \robplus{} asynchronous temporal robustness is $\thetap_{\varphi_{\text{MAS}}}(\trajasyn,0) = 4$. For a predicate $p_2:=\pos\coordup{2}\in Stadium$, the five points depicted in green show that only four time steps into the future can be done from time $t=7$ before $p_2$ satisfaction changes, therefore, $\thetap_{\varphi_{\text{MAS}}}(\trajasyn, 0)=\thetap_{p_2}(\trajasyn, 7)=4$.}
	\label{fig:penn_pred}
\end{figure}

Next, we solved Problem \ref{prob:synthesis} with the \robplus{} asynchronous temporal robustness objective, $\objs\defeq\thetap_{\varphi_{\text{MAS}}}$, same horizon $H\defeq 50$ and the same desired temporal robustness lower bound $\objs^*\defeq1$.
It led to the optimal signal $\trajasyn$ presented in Fig.~\ref{fig:penn_map_theta} and the \robplus{} temporal robustness value of $\thetap_{\varphi_{\text{MAS}}}(\trajasyn,0) = 4$ time steps, see Fig.~\ref{fig:penn_pred} for the evaluation of the predicate signals. Simulation is available at \href{https://tinyurl.com/asynrob}{https://tinyurl.com/asynrob}. 
In Table~\ref{tab:experiments}, we show
that the maximization of $\thetap_{\varphi_{\text{MAS}}}(\sstraj,0)$ is more computationally challenging than the previous maximization of $\etap_{\varphi_{\text{MAS}}}(\sstraj,0)$. It results in 2026 Boolean and 2095 integer variables and the solver
needs $69.6$ seconds on average to solve Problem~\ref{prob:synthesis}.  This result is expected as previously analyzed in Propositions \ref{thm:milp_sync_encoding} and \ref{thm:milp_asyn_encoding} where the number of binary and integer variables required for the MILP encoding of $\thetap_{\varphi_{\text{MAS}}}(\sstraj)$ is a function of $O(H\cdot(|AP|+|\varphi_{\text{MAS}}|))$, while the MILP encoding of $\etap_{\varphi_{\text{MAS}}}(\sstraj)$ is computationally more tractable and requires only $O(H\cdot|AP|)$ binary and $O(H\cdot|\varphi_{\text{MAS}}|)$ integer variables.

Let us now look at Fig.~\ref{fig:penn_map_theta} and
denote the predicates $p_1 := \pos\coordup{1}\in Stadium$ and $p_2:=\pos\coordup{2}\in Stadium$. Then from  Fig.\ref{fig:penn_map_theta}
(a) one can see that $\chi_{p_1}(\trajasyn, t)=-1$ for all time points $t=0,\ldots, 7$ but 
$\chi_{p_2}(\trajasyn, t)=\po$ for $t=0,\ldots,11$. Therefore, $\thetap_{p_1 \vee p_2}(\trajasyn, 7)=\max(\thetap_{p_1}(\trajasyn,7), \thetap_{p_2}(\trajasyn, 7)) = \max(0, 4) = 4$, and thus, $\thetap_{\varphi_{\text{MAS}}}(\trajasyn, 0) = \thetap_{\varphi_{\text{surveil}}}(\trajasyn, 0) = 4$.

The fact, that $\thetap_{\varphi_{\text{MAS}}}(\sstraj,0)<10$ is expected due to Theorem~\ref{thm:theta_bounded}. Adding the constraint $\thetap_{\varphi_{\text{MAS}}}(\sstraj,0)<10$ to the overall MILP implementation of Problem~\ref{prob:synthesis} with the \robplus{} asynchronous temporal robustness objective leads to a problem with $11144$ constraints but faster solver computation time, $61.6$ seconds on average, see Table~\ref{tab:experiments}.

%$0.65\pm0.1 $             & $0.7\pm 0.15$  

%%%%%%%%%%%%%%%%%%%%%%%%%%%%%%%%%%%%%%%%%

\begin{table}[t!]
	\renewcommand{\arraystretch}{1.2}
	\setlength{\tabcolsep}{0.9pt}
	\centering
	\begin{tabular}{l|c|c|cc|cc|}
		\multicolumn{1}{c|}{\multirow{2}{*}{\textbf{Mission}}} &
		\multirow{2}{*}{\begin{tabular}[c]{@{}c@{}}Temp. Robustness\\ (time units)\end{tabular}} &
		\multirow{2}{*}{\# Constraints} &
		\multicolumn{2}{c|}{\# Variables} &
		\multicolumn{2}{c|}{Computation time (s)} \\
		\multicolumn{1}{c|}{} &  &  & Boolean & Integer & YALMIP & Solver \\ \hline
		\multirow{2}{*}{$\varphi_{\text{MAS}}$}  & $\etap_{\varphi_{\text{MAS}}}(\trajsync,0) = 10$ & 4,597 & 1165                 &   226              &  0.65$\pm$0.1            & 0.7$\pm$0.15           \\
		& $\thetap_{\varphi_{\text{MAS}}}(\trajasyn,0) = 4$ & 11,143 &          2026        &            2095      &          1$\pm$0.14      & 69.6$\pm$37.2   \\\hline
		$\varphi_{\text{MAS}}$ $\thetap_{\varphi_{\text{MAS}}}(\sstraj,0)<10$& 
		$\thetap_{\varphi_{\text{MAS}}}(\trajasyn,0) = 4$
		& 11,144& 2026& 2095& 1$\pm$0.16 &     61.6$\pm$31.9       
	\end{tabular}
	\caption{\small Computational complexity report. \textit{Computation time} includes mean $\pm$ standard deviation of \textit{YALMIP time} (the time used to build the MILP and convert it into appropriate format for the solver) and \textit{Solver time} (the time taken by Gurobi to solve the problem). Results obtained from 100
		runs of Problem \ref{prob:synthesis} with initial positions chosen at random in small proximity of $\pos_{0}\coordup{1}$ and $\pos_{0}\coordup{2}$.}
	\label{tab:experiments}
%	\vspace{-6mm}%
\end{table}

%\begin{table}[t!]
%	\renewcommand{\arraystretch}{1.2}
%	\setlength{\tabcolsep}{1.2pt}
%	\centering
%	\begin{tabular}{l|c|c|cc|cc|}
%		\multicolumn{1}{c|}{\multirow{2}{*}{\textbf{Mission}}} &
%		\multirow{2}{*}{\begin{tabular}[c]{@{}c@{}}Time Robustness\\ (time units)\end{tabular}} &
%		\multirow{2}{*}{\# Constraints} &
%		\multicolumn{2}{c|}{\# Variables} &
%		\multicolumn{2}{c|}{Computation time (s)} \\
%		\multicolumn{1}{c|}{} &  &  & Boolean & Integer & YALMIP & Solver \\ \hline
%		\multirow{2}{*}{$\varphi_{\text{MAS}}$}  & $\etap_{\varphi_{\text{MAS}}}(\trajsync,0) = 10$ & 4,597 & 1165                 &   226              &  0.7             & 0.8           \\
%		& $\thetap_{\varphi_{\text{MAS}}}(\trajasyn,0) = 4$ & 11,143 &          2026        &            2095      &           1.4     & 120.1   \\\hline
%		$\varphi_{\text{MAS}}$, $\thetap_{\varphi_{\text{MAS}}}(\sstraj,0)<10$& 
%		$\thetap_{\varphi_{\text{MAS}}}(\sstraj_3,0) = 4$
%		& 11,144& 2026& 2095& 1.1 &     82.3       
%	\end{tabular}
%	\caption{\small Computational complexity report. \textit{Computation time} includes \textit{YALMIP time} (the time used to build the MILP and convert it into appropriate format for the solver) and \textit{Solver time} (the time taken by Gurobi to solve the problem).}
%	\label{tab:experiments}
%	\vspace{-6mm}%
%\end{table}
%%%%%%%%%%%%%%%%%%%
\section{Conclusions and future work}
\label{sec:conclusions}

This work presents a theoretical framework for  temporal robustness of Signal Temporal Logic (STL) specifications which are interpreted over continuous and discrete time signals. 
%The robustness function that we consider is with respect to 
In particular we defined synchronous and asynchronous temporal robustness and showed that these notions quantify the robustness with respect to synchronous and asynchronous time shifts in the predicates of signal temporal logic specifications. 

For both notions of temporal robustness, we analyzed their desirable properties and showed that the asynchronous temporal robustness is upper bounded by the synchronous temporal robustness in its absolute value. We further showed two particular STL fragments for which the two robustness notions are equivalent.

We further addressed the control synthesis problem in which we aim to design a control law that maximizes the temporal robustness of a dynamical system. We presented Mixed-Integer Linear Programming (MILP) encodings for the synchronous and asynchronous temporal robustness that solve the control synthesis problem. 

We are currently exploring several future research directions. First, we are interested in the control synthesis problem when considering temporal robustness of stochastic dynamical systems. We also explore STL robustness notions that combine spatial and temporal robustness. Third, we are interested in computationally more tractable solutions to the control synthesis problem with temporal robustness objectives. 

%%
%% The acknowledgments section is defined using the "acks" environment
%% (and NOT an unnumbered section). This ensures the proper
%% identification of the section in the article metadata, and the
%% consistent spelling of the heading.
\begin{acks}
This work was supported by the AFOSR under grant FA9550-19-1-0265 (Assured Autonomy in Contested Environments) and
by the ARL under grant DCIST CRA W911NF-17-2-0181.
\end{acks}

%%
%% The next two lines define the bibliography style to be used, and
%% the bibliography file.
\bibliographystyle{ACM-Reference-Format}
\bibliography{root_timerob}

%%
%% If your work has an appendix, this is the place to put it.
\appendix

%%%%%%%%%%%%%%%%%%%%%%%%%%%%%%%%%%%%%%%%%%%%%%%%%%%%%%%%%%%%%%%%%%%%%%%%%%%%%%%%%%%%%
% Section Sync Rob
\section{Proofs of Section~\ref{sec:prop_sound}}
\label{appndx:soundness}

\subsection{Proof of Theorem \ref{thm:sat_eta}}
Follows directly from Def.~\ref{def:time_rob_all}.
We are going to prove items \ref{ti1}) and \ref{ti3}). Items \ref{ti2}) and \ref{ti4}) can be proven analogously. 

Item \ref{ti1}) states that $\eta^{\pm}_\varphi(\sstraj,t) > 0 \ \Longrightarrow\ \chi_\varphi(\sstraj,t)= 1$.
By Def.~\ref{def:time_rob_all}, 
$\eta^{\pm}_\varphi(\sstraj,t) = \chi_\varphi(\sstraj, t)\cdot
\sup\{\tau\geq0 \ :\ \forall t'\in t\pm [0,\tau],\ \chi_\varphi(\sstraj,t')=\chi_\varphi(\sstraj,t)\}$. Since $\eta^{\pm}_\varphi(\sstraj,t) > 0$, $\tau \geq 0$ and $\chi\in\{\pm 1\}$, then $\chi_\varphi(\sstraj, t) = 1$.

Item \ref{ti3}) states that $\chi_\varphi(\sstraj,t)= 1 \  \Longrightarrow\ 
\eta^{\pm}_\varphi(\sstraj,t) \geq 0$.
Since $\chi_\varphi(\sstraj,t)= 1$ then
by Def.~\ref{def:time_rob_all}, 
$\eta^{\pm}_\varphi(\sstraj,t) = 
\sup\{\tau\geq0 \ :\ \forall t'\in t\pm[0,\tau],\ \chi_\varphi(\sstraj,t')=1\} \geq 0$.
\subsection{Proof of Corollary \ref{cor:eta_lr}}
Corollary \ref{cor:eta_lr} is a direct consequence of Theorem \ref{thm:sat_eta}. Let $\varphi$ be an STL formula, $\sstraj:\TDom \rightarrow X$ be a signal and $t\in\TDom$ be a time point.
\begin{enumerate}
	\item 
	First, we are going to prove 
	$\eta^{\pm}_\varphi(\sstraj,t)> 0\ \Longrightarrow\  \eta^{\mp}_\varphi(\sstraj,t)\geq 0$. 
	The following sequence holds:
	
	$\eta^{\pm}_\varphi(\sstraj,t)> 0\quad
	\overset{\text{Thm.}\ref{thm:sat_eta}(\ref{ti1})}{\Longrightarrow}
	\chi_\varphi(\sstraj,t)= +1
	\overset{\text{Thm.}\ref{thm:sat_eta}(\ref{ti3})}{\Longrightarrow}
	\quad \eta^{\mp}_\varphi(\sstraj,t)\geq 0$.
	\item 
	Second, we are going to prove $\eta^{\pm}_\varphi(\sstraj,t)< 0\ \Longrightarrow\  \eta^{\mp}_\varphi(\sstraj,t)\leq 0$. The following sequence holds:
	
	$\eta^{\pm}_\varphi(\sstraj,t)< 0\quad
	\overset{\text{Thm.}\ref{thm:sat_eta}(\ref{ti2})}{\Longrightarrow}
	\chi_\varphi(\sstraj,t)= -1
	\overset{\text{Thm.}\ref{thm:sat_eta}(\ref{ti4})}{\Longrightarrow}
	\quad \eta^{\mp}_\varphi(\sstraj,t)\leq 0$.
\end{enumerate}

\subsection{Proof of Theorem \ref{thm:chi_sync_stronger}}
%
%	\begin{align*}
%	|\eta^{\pm}_\varphi(\sstraj, t)|=i\quad\Longleftrightarrow\quad
%	&\forall t'\in t\pm [0, i),\ \chi_{\varphi}(\sstraj,t')=\chi_\varphi(\sstraj,t) \qquad\text{and}\\
%	[&
%	\forall\epsilon>0,\ \exists \tstep\in[i, i+\epsilon),\ \chi_{\varphi}(\sstraj,t \pm \tstep)\not=\chi_\varphi(\sstraj,t)
%	\quad\text{or}\quad 
%	%	&\chi_{\varphi}(\sstraj,t \pm i)\not=\chi_\varphi(\sstraj,t)\quad \vee\\ 
%	i=+\infty
%	].
%\end{align*}

Let $\varphi$ be an STL formula, $\sstraj:\Re \rightarrow X$ be a continuous-time signal and $t\in\Re$ be a time point. For any value $\trv\in\CoRe_{\geq 0}$, we want to prove that the synchronous temporal robustness $|\eta^{\pm}_\varphi(\sstraj, t)|=\trv$ if and only if $\forall t'\in t\pm [0, \trv)$, $\chi_{\varphi}(\sstraj,t')=\chi_\varphi(\sstraj,t)$ and if $\trv<\infty$ then $\forall\epsilon>0$, $\exists \tstep\in[\trv, \trv+\epsilon)$, $\chi_{\varphi}(\sstraj,t \pm \tstep)\not=\chi_\varphi(\sstraj,t)$.

Let $|\eta^{\pm}_\varphi(\sstraj, t)|=\trv$.
Denote the set $\varUpsilon=\{\tau\geq0 \ :\ \forall t'\in t\pm [0,\tau],\ \chi_{\varphi}(\sstraj,t')=\chi_\varphi(\sstraj,t)\}$.
 To prove the result we distinguish between the cases of $\trv=\infty$ and $\trv<\infty$ as following:

$(1)$ Let $\trv=\infty$. Then by Def.~\ref{def:time_rob_all}, $\sup \varUpsilon=|\eta^{\pm}_\varphi(\sstraj, t)|=\infty$. Therefore, we have to show that
	\begin{equation*}
	\sup\varUpsilon =\infty\ \Longleftrightarrow\ \forall t'\in t\pm [0, \infty),\ \chi_{\varphi}(\sstraj,t')=\chi_\varphi(\sstraj,t)
	\end{equation*}
		which follows from the definition of the supremum for the unbounded set.
%	$\Longrightarrow$ 
%	If $\sup\varUpsilon =+\infty$, i.e. $\varUpsilon$ is unbounded then $\forall t'\in t\pm [0,+\infty), \chi_{\varphi}(\sstraj,t')=\chi_{\varphi}(\sstraj,t)$.
%	
%	$\Longleftarrow$ 
%	If $\forall t'\in t\pm [0, +\infty),\ \chi_{\varphi}(\sstraj,t')=\chi_\varphi(\sstraj,t)$ then  $\sup \varUpsilon=+\infty$.
%	
	
%----------------------------------------------------------------

$(2)$ Let $\trv<\infty$.  Then we must show that
	\begin{equation}
		\label{eq:sufnes}
		\begin{aligned}
			\sup\varUpsilon=\trv\quad\Longleftrightarrow\quad
			&\forall t'\in t\pm [0, \trv),\ \chi_{\varphi}(\sstraj,t')=\chi_\varphi(\sstraj,t) \qquad\text{and}\\
			&
			\forall\epsilon>0,\ \exists \tstep\in[\trv, \trv+\epsilon),\ \chi_{\varphi}(\sstraj,t \pm \tstep)\not=\chi_\varphi(\sstraj,t).
		\end{aligned}
	\end{equation}
We next prove sufficiency ($\Longrightarrow$) and necessity ($\Longleftarrow$) of \eqref{eq:sufnes} as following: 

	$\Longrightarrow$
	Let $\sup\varUpsilon=\trv$. Since $\trv<\infty$ we can apply the $\epsilon$ definition of supremum: 
	$\forall \epsilon>0$, $\exists \tau^*\in \varUpsilon$ such that $\tau^*> \trv-\epsilon$.
	i.e. the following holds:
	\begin{equation*}
		\label{eq:eps0}
		\forall \epsilon>0,\ \exists \tau^*\geq 0,\ \tau^*>\trv-\epsilon\ 
		\text{such that}\  \forall t'\in t\pm [0,\tau^*],\  \chi_\varphi(\sstraj,t')=\chi_\varphi(\sstraj,t).
	\end{equation*} 
	Since $[0,\trv-\epsilon]\subset[0,\tau^*]$ then it holds that
	\begin{equation}
		\label{eq:eps1}
		\forall \epsilon>0,\ \forall t'\in t\pm [0,\trv-\epsilon],\  \chi_\varphi(\sstraj,t')=\chi_\varphi(\sstraj,t).
	\end{equation} 
	%Now take any $t^*\in t\pm [0,i)$. We want to prove that $\chi_\varphi(\sstraj,t^*)=\chi_\varphi(\sstraj,t)$ holds.
	%Since $t^*\in t\pm [0,i)$, there exists $\epsilon=\epsilon^*>0$ such that $t^*=t \pm (i - \epsilon^*)$. Now take another $\epsilon= \frac{\epsilon^*}{2}>0$ then due to \eqref{eq:eps1}, 
	%$\forall t'\in t\pm \left[0,i-\frac{\epsilon^*}{2}\right)$,  $\chi_\varphi(\sstraj,t')=\chi_\varphi(\sstraj,t)$.
	%Since $i-\epsilon^* \in [0,i-\frac{\epsilon^*}{2})$, then 
	%for $t'=t \pm (i-\epsilon^*)$, $\chi_\varphi(\sstraj,t')=\chi_\varphi(\sstraj,t)$, thus, 
	%$\chi_\varphi(\sstraj,t^*)=\chi_\varphi(\sstraj,t)$.
	\begin{itemize}
		\item 
		First, we want to prove that the first line of the RHS of \eqref{eq:sufnes} holds, i.e., we want to show that $\forall t'\in t\pm [0, \trv),\ \chi_{\varphi}(\sstraj,t')=\chi_\varphi(\sstraj,t)$.
		Assume the opposite, i.e. assume that $\exists \tstep^*\in [0, \trv)$ such that $\chi_\varphi(\sstraj,t\pm \tstep^*)\not= \chi_\varphi(\sstraj,t)$. 
		We can represent such $\tstep^*=\trv-\epsilon^*$ for some
		$\epsilon^*$, $0<\epsilon^*\leq \trv$.
		Then due to \eqref{eq:eps1},  
		$\forall t'\in t\pm [0,\trv-\epsilon^*]$, $\chi_\varphi(\sstraj,t')=\chi_\varphi(\sstraj,t)$, which contradicts with the above $\chi_\varphi(\sstraj,t\pm \tstep^*)\not= \chi_\varphi(\sstraj,t)$. Thus, it indeed holds that:
		\begin{equation}
			\label{eq:t1}
		\forall t'\in t\pm [0, \trv),\ \chi_{\varphi}(\sstraj,t')=\chi_\varphi(\sstraj,t).
	\end{equation}
		\item Second, we want to prove that the second line of the RHS of \eqref{eq:sufnes} holds, i.e., we want to show that 
		$\forall\epsilon>0,\ \exists \tstep\in[\trv, \trv+\epsilon),\ \chi_{\varphi}(\sstraj,t \pm \tstep)\not=\chi_\varphi(\sstraj,t)$.
		Assume the opposite,
		 i.e. assume $\exists \epsilon^*>0$, $\forall \tstep\in[\trv, \trv+\epsilon^*)$, $\chi_{\varphi}(\sstraj,t \pm \tstep)=\chi_\varphi(\sstraj,t)$, which can be rewritten as $\exists \epsilon^*>0$, $\forall t'\in t\pm [\trv, \trv+\epsilon^*)$, $\chi_{\varphi}(\sstraj,t')=\chi_\varphi(\sstraj,t)$.			
			 In combination with $\eqref{eq:t1}$, we get $\exists \epsilon^*>0$, $\forall t'\in t\pm [0, \trv+\epsilon^*),\ \chi_{\varphi}(\sstraj,t')=\chi_\varphi(\sstraj,t)$. 
			 Since $\epsilon^*>0$ then $\tau^*=\trv+\frac{\epsilon^*}{2}\in [0, \trv+\epsilon^*)$ and $\tau^*\geq 0$, therefore, $\tau^* \in \varUpsilon$ but also, $\tau^* > \trv$ which is a contradiction, since any value $\tau\in \varUpsilon$ should be $\tau \leq \trv$ (since $\sup\varUpsilon = \trv$). 
			 Thus, it indeed holds that 
			 $\forall\epsilon>0$, $\exists \tstep\in[\trv, \trv+\epsilon)$, $\chi_{\varphi}(\sstraj,t \pm \tstep)\not=\chi_\varphi(\sstraj,t)$.
	\end{itemize}

	$\Longleftarrow$ 
		Let $\forall t'\in t\pm [0, \trv),\ \chi_{\varphi}(\sstraj,t')=\chi_\varphi(\sstraj,t)$ and $\forall\epsilon>0,\ \exists \tstep\in[\trv, \trv+\epsilon),\ \chi_{\varphi}(\sstraj,t \pm \tstep)\not=\chi_\varphi(\sstraj,t)$, where $\trv<\infty$.
	Blow we show that $\sup\varUpsilon=\trv$. 
	\begin{itemize}
		\item First, we are going to show that $\forall s\in \varUpsilon$, $s\leq \trv$. Assume the opposite, assume
		$\exists s=s^*\in\varUpsilon$ such that $s^* > \trv$. From the definition of $\varUpsilon$ this can be rewritten as $\exists s^* (s^*> \trv\ \wedge\ \forall t'\in t\pm [0,s^*],\  \chi_\varphi(\sstraj,t')=\chi_\varphi(\sstraj,t))$.
		Take $\epsilon = s^*-\trv$. Since $s^*>\trv$, $\epsilon>0$, therefore, from what is given, $\exists \tstep\in[\trv, \trv+\epsilon)=[\trv,s^*),\ \chi_{\varphi}(\sstraj,t \pm \tstep)\not=\chi_\varphi(\sstraj,t)$. But since $[\trv,s^*)\subset [0, s^*]$ then $\tstep\in [0, s^*]$ which is a contradiction with the assumed. 
		Therefore, it indeed holds that $\forall s\in \varUpsilon$, $s\leq \trv$.
		\item Second, we are going to show that $\forall \epsilon>0$, $\exists \tau^*\in \varUpsilon$ such that $\tau^*> \trv-\epsilon$. Take any $\epsilon>0$.
		
		\begin{itemize}
			\item If $\trv < \frac{\epsilon}{2}$, i.e. 
			$\trv-\epsilon < \trv-\frac{\epsilon}{2}< 0$
			then let $\tau^*=0$. In this case, $\tau^*\in \varUpsilon$ and $\tau^*> \trv-\epsilon$. 
			\item If $\trv \geq \frac{\epsilon}{2}$, then let $\tau^*=\trv-\frac{\epsilon}{2}$. In this case, $\tau^*\geq 0$ and $\tau^*=\trv-\frac{\epsilon}{2} > \trv-\epsilon$. Also, we know that  $\forall t'\in t\pm [0, \trv),\ \chi_{\varphi}(\sstraj,t')=\chi_\varphi(\sstraj,t)$. Since $[0, \trv-\frac{\epsilon}{2}] \subset [0, \trv)$ then $\forall t'\in t\pm [0, \tau^*]$, $\chi_{\varphi}(\sstraj,t')=\chi_\varphi(\sstraj,t)$, i.e. $\tau^*\in\varUpsilon$. 
		\end{itemize}
		
	\end{itemize}
 Thus, $\sup \varUpsilon=\trv$ which concludes the proof.

\subsection{Proof of Eq.\eqref{eq:chi_eta_disc} (Theorem \ref{thm:chi_sync_stronger} for discrete-time)} 

Let $\varphi$ be an STL formula, $\sstraj:\Ze \rightarrow X$ be a discrete-time signal and $t\in\Ze$ be a time point. For any finite value $\trv\in\Ze_{\geq 0}$, we want to prove that the synchronous temporal robustness $|\eta^{\pm}_\varphi(\sstraj, t)|=\trv$ if and only if $\forall t'\in t\pm [0, \trv]$, $\chi_{\varphi}(\sstraj,t')=\chi_\varphi(\sstraj,t)$ and
$\chi_{\varphi}(\sstraj,t \pm (\trv+1))\not=\chi_\varphi(\sstraj,t)$.

Let $|\eta^{\pm}_\varphi(\sstraj,t)| =\trv$. 
Denote the set $\varUpsilon=\{\tau\geq0 \ :\ \forall t'\in t\pm [0,\tau],\ \chi_{\varphi}(\sstraj,t')=\chi_\varphi(\sstraj,t)\}$.
Then, by Def. \ref{def:time_rob_all}, 
$|\eta^{\pm}_\varphi(\sstraj,t)|=\sup \varUpsilon=\trv$.
Since $\trv$ is finite and $t\in\Ze$ then it holds that
$\sup\varUpsilon=\max\varUpsilon=\trv$,
which holds if and only if
$\forall t'\in t\pm [0, \trv]$, $\chi_{\varphi}(\sstraj,t')=\chi_\varphi(\sstraj,t)$ and
$\chi_{\varphi}(\sstraj,t \pm (\trv+1))\not=\chi_\varphi(\sstraj,t)$.

\subsection{Proof of Theorem \ref{thm:eta_decrease}}

Let $\varphi$ be an STL formula, $\sstraj:\Re \rightarrow X$ be a continuous-time signal and $t\in\Re$ be a time point. For any value $\trv\in\CoRe_{\geq 0}$, we want to prove that
if the synchronous temporal robustness 
$|\eta^{\pm}_\varphi(\sstraj, t)|= \trv$ then $\forall \tstep\in [0, \trv)$, $|\eta^{\pm}_{\varphi}(\sstraj,t \pm \tstep)| = \trv -\tstep$.	

	Let $|\eta^{\pm}_\varphi(\sstraj, t)|= \trv\geq 0$. 
	To prove the result we distinguish between the cases of $\trv=\infty$ and $\trv<\infty$:
		\begin{enumerate}
			\item Let $\trv=\infty$. Then according to Thm.~\ref{thm:chi_sync_stronger}, $\forall \tstep\in [0, \infty),\ \chi_{\varphi}(\sstraj,t \pm \tstep)=\chi_\varphi(\sstraj,t)$. 
			Take any $\tstep\in [0, \infty)$, then it holds that $\forall \tstep'\in [0, \infty)$, $\chi_{\varphi}(\sstraj,t \pm \tstep\pm \tstep')=\chi_\varphi(\sstraj,t \pm \tstep)$. Thus, according to Thm.~\ref{thm:chi_sync_stronger}, 
			$|\eta^{\pm}_\varphi(\sstraj, t \pm \tstep)|=\infty= \infty- \tstep$.

			\item Let $\trv<\infty$.
			Then according to Thm.~\ref{thm:chi_sync_stronger}, $\forall \tstep\in [0, \trv),\ \chi_{\varphi}(\sstraj,t\pm \tstep)=\chi_\varphi(\sstraj,t)$ and 
			$\forall\epsilon>0,\ \exists \tilde{\tstep}\in[\trv, \trv+\epsilon),\ \chi_{\varphi}(\sstraj,t \pm \tilde{\tstep})\not=\chi_\varphi(\sstraj,t)$. 
			Take any $\tstep\in [0, \trv)$. Then 
			the first property leads to 
			$\forall \tstep' \in [0, \trv-\tstep),\  \chi_{\varphi}(\sstraj,t\pm \tstep\pm \tstep')=\chi_\varphi(\sstraj,t\pm \tstep)=\chi_{\varphi}(\sstraj,t)$.
			The second property leads to 
			$\forall\epsilon>0,\ \exists \tilde{\tstep}\in[\trv-\tstep, \trv+\epsilon-\tstep),\ \chi_{\varphi}(\sstraj,t \pm \tstep\pm \tilde{\tstep})\not=\chi_\varphi(\sstraj,t)=\chi_\varphi(\sstraj,t\pm \tstep)$.
			The combination of these two due to Thm.~\ref{thm:chi_sync_stronger} leads to $|\eta^{\pm}_\varphi(\sstraj, t\pm \tstep)|=\trv-\tstep$.
			%			Since $\chi_\varphi(\sstraj,t\pm \tstep)=\chi_\varphi(\sstraj, t)$ then
			%			\begin{align*}
				%				\eta^{\pm}_\varphi(\sstraj, t\pm \tstep) \overset{\text{Rem.}~\ref{rem:abs}}{=}
				%				&\chi_\varphi(\sstraj, t\pm \tstep)\cdot|\eta^{\pm}_\varphi(\sstraj, t\pm \tstep)|= 
				%				\chi_\varphi(\sstraj, t\pm \tstep)\cdot (|i|-\tstep) \\ 
				%				=\ \ \ &\chi_\varphi(\sstraj, t)\cdot (|i|-\tstep) =
				%				\chi_\varphi(\sstraj, t) \cdot|i| - \chi_\varphi(\sstraj, t)\cdot h \\
				%				\overset{\text{Rem.}~\ref{rem:abs}}{=}& i - \chi_\varphi(\sstraj, t)\cdot h.
				%			\end{align*}
		\end{enumerate}
\subsection{Proof of Eq.\eqref{eq:eta_decrease_disc} (Theorem \ref{thm:eta_decrease} for discrete-time)} 

Let $\varphi$ be an STL formula, $\sstraj:\Ze \rightarrow X$ be a discrete-time signal and $t\in\Ze$ be a time point. For any finite value $\trv\in\Ze_{\geq 0}$, we want to prove that
if $|\eta^{\pm}_\varphi(\sstraj, t)|= \trv$ then $\forall \tstep\in [0, \trv]$, $|\eta^{\pm}_{\varphi}(\sstraj,t \pm \tstep)| = \trv -\tstep$.	

	Let $|\eta^{\pm}_\varphi(\sstraj, t)|= \trv$.
	Then according to \eqref{eq:chi_eta_disc}, it holds that $\forall \tstep\in [0, \trv],\ \chi_{\varphi}(\sstraj,t\pm \tstep)=\chi_\varphi(\sstraj,t)$ and 
	$\chi_{\varphi}(\sstraj,t \pm (\trv+1))\not=\chi_\varphi(\sstraj,t)$. 
	Take any $\tstep\in [0, \trv]$. Then 
	the first property leads to 
	$\forall \tstep' \in [0, \trv-\tstep],\  \chi_{\varphi}(\sstraj,t\pm \tstep\pm \tstep')=\chi_\varphi(\sstraj,t\pm \tstep)=\chi_{\varphi}(\sstraj,t)$.
	The second property leads to 
	$\chi_{\varphi}(\sstraj,t \pm \tstep\pm (\trv- \tstep+1))\not=\chi_\varphi(\sstraj,t)=\chi_\varphi(\sstraj,t\pm \tstep)$.
	The combination of these two due to \eqref{eq:eta_decrease_disc} leads to $|\eta^{\pm}_\varphi(\sstraj, t\pm \tstep)|=\trv-\tstep$.

%%%%%%%%%%%%%%%%%%%%%%%%%%%%%%%%%%%%%%%%%%%%%%%%%%%%%%%%%%%%%%%%%%%%%%%%%%%%%%%%%%%%%
% Section: Async Rob
%\section{Proofs of Section~\ref{sec:time_rob_async}}

\subsection{Proof of Theorem \ref{thm:sat_theta}}

In this proof, we will use the following lemma.

\begin{lemma}
	\label{rem:sup0}
	For a set $S\subseteq \CoRe$ if $\sup S>0$ then $\exists s\in S$ such that $s>0$.	
\end{lemma}
\begin{proof}
	Let a set $S\subseteq \CoRe$ be such that $\sup S>0$. Assume the opposite of what we want to prove, i.e. assume that $\forall s\in S$, $s\leq 0$. Then 0 is an upper bound of $S$. 
	So $\sup S \leq 0$ but we are given $\sup S>0$ which is a contradiction.
	Therefore, $\exists s\in S$ such that $s>0$.
\end{proof}

To prove Thm. \ref{thm:sat_theta}, we are going to prove items \ref{i1}) and \ref{i3}). Items \ref{i2}) and \ref{i4}) can be proven analogously. 
%The rest of the proof is by induction on the structure of $\varphi$.  

1) We want to prove that $\theta^{\pm}_\varphi(\sstraj,t) > 0 \Longrightarrow \chi_\varphi(\sstraj,t)= \po$. Thus, let $\theta^{\pm}_\varphi(\sstraj,t) > 0$. The rest of the proof is by induction on the structure of formula $\varphi$.  
\begin{enumerate}[label={\textbf{Case}}]
	\item $\varphi=p$. 
	Since we are given that $\theta^{\pm}_p(\sstraj,t) > 0$, then due to 
	Corollary~\ref{rem:eq}, $\eta^{\pm}_p(\sstraj,t) > 0$ which due to Thm.~\ref{thm:sat_eta}
	leads to $\chi_p(\sstraj,t)= \po$.
	\item $\varphi=\neg \varphi_1$.
	Since $\theta^{\pm}_{\neg\varphi_1}(\sstraj,t) > 0$ then due to \eqref{eq:t_neg} it holds that $\theta^{\pm}_{\varphi_1}(\sstraj,t)< 0$. By the induction hypothesis for $\varphi_1$ we get that $\chi_{\varphi_1}(\sstraj,t)=-1$, and thus by Def.~\ref{def:char_func}, $\chi_{\neg\varphi_1}(\sstraj,t)=-\chi_{\varphi_1}(\sstraj,t) =\po$.	
	\item $\varphi=\varphi_1\wedge\varphi_2$.
	Since $\theta^{\pm}_{\formula_1 \land \formula_2}(\sstraj,t) = \theta^{\pm}_{\formula_1}(\sstraj,t) \ \sqcap\ \theta^{\pm}_{\formula_2}(\sstraj,t) > 0$,
	both terms are positive: $\theta^{\pm}_{\formula_1}(\sstraj,t) > 0$ and $\theta^{\pm}_{\formula_2}(\sstraj,t) > 0$.
	By the induction hypothesis we get that $\chi_{\formula_1}(\sstraj,t)=\chi_{\formula_2}(\sstraj,t)=\po$, and thus by Def.~\ref{def:char_func}, $\chi_{\formula}(\sstraj,t)= \chi_{\formula_1}(\sstraj,t) \sqcap \chi_{\formula_2}(\sstraj,t)=\po$.
	\item $\varphi=\varphi_1 \until_I\varphi_2$. 
	Since $\theta^{\pm}_\varphi(\sstraj,t) > 0$ then due to Lemma~\ref{rem:sup0} and Def.~\ref{def:time_rob_rec} for Until operator, $\exists t'\in t+ I$ such that $\theta^{\pm}_{\formula_2}(\sstraj,t') \ \sqcap \
	\bigsqcap_{t'' \in [t,t')} \theta^{\pm}_{\formula_1}(\sstraj,t'')>0$.
	Now due to the infimum operators, $\theta^{\pm}_{\formula_2}(\sstraj,t') >0$ and $\forall t'' \in [t,t')$, $\theta^{\pm}_{\formula_1}(\sstraj,t'')>0$.
	Therefore, from the induction hypothesis, $\exists t'\in t+I$, $\chi_{\formula_2}(\sstraj,t')=\po$ and 
	$\forall t'' \in [t,t')$, $\chi_{\formula_1}(\sstraj,t'')=\po$. And thus, 
	$\chi_{\formula}(\sstraj,t) = \po$.
\end{enumerate}
%%%%%%%%%%%%%%%%%%%%%%%%%%%%%

3) We want to prove that  $\chi_\varphi(\sstraj,t)= \po \Longrightarrow
\theta^{\pm}_\varphi(\sstraj,t) \geq 0$. 
Thus, let $\chi_\varphi(\sstraj,t)= \po$. The rest of the proof is by induction on the structure of formula $\varphi$.  
\begin{itemize}[label={\textbf{Case}}]
	\item $\varphi=p$. 
	Since we are given that $\chi_p(\sstraj,t) = \po$ then due to Thm.~\ref{thm:sat_eta}(\ref{ti3}) it holds that
	$\eta^{\pm}_p(\sstraj,t)\geq 0$ and thus, due to Corollary~\ref{rem:eq}, 
	$\theta^{\pm}_p(\sstraj,t)=\eta^{\pm}_p(\sstraj,t) \geq 0$. 
		
	\item $\varphi=\neg \varphi_1$.
	Since we are given that $\chi_{\neg\varphi_1}(\sstraj,t) = \po$ then it holds that $\chi_{\varphi_1}(\sstraj,t)=-1$. By the induction hypothesis for $\varphi_1$, $\theta^{\pm}_{\varphi_1}(\sstraj,t) \leq 0$, and thus due to \eqref{eq:t_neg}, $\theta^{\pm}_{\neg\varphi_1}(\sstraj,t) = - \theta^{\pm}_{\varphi_1}(\sstraj,t) \geq 0$. 
	\item $\varphi=\varphi_1\wedge\varphi_2$. Since $\chi_\varphi(\sstraj, t)=\chi_{\formula_1}(\sstraj, t) \sqcap \chi_{\formula_2}(\sstraj,t) = \po$ then both $\chi_{\formula_1}(\sstraj,t)=\po$ and $\chi_{\formula_2}(\sstraj,t)=\po$. By the induction hypothesis we get that $\theta^{\pm}_{\formula_1}(\sstraj,t)\geq 0$ and $\theta^{\pm}_{\formula_2}(\sstraj,t)\geq 0$ and thus, 
	$\theta^{\pm}_\formula(\sstraj,t)= \theta^{\pm}_{\formula_1}(\sstraj,t)\sqcap \theta^{\pm}_{\formula_2}(\sstraj,t) \geq 0$.	
	\item $\varphi=\varphi_1 \until_I\varphi_2$. Since $\chi_\varphi(\sstraj,t)= \po$ then from the definition of the characteristic function for Until operator,  $\exists t'\in t+ I$ such that $\chi_{\formula_2}(\sstraj,t')=\po$ and $\forall t'' \in [t,t')$, $\chi_{\formula_1}(\sstraj,t'') = \po$. By the induction hypothesis we obtain that $\exists t'\in t+ I$ such that $\theta^{\pm}_{\formula_2}(\sstraj,t')\geq 0$ and
	$\forall t'' \in [t,t')$, $\theta^{\pm}_{\formula_1}(\sstraj,t'')\geq 0$
	and thus, by \eqref{eq:t_until} we conclude that
	$\theta^{\pm}_\formula(\sstraj,t)\geq 0.$
\end{itemize}
\subsection{Proof of Corollary \ref{cor:theta_lr}}
Corollary \ref{cor:theta_lr} is a direct consequence of Theorem \ref{thm:sat_theta}.
Let $\varphi$ be an STL formula, $\sstraj:\TDom \rightarrow X$ be a signal and $t\in\TDom$ be a time point.
\begin{enumerate}
	\item 
	First, we are going to prove $\theta^{\pm}_\varphi(\sstraj,t)> 0\ \Longrightarrow\ \theta^{\mp}_\varphi(\sstraj,t)\geq 0$. 
	The following sequence holds:

	$\theta^{\pm}_\varphi(\sstraj,t)> 0\quad
	\overset{\text{Thm.}\ref{thm:sat_theta}(\ref{i1})}{\Longrightarrow}
	\chi_\varphi(\sstraj,t)= +1
	\overset{\text{Thm.}\ref{thm:sat_theta}(\ref{i3})}{\Longrightarrow}
	\quad \theta^{\mp}_\varphi(\sstraj,t)\geq 0$.
	
	\item Second, we are going to prove $\theta^{\pm}_\varphi(\sstraj,t)< 0\ \Longrightarrow\ \theta^{\mp}_\varphi(\sstraj,t)\leq 0$.
	The following sequence holds:
	
	$\theta^{\pm}_\varphi(\sstraj,t)< 0\quad
	\overset{\text{Thm.}\ref{thm:sat_theta}(\ref{i2})}{\Longrightarrow}
	\chi_\varphi(\sstraj,t)= -1
	\overset{\text{Thm.}\ref{thm:sat_theta}(\ref{i4})}{\Longrightarrow}
	\quad \theta^{\mp}_\varphi(\sstraj,t)\leq 0$.
\end{enumerate}

\subsection{Proof of Theorem \ref{thm:chi_async}}
\label{sec:chi_async}

Let $\varphi$ be an STL formula, $\sstraj:\Re \rightarrow X$ be a continuous-time signal and $t\in\Re$ be a time point. For any value $\trv\in\CoRe_{\geq 0}$, we want to prove that
if 
$|\theta^{\pm}_\varphi(\sstraj,t)| = \trv$  then $\forall t'\in t \pm \,[0,\trv)$, $\chi_\varphi(\sstraj,t')=\chi_\varphi(\sstraj,t)$.

Note that the result is trivial for $\trv=0$. 
Therefore, let $|\theta^{\pm}_\varphi(\sstraj, t)|= \trv>0$. 
		The rest of the proof is by induction on the structure of the formula $\varphi$.  	
		\begin{enumerate}[label={\textbf{Case}}]
			\item $\varphi=p$. 
			Since $|\theta^{\pm}_p(\sstraj, t)|= \trv$ then due to Corollary \ref{rem:eq}, $|\eta^{\pm}_p(\sstraj,t)|=\trv$. From Theorem \ref{thm:chi_sync_stronger} we get that
			$\forall t'\in t\pm [0,\trv),\ \chi_p(\sstraj,t')=\chi_p(\sstraj,t)$.
			\item $\varphi = \neg\varphi_1$.
			Since $|\theta^{\pm}_{\neg\varphi_1}(\sstraj,t)| =r$ then due to \eqref{eq:t_neg} it holds that $|\theta^{\pm}_{\varphi_1}(\sstraj,t)|=r$. By the induction hypothesis for $\varphi_1$ we get that
			$\forall t'\in t\pm [0,\trv),\ \chi_{\varphi_1}(\sstraj,t')=\chi_{\varphi_1}(\sstraj,t)$. Therefore, due to Def.~\ref{def:char_func}, it holds that
			$\forall t'\in t\pm [0,\trv)\quad  \chi_{\neg\varphi_1}(\sstraj,t')
				=
				-\chi_{\varphi_1}(\sstraj,t') 
				=
				-\chi_{\varphi_1}(\sstraj,t) 
				=
				\chi_{\neg\varphi_1}(\sstraj,t)$.
			%%%%%%%%%%%%%%%%%%%%%%%%%%%%%%%%%%%%%%%%%%%%%
			\item $\varphi=\varphi_1 \wedge \varphi_2$.
			We consider the two separate cases of $\chi_{\varphi}(\sstraj,t)= 1$ and $\chi_{\varphi}(\sstraj,t)= -1$ as follows:			
		
		\textbf{1.} Let $\chi_{\varphi}(\sstraj,t)= 1$. 
				Since $|\theta^{\pm}_\varphi(\sstraj, t)|= \trv>0$
				then due to Thm.~\ref{thm:sat_theta}(\ref{i3}), $\theta^{\pm}_{\formula}(\sstraj,t) = \trv> 0$. WLOG assume that $\theta^{\pm}_{\varphi_1}(\sstraj,t) \leq \theta^{\pm}_{\varphi_2}(\sstraj,t)$. Therefore,
				$	|\theta^{\pm}_\varphi(\sstraj,t)|=
					\theta^{\pm}_\varphi(\sstraj,t)\overset{\eqref{eq:t_and}}{=}
					\theta^{\pm}_{\formula_1}(\sstraj,t) \ \sqcap\ \theta^{\pm}_{\formula_2}(\sstraj,t)=
					\theta^{\pm}_{\varphi_1}(\sstraj,t)=\trv$.
			If we denote $\theta^{\pm}_{\formula_2}(\sstraj,t)=j$ then 
				$0<\trv=\theta^{\pm}_{\varphi_1}(\sstraj,t) \leq \theta^{\pm}_{\varphi_2}(\sstraj,t)=j$.
			From $\trv\leq j$ and by the induction hypothesis for $\varphi_k$ for $k\in\{1,2\}$ it holds that
			$\forall t'\in t\pm [0,\trv)$,  $\chi_{\varphi_k}(\sstraj,t')=\chi_{\varphi_k}(\sstraj,t)$.
			Thus, using Def.~\ref{def:char_func}, $\forall t'\in t\pm [0,\trv)$ it holds that
				$\chi_{\varphi}(\sstraj,t') =
				\chi_{\varphi_1}(\sstraj,t') \sqcap 
				\chi_{\varphi_2}(\sstraj,t')=
				\chi_{\varphi_1}(\sstraj,t) \sqcap
				\chi_{\varphi_2}(\sstraj,t)=
				\chi_\varphi(\sstraj,t)$.
			
		 \textbf{2.} Let $\chi_{\varphi}(\sstraj,t)= -1$.
				Since $|\theta^{\pm}_\varphi(\sstraj, t)|= \trv>0$
				then due to Thm.~\ref{thm:sat_theta}(\ref{i4}), $\theta^{\pm}_{\formula}(\sstraj,t) = -\trv< 0$.
		WLOG assume that $\theta^{\pm}_{\varphi_1}(\sstraj,t) \leq \theta^{\pm}_{\varphi_2}(\sstraj,t)$. Therefore,
		$
			-|\theta^{\pm}_\varphi(\sstraj,t)|=
			\theta^{\pm}_\varphi(\sstraj,t)\overset{\eqref{eq:t_and}}{=}
			\theta^{\pm}_{\formula_1}(\sstraj,t) \ \sqcap\ \theta^{\pm}_{\formula_2}(\sstraj,t)=
			\theta^{\pm}_{\varphi_1}(\sstraj,t)=-\trv<0
		$.
	Since $\theta^{\pm}_{\varphi_1}(\sstraj,t)<0$ then $\chi_{\varphi_1}(\sstraj,t)=-1$.
			From the induction hypothesis for $\varphi_1$ we get that
			$\forall t'\in t\pm [0,\trv)$,  $\chi_{\varphi_1}(\sstraj,t')=\chi_{\varphi_1}(\sstraj,t)=-1$.
		Thus, by Def.~\ref{def:char_func}, $\forall t'\in t\pm [0,\trv)$ it holds that
		$\chi_{\varphi}(\sstraj,t') =
		\chi_{\varphi_1}(\sstraj,t') \sqcap \chi_{\varphi_2}(\sstraj,t')=
			-1 \sqcap \chi_{\varphi_2}(\sstraj,t')=
				-1=	\chi_\varphi(\sstraj,t)$.
			%%%%%%%%%%%%%%%%%%%%%%%%%%%%%%%%%%%%%%%%%%%%%
			\item $\varphi = \varphi_1 \until_I \varphi_2$.
			We consider the two separate cases of $\chi_{\varphi}(\sstraj,t)= 1$ and $\chi_{\varphi}(\sstraj,t)= -1$ as follows:		
			
		\textbf{1.} Let $\chi_{\varphi}(\sstraj,t)= \po$. 
				%	Therefore, 
				%	\begin{equation}
				%	\etam_\varphi(\sstraj,t) =
				%	\min\{\tau\geq0 \ :\ \forall t'\in[t,t+\tau],\ \chi_{\varphi_1 \until_I \varphi_2}(\sstraj,t')=\po\} \geq 0.	
				%	\end{equation}
				Since $|\theta^{\pm}_\varphi(\sstraj, t)|= \trv>0$
				then due to Thm.~\ref{thm:sat_theta}(\ref{i3}), $\theta^{\pm}_{\formula}(\sstraj,t) = \trv > 0$, therefore, due to \eqref{eq:t_until}, it holds that
				$\bigsqcup_{t'\in t+I} \left(
					\theta^{\pm}_{\formula_2}(\sstraj,t') \ \sqcap \
					\bigsqcap_{t'' \in [t,t')} \theta^{\pm}_{\formula_1}(\sstraj,t'')
					\right) = \trv > 0$.
				To prove the rest we distinguish between the cases of $\trv<\infty$ and $\trv=\infty$ as following:
				
			\textbf{1.1.} Let $\trv<\infty$. Then due to the $\epsilon$ definition of the supremum the following holds:
				\begin{equation}
					\label{eq:11}
					\forall \epsilon>0,\ \exists t'\in t+I,\quad 
					 \theta^{\pm}_{\formula_2}(\sstraj,t') \ \sqcap \
					 \bigsqcap_{t'' \in [t,t')} \theta^{\pm}_{\formula_1}(\sstraj,t'') > \trv-\epsilon 
				\end{equation} 
			which by the definition of the infimum ($\forall x,\ \inf f(x) \leq f(x)$) further leads to:
			\begin{equation}
				\label{eq:ref2}
				\forall \epsilon>0,\ \exists t'\in t+I\ \left(
				\theta^{\pm}_{\formula_2}(\sstraj,t')>\trv-\epsilon
				\quad \wedge\quad 
				\forall t'' \in [t,t'),\ \theta^{\pm}_{\formula_1}(\sstraj,t'') > \trv-\epsilon
				\right)
			\end{equation}
			Note that since $\trv>0$ then using Thm.~\ref{thm:sat_theta}, $\chi_{\varphi_2}(\sstraj,t')=\chi_{\varphi_1}(\sstraj,t'')=\po$. By the induction hypothesis we get that
			\begin{equation*}
				\begin{aligned}
			\forall \epsilon>0,\ &\exists t'\in t+I\ ( \forall \tstep\in [0,\theta^{\pm}_{\formula_2}(\sstraj,t')),\  \chi_{\varphi_2}(\sstraj,t' \pm \tstep)=\po	\ \wedge\\
			&\forall t'' \in [t,t'), \forall \tstep\in [0,\theta^{\pm}_{\formula_1}(\sstraj,t'')),\  \chi_{\varphi_1}(\sstraj,t'' \pm \tstep)=\po
			)
				\end{aligned}			
		\end{equation*}
	where $\theta^{\pm}_{\formula_2}(\sstraj,t') > \trv-\epsilon$ and $\theta^{\pm}_{\formula_1}(\sstraj,t'') > \trv-\epsilon$. 
 Therefore, $[0,\theta^{\pm}_{\formula_2}(\sstraj,t')) \supset [0, \trv-\epsilon)$ and  $[0,\theta^{\pm}_{\formula_1}(\sstraj,t'')) \supset [0, \trv-\epsilon)$.
 Thus,
	\begin{equation*}
		\begin{aligned}
			\forall \epsilon>0,\ &\exists t'\in t+I\ ( \forall \tstep\in [0,\trv-\epsilon),\  \chi_{\varphi_2}(\sstraj,t' \pm \tstep)=\po	\ \wedge\\
			&\forall t'' \in [t,t'), \forall \tstep\in [0,\trv-\epsilon),\  \chi_{\varphi_1}(\sstraj,t'' \pm \tstep)=\po
			)
		\end{aligned}			
	\end{equation*}
	Therefore, due to Lemma \ref{lemma:quant_laws_set}(\ref{i:quant_scope_set}), (\ref{i:forall_exists_set}) and (\ref{i:quant_set}) it holds that
	$\forall \epsilon>0$, $\forall \tstep\in[0,\trv-\epsilon)$, $\exists t'\in t+I$,
	\begin{equation*}
		\chi_{\varphi_2}(\sstraj,t' \pm \tstep)=\po	\ \wedge\ 
			\forall t'' \in [t,t'),\  \chi_{\varphi_1}(\sstraj,t'' \pm \tstep)=\po
	\end{equation*}
Thus, by Def.~\ref{def:char_func},
	$\forall \epsilon>0,\ \forall \tstep\in[0,\trv-\epsilon)$,   
%	the following holds:
%\begin{align*}
%	\chi_{\varphi}(\sstraj,t \pm \tstep)
%	\overset{\text{Def.}~\ref{def:char_func}}{=} 
%	&\bigsqcup_{t'\in t \pm \tstep+ I} \left(
%	\chi_{\formula_2}(\sstraj,t') \ \sqcap \
%	\bigsqcap_{t'' \in [t\pm \tstep,t')} \chi_{\formula_1}(\sstraj,t'') 
%	\right) =\\
%	&\bigsqcup_{t'\in t+ I} \left(
%	\chi_{\formula_2}(\sstraj,t' \pm \tstep) \ \sqcap \
%	\bigsqcap_{t'' \in [t,t')} \chi_{\formula_1}(\sstraj,t'' \pm \tstep) 
%	\right)=\po.
%\end{align*}
%We obtained that $\forall \epsilon>0$, $\forall \tstep\in[0,\trv-\epsilon)$, 
$\chi_{\varphi}(\sstraj,t \pm \tstep)=\po$,  
thus\footnote{\label{footnote_eps} If $\forall \epsilon>0$, $\forall \tstep\in [0,\ \trv-\epsilon)$, $A(h)$ then $\forall \tstep\in [0,\ \trv)$, $A(h)$. Assume the opposite, assume $\exists \tstep^*\in [0,\trv)$ such that $\neg A(\tstep^*)$. Therefore, $\exists \epsilon^*>0$ such that we can write $\tstep^*=\trv-\epsilon^*$ and $\neg A(\tstep^*)$. Let $\epsilon=\frac{\epsilon^*}{2}$ then from the given it follows that 
	$\forall \tstep\in [0,\ \trv-\frac{\epsilon^*}{2})$, $A(\tstep^*)$. But $\tstep^*=\trv-\epsilon^* \in [0, \trv-\frac{\epsilon^*}{2})$, thus, $A(\tstep^*)$ which is a contradiction with $\neg A(\tstep^*)$.}, 
$\forall \tstep\in[0,\trv), \chi_{\varphi}(\sstraj,t \pm \tstep)=\po$.

	\textbf{1.2.} Let $\trv=\infty$. Then the following holds:
	\begin{equation}
		\label{eq:ref3}
		\forall M\in\Re,\ \exists t'\in t+I,\quad
		\theta^{\pm}_{\formula_2}(\sstraj,t') \ \sqcap \
		\bigsqcap_{t'' \in [t,t')} \theta^{\pm}_{\formula_1}(\sstraj,t'') > M 
	\end{equation} 
Which is similar to \eqref{eq:11} but uses $M$ instead of $\trv-\epsilon$. Following the above steps, one can obtain that $\forall M\in\Re$, $\forall \tstep\in[0,M)$, $\chi_{\varphi}(\sstraj,t \pm \tstep)=\po$,  i.e. $\forall \tstep\in[0,\infty)$, $\chi_{\varphi}(\sstraj,t \pm \tstep)=\po$.

			\textbf{2.} Let $\chi_{\varphi}(\sstraj,t)= -1$. 	
			Since $|\theta^{\pm}_\varphi(\sstraj, t)|= \trv>0$
			then due to Thm.~\ref{thm:sat_theta}(\ref{i4}), $\theta^{\pm}_{\formula}(\sstraj,t) = -\trv < 0$, therefore, 
			due to \eqref{eq:t_until}, it holds that 
			$\bigsqcup_{t'\in t+I} \left(
				\theta^{\pm}_{\formula_2}(\sstraj,t') \ \sqcap \
				\bigsqcap_{t'' \in [t,t')} \theta^{\pm}_{\formula_1}(\sstraj,t'')
				\right) = -\trv < 0$.
			To prove the rest we distinguish between the cases of $\trv=\infty$ and $\trv<\infty$ as following:		
			
			\textbf{2.1} Let $\trv=\infty$.
			Then we conclude that $\forall t'\in t+I$, $\theta^{\pm}_{\formula_2}(\sstraj,t')= -\infty$  or 
			$\bigsqcap_{t'' \in [t,t')} \theta^{\pm}_{\formula_1}(\sstraj,t'') = -\infty$.
			Therefore, by the definition of an infimum (unbounded set), the following holds:
			\begin{equation}
				\label{eq:ref15}
				\forall t'\in t+I\ 
				\left(\theta^{\pm}_{\formula_2}(\sstraj,t')= -\infty \ \vee \
				\forall M\in\Re_+,\ \exists t''\in[t,t'),\ 
				\theta^{\pm}_{\formula_1}(\sstraj,t'') < -M
				\right)
			\end{equation}
			By the induction hypothesis and Lemma \ref{lemma:quant_laws_set}(\ref{i:quant_set}), (\ref{i:forall_set}) and (\ref{i:quant_scope_set}), it holds that $\forall M\in\Re_+$, $\forall \tstep\in[0,M)$, $\chi_{\varphi}(\sstraj,t \pm \tstep)=-1$,  i.e. $\forall \tstep\in[0,\infty)$, $\chi_{\varphi}(\sstraj,t \pm \tstep)=-1$.  	
				
		\textbf{2.2} Let $r<\infty$.
		The supremum operator over $t+I$ leads to each term being some $-j_{t'}\leq -\trv$, and then the infimum operator leads to 
		the following:
			\begin{equation}
				\label{eq:j}
				\forall t'\in t+I,\ \exists j_{t'}\geq \trv >0\  \left(\theta^{\pm}_{\formula_2}(\sstraj,t')= -j_{t'}  \vee
				\bigsqcap_{t'' \in [t,t')} \theta^{\pm}_{\formula_1}(\sstraj,t'') = -j_{t'}\right)	
			\end{equation}
	Take any $t'\in t+I$, we will work with each $j_{t'}$ separately. We will distinguish between the cases of $j_{t'}<\infty$ and $j_{t'}=\infty$ as following:		
		\begin{enumerate}
			\item Let $j_{t'} <\infty$. Then from \eqref{eq:j} by the definition of infimum, the following holds:
			\begin{equation}
				\label{eq:ref4}
				\theta^{\pm}_{\formula_2}(\sstraj,t')= -j_{t'} \quad \vee\quad
				\forall \epsilon>0, \ \exists t''\in [t,t'),\ 
				\theta^{\pm}_{\formula_1}(\sstraj,t'') < -j_{t'}+\epsilon
			\end{equation}
		Note, that $|\theta^{\pm}_{\formula_2}(\sstraj,t')| = -\theta^{\pm}_{\formula_2}(\sstraj,t')= j_{t'} \geq \trv> \trv-\epsilon$ and $|\theta^{\pm}_{\formula_1}(\sstraj,t'')| =-\theta^{\pm}_{\formula_1}(\sstraj,t'') > j_{t'}-\epsilon \geq \trv-\epsilon$. 
		Therefore, by the induction hypothesis and Lemma \ref{lemma:quant_laws_set}(\ref{i:quant_set}),(\ref{i:forall_set}) 
%		\begin{align*}
%			&\forall \epsilon>0,\ \forall \tstep\in [0,\ \trv-\epsilon),\ \chi_{\formula_2}(\sstraj,t' \pm \tstep)= -1 \\&\qquad\vee\quad
%			\forall \epsilon>0,\ \exists t''\in [t,t'),\  
%			\forall \tstep\in [0, \trv-\epsilon),\ 
%			\chi_{\formula_1}(\sstraj,t'' \pm \tstep) =-1. 
%		\end{align*} 
%	Due to \ref{lemma:quant_laws_set}, items \ref{i:quant_set} and \ref{i:forall_set}
%		it follows that
it holds that $\forall \epsilon>0$, $\forall \tstep\in [0,\ \trv-\epsilon)$, 			$\chi_{\formula_2}(\sstraj,t' \pm \tstep)\ \sqcap
\bigsqcap_{t'' \in [t,t')}   
\chi_{\formula_1}(\sstraj,t'' \pm \tstep)=-1$, thus\footref{footnote_eps}, it holds that
	\begin{equation}
		\label{eq:j_finite}
		\forall \tstep\in[0, \trv) \qquad 
		\chi_{\formula_2}(\sstraj,t' \pm \tstep)\ \sqcap
		\bigsqcap_{t'' \in [t,t')}   
		\chi_{\formula_1}(\sstraj,t'' \pm \tstep)=-1. 
	\end{equation}
		\item Let $j_{t'} =\infty$. Then from \eqref{eq:j} by the definition of infimum, the following holds:
		\begin{equation}
			\label{eq:ref5}
			\theta^{\pm}_{\formula_2}(\sstraj,t')= -\infty \ \vee \
			\forall M\in\Re_+,\ \exists t''\in[t,t'),\ 
			\theta^{\pm}_{\formula_1}(\sstraj,t'') < -M
		\end{equation}
	Since $|\theta^{\pm}_{\formula_1}(\sstraj,t'')| = -\theta^{\pm}_{\formula_1}(\sstraj,t'')>M$ then applying the induction hypothesis leads to:
	\begin{align*}
		&\forall \tstep\in [0,\ \infty),\ \chi_{\formula_2}(\sstraj,t' \pm \tstep)= -1 \\&\qquad\vee\quad
		\forall M\in\Re_+,\ \exists t''\in [t,t'),\  
		\forall \tstep\in [0,\ M),\ 
		\chi_{\formula_1}(\sstraj,t'' \pm \tstep) =-1. 
	\end{align*}
Since $\trv$ is finite, $[0, \trv)\subset [0,\infty)$ and the above property holds for any $M \geq \trv$. Now due to Lemma \ref{lemma:quant_laws_set}(\ref{i:quant_set}), (\ref{i:forall_set}) we conclude that the following holds:
%		\begin{align*}
%			&\forall \tstep\in [0,\ \trv),\ \chi_{\formula_2}(\sstraj,t' \pm \tstep)= -1 \\&\qquad\vee\quad
%			\exists t''\in [t,t'),\  
%			\forall \tstep\in [0,\ \trv),\ 
%			\chi_{\formula_1}(\sstraj,t'' \pm \tstep) =-1. 
%		\end{align*}
%		Due to \ref{lemma:quant_laws_set}, items \ref{i:quant_set} and \ref{i:forall_set} it follows that 
		\begin{equation}
			\label{eq:j_inf}
		\forall \tstep\in[0,\trv) \qquad 
		\chi_{\formula_2}(\sstraj,t' \pm \tstep)\ \sqcap
		\bigsqcap_{t'' \in [t,t')}   
		\chi_{\formula_1}(\sstraj,t'' \pm \tstep)=-1. 
	\end{equation}
		\end{enumerate}
		 Note that \eqref{eq:j_finite} and \eqref{eq:j_inf} are equivalent, therefore, they hold regardless of each $j_{t'}$ value. Thus, it holds that 
	$\forall t'\in t+I$, 
		  $\forall \tstep\in[0, \trv)$,  
		  	$\chi_{\formula_2}(\sstraj,t' \pm \tstep)\ \sqcap
		  	\bigsqcap_{t'' \in [t,t')}   
		  	\chi_{\formula_1}(\sstraj,t'' \pm \tstep)=-1$,
	  	which due to Lemma \ref{lemma:quant_laws_set}(\ref{i:quant_scope_set}) leads to $\forall \tstep\in[0, \trv)$, $\chi_{\varphi}(\sstraj,t \pm \tstep)=-1$.
\end{enumerate}
%\subsection{Proof of Theorem \ref{thm:chi_async_disc}}
\subsection{Proof of Eq.\eqref{eq:chi_theta_disc} (Theorem \ref{thm:chi_async} for discrete-time)} 

Let $\varphi$ be an STL formula, $\sstraj:\Ze \rightarrow X$ be a discrete-time signal and $t\in\Ze$ be a time point. For any value $\trv\in\Ze_{\geq 0}$, we want to prove that
if 
$|\theta^{\pm}_\varphi(\sstraj,t)| = \trv$  then $\forall t'\in t \pm \,[0,\trv]$, $\chi_\varphi(\sstraj,t')=\chi_\varphi(\sstraj,t)$.
Note that the result is trivial for $\trv=0$. 
Therefore, let $|\theta^{\pm}_\varphi(\sstraj, t)|= \trv>0$. 
The rest of the proof is by induction on the structure of the formula $\varphi$ and follows the similar steps as Section \ref{sec:chi_async}, therefore, we only present the base case.

\textbf{Base case} $\varphi=p$. Since $|\theta^{\pm}_p(\sstraj, t)|= \trv$ then due to Corollary \ref{rem:eq}, $|\eta^{\pm}_p(\sstraj,t)|=\trv$. From \eqref{eq:chi_eta_disc}, i.e., Theorem \ref{thm:chi_sync_stronger} for discrete-time, we get that
$\forall t'\in t\pm [0,\trv],\ \chi_p(\sstraj,t')=\chi_p(\sstraj,t)$.

\subsection{Proof of Theorem \ref{thm:theta_bound}}
\label{sec:theta_bound}

Let $\varphi$ be an STL formula, $\sstraj:\Re \rightarrow X$ be a continuous-time signal and $t\in\Re$ be a time point. For any value $\trv\in\CoRe_{\geq 0}$, we want to prove that
if 
$|\theta^{\pm}_\varphi(\sstraj,t)| = \trv$  then $\forall \tau\in [0,\trv)$, $|\theta^{\pm}_{\varphi}(\sstraj,t \pm \tstep)| \geq \trv -\tstep$.

Note that the result is trivial for $\trv=0$. 
Therefore, let $|\theta^{\pm}_\varphi(\sstraj, t)|= \trv>0$. 
The rest of the proof is by induction on the structure of the formula $\varphi$.  	
\begin{enumerate}[label={\textbf{Case}}]
		\item $\varphi=p$. 
		Let  $|\theta^{\pm}_p(\sstraj, t)|= \trv$. Then due to Cor.~\ref{rem:eq}, 
		$|\eta^{\pm}_p(\sstraj, t)|= \trv$ and by Theorem~\ref{thm:eta_decrease}, 
		$\forall \tstep\in [0, \trv)$, $|\eta^{\pm}_{p}(\sstraj,t \pm \tstep)| = \trv -\tstep$. Which again due to Cor.~\ref{rem:eq} leads to  $\forall \tstep\in [0, \trv)$, $|\theta^{\pm}_{p}(\sstraj,t \pm \tstep)| = \trv -\tstep$.
		
		\item $\varphi = \neg\varphi_1$.
		Since $|\theta^{\pm}_{\neg\varphi_1}(\sstraj,t)| =r$ then due to \eqref{eq:t_neg} it holds that $|\theta^{\pm}_{\varphi_1}(\sstraj,t)|=r$. By the induction hypothesis for $\varphi_1$ we get that
		$\forall \tstep\in [0, \trv)$, $|\theta^{\pm}_{\varphi_1}(\sstraj,t \pm \tstep)| \geq \trv -\tstep$. 
		 Therefore, due to Def.~\ref{def:time_rob_rec}, it holds that
	$\forall \tstep\in [0,\trv)$, $|\theta^{\pm}_{\neg\varphi_1}(\sstraj,t \pm \tstep)|=
	|\theta^{\pm}_{\varphi_1}(\sstraj,t \pm \tstep)| \geq \trv -\tstep$.
		%%%%%%%%%%%%%%%%%%%%%%%%%%%%%%%%%%%%%%%%%%%%%
		\item $\varphi=\varphi_1 \wedge \varphi_2$.
			We consider the two separate cases of $\chi_{\varphi}(\sstraj,t)= 1$ and $\chi_{\varphi}(\sstraj,t)= -1$ as follows:	
		
		\textbf{1.} Let $\chi_{\varphi}(\sstraj,t)= 1$. 
			Since $|\theta^{\pm}_\varphi(\sstraj, t)|= \trv>0$
			then due to Thm.~\ref{thm:sat_theta}(\ref{i3}), $\theta^{\pm}_{\formula}(\sstraj,t) = \trv> 0$. 
		WLOG assume that $\theta^{\pm}_{\varphi_1}(\sstraj,t) \leq \theta^{\pm}_{\varphi_2}(\sstraj,t)$. Therefore,
			$	|\theta^{\pm}_\varphi(\sstraj,t)|=
			\theta^{\pm}_\varphi(\sstraj,t)\overset{\eqref{eq:t_and}}{=}
			\theta^{\pm}_{\formula_1}(\sstraj,t) \ \sqcap\ \theta^{\pm}_{\formula_2}(\sstraj,t)=
			\theta^{\pm}_{\varphi_1}(\sstraj,t)=\trv$.
			If we denote $\theta^{\pm}_{\formula_2}(\sstraj,t)=j$ then 
			$0<\trv=\theta^{\pm}_{\varphi_1}(\sstraj,t) \leq \theta^{\pm}_{\varphi_2}(\sstraj,t)=j$.
		Since $\chi_{\varphi}(\sstraj,t)=\po$ then due to Def.~\ref{def:char_func}, $\chi_{\varphi_1}(\sstraj,t)=\chi_{\varphi_2}(\sstraj,t)=\po$ and since $\trv\leq j$ then according to Thm.~\ref{thm:chi_async}, 
				 $\forall t'\in t\pm [0, \trv),\ \chi_{\varphi_k}(\sstraj,t')=\po$ for both $k\in\{1,2\}$. Due to Thm.~\ref{thm:sat_theta}, item~\ref{i3}, $\forall t'\in t\pm [0, \trv)$, $\theta^{\pm}_{\varphi_k}(\sstraj, t') \geq 0$ and thus, $\theta^{\pm}_{\varphi}(\sstraj, t') \geq 0$.
			From $\trv\leq j$ and the induction hypothesis we get that
				$\forall \tstep\in [0, \trv)$, $|\theta^{\pm}_{\varphi_1}(\sstraj,t \pm \tstep)| \geq \trv -\tstep$
				and $\forall \tstep\in [0, j)$, $|\theta^{\pm}_{\varphi_2}(\sstraj,t \pm \tstep)| \geq j -\tstep \geq \trv-\tstep$.	
			Since $[0, \trv)\subseteq[0, j)$, using Def.~\ref{def:time_rob_rec} we can conclude that $\forall \tstep\in [0,\trv)$,
				$|\theta^{\pm}_{\varphi}(\sstraj,t \pm \tstep)| =
				\theta^{\pm}_{\varphi}(\sstraj,t \pm \tstep) =
				\theta^{\pm}_{\varphi_1}(\sstraj,t \pm \tstep) \sqcap 
				\theta^{\pm}_{\varphi_2}(\sstraj,t \pm \tstep) \geq\trv-\tstep$.
		
		\textbf{2.} Let $\chi_{\varphi}(\sstraj,t)= -1$.
			Since $|\theta^{\pm}_\varphi(\sstraj, t)|= \trv>0$
			then due to Thm.~\ref{thm:sat_theta}(\ref{i4}), $\theta^{\pm}_{\formula}(\sstraj,t) = -\trv< 0$.
			WLOG assume that $\theta^{\pm}_{\varphi_1}(\sstraj,t) \leq \theta^{\pm}_{\varphi_2}(\sstraj,t)$. Therefore,
			$
			-|\theta^{\pm}_\varphi(\sstraj,t)|=
			\theta^{\pm}_\varphi(\sstraj,t)\overset{\eqref{eq:t_and}}{=}
			\theta^{\pm}_{\formula_1}(\sstraj,t) \ \sqcap\ \theta^{\pm}_{\formula_2}(\sstraj,t)=
			\theta^{\pm}_{\varphi_1}(\sstraj,t)=-\trv<0$, i.e. $|\theta^{\pm}_{\varphi_1}(\sstraj,t)| = -\theta^{\pm}_{\varphi_1}(\sstraj,t)$.
			Now from the induction hypothesis for $\varphi_1$ we get that
			$\forall \tstep\in [0, \trv)$, $|\theta^{\pm}_{\varphi_1}(\sstraj,t \pm \tstep)| \geq \trv -\tstep$.
		Note that since $\theta^{\pm}_{\varphi}(\sstraj,t)=-\trv$ then due to Thm.~\ref{thm:chi_async}, $\forall t'\in t\pm [0,\trv)$, $\chi_{\varphi}(\sstraj, t') =-1$ and thus,  
		due to Thm.~\ref{thm:sat_theta}, item~\ref{i4}, and thus, $\theta^{\pm}_{\varphi}(\sstraj, t') \leq 0$.
			Thus, using Def.~\ref{def:time_rob_rec} we can conclude that $\forall \tstep\in [0,\trv)$,
				$|\theta^{\pm}_{\varphi}(\sstraj,t \pm \tstep)| =
				-\theta^{\pm}_{\varphi}(\sstraj,t \pm \tstep) 
				=-\theta^{\pm}_{\varphi_1}(\sstraj,t \pm \tstep) \sqcup 
				-\theta^{\pm}_{\varphi_2}(\sstraj,t \pm \tstep)
				=\  
				|\theta^{\pm}_{\varphi_1}(\sstraj,t \pm \tstep)| \sqcup 
				-\theta^{\pm}_{\varphi_2}(\sstraj,t \pm \tstep) 
				\geq \trv-\tstep$.
		%%%%%%%%%%%%%%%%%%%%%%%%%%%%%%%%%%%%%%%%%%%%%
		\item $\varphi = \varphi_1\until_I \varphi_2$. The proof is analogous to Section \ref{sec:chi_async}.
	We consider the two separate cases of $\chi_{\varphi}(\sstraj,t)= 1$ and $\chi_{\varphi}(\sstraj,t)= -1$ as follows:	

\textbf{1.} Let $\chi_{\varphi}(\sstraj,t)= \po$. Since 
$\chi_{\varphi}(\sstraj,t)= \po$ then due to Thm.~\ref{thm:chi_async}, $\forall \tstep\in [0, \trv)$, $\chi_\formula(\sstraj,t \pm \tstep)=\po$ and thus, due to Thm.~\ref{thm:sat_theta}, 
$\forall \tstep\in [0, \trv)$, $\theta^{\pm}_\formula(\sstraj,t \pm \tstep) \geq 0$.
Same as in Section \ref{sec:chi_async} to prove the rest we distinguish between the cases of $\trv<\infty$ and $\trv=\infty$ as following:

	%	Therefore, 
	%	\begin{equation}
		%	\etam_\varphi(\sstraj,t) =
		%	\min\{\tau\geq0 \ :\ \forall t'\in[t,t+\tau],\ \chi_{\varphi_1 \until_I \varphi_2}(\sstraj,t')=\po\} \geq 0.	
		%	\end{equation}

	\textbf{1.1.} Let $\trv<\infty$. Then \eqref{eq:ref2} holds.
	Note that since $\trv>0$ then 
	$|\theta^{\pm}_{\formula_2}(\sstraj,t')|= \theta^{\pm}_{\formula_2}(\sstraj,t')>0$ and $|\theta^{\pm}_{\formula_1}(\sstraj,t'')|=\theta^{\pm}_{\formula_1}(\sstraj,t'')>0$ and then according to Thm.~\ref{thm:chi_async} and Thm.~\ref{thm:sat_theta}, item~\ref{i3},
	$\forall \tstep\in [0, \theta^{\pm}_{\formula_2}(\sstraj,t'))$, 
	$\theta^{\pm}_{\varphi_2}(\sstraj, t' \pm \tstep) \geq 0$ and 
	$\forall \tstep\in [0, \theta^{\pm}_{\formula_1}(\sstraj,t''))$, 
	 $\theta^{\pm}_{\varphi_1}(\sstraj, t'' \pm \tstep) \geq 0$.
	Applying the induction hypothesis we get:
	\begin{equation*}
		\begin{aligned}
			\forall \epsilon>0,\ &\exists t'\in t+I\ ( \forall \tstep\in [0,\theta^{\pm}_{\formula_2}(\sstraj,t')),\  \theta^{\pm}_{\formula_2}(\sstraj,t' \pm \tstep)\geq \theta^{\pm}_{\formula_2}(\sstraj,t')-\tstep	\ \wedge\\
			&\forall t'' \in [t,t'), \forall \tstep\in [0,\theta^{\pm}_{\formula_1}(\sstraj,t'')),\  \theta^{\pm}_{\varphi_1}(\sstraj,t'' \pm \tstep)\geq \theta^{\pm}_{\varphi_1}(\sstraj,t'') - \tstep
			)
		\end{aligned}			
	\end{equation*}
	where $\theta^{\pm}_{\formula_2}(\sstraj,t') > \trv-\epsilon$ and $\theta^{\pm}_{\formula_1}(\sstraj,t'') > \trv-\epsilon$. 
	Therefore, $[0,\theta^{\pm}_{\formula_2}(\sstraj,t')) \supset [0,\trv-\epsilon)$ and  $[0,\theta^{\pm}_{\formula_1}(\sstraj,t'')) \supset [0, \trv-\epsilon)$.
	Next, applying Lemma \ref{lemma:quant_laws_set}(\ref{i:quant_scope_set}),(\ref{i:forall_exists_set}),(\ref{i:quant_set}) and Def.~\ref{def:time_rob_rec}, it holds that $\forall \epsilon>0$, $\forall \tstep\in[0,\trv-\epsilon)$, $\theta^{\pm}_{\varphi}(\sstraj,t \pm \tstep) >\trv-\epsilon-\tstep$.
%	\begin{equation*}
%		\begin{aligned}
%			\forall \epsilon>0,\ &\exists t'\in t+I\ ( \forall \tstep\in [0,\trv-\epsilon),\  \theta^{\pm}_{\formula_2}(\sstraj,t' \pm \tstep)>\trv-\epsilon-\tstep	\ \wedge\\
%			&\forall t'' \in [t,t'), \forall \tstep\in [0,\trv-\epsilon),\  \theta^{\pm}_{\varphi_1}(\sstraj,t'' \pm \tstep) >\trv-\epsilon-\tstep
%			)
%		\end{aligned}			
%	\end{equation*} 
%\begin{equation*}
%		\theta^{\pm}_{\formula_2}(\sstraj,t' \pm \tstep)>\trv-\epsilon-\tstep	\ \wedge\ 
%		\forall t'' \in [t,t'),\  \theta^{\pm}_{\varphi_1}(\sstraj,t'' \pm \tstep) >\trv-\epsilon-\tstep		
%\end{equation*}
%Thus, due to , the following holds:
Thus\footnote{\label{footnote}
	Assume the opposite, assume $\exists \tstep=\tstep^*\in [0,\trv)$, $\theta^{\pm}_{\formula}(\sstraj,t \pm \tstep^*) < \trv-\tstep^*$.
	Thus, if we set $\epsilon^* = \frac{\trv-\tstep^*-\theta^{\pm}_{\formula}(\sstraj,t\pm \tstep^*)}{2}>0$ (in other words, $\epsilon^*+\theta^{\pm}_{\formula}(\sstraj,t\pm \tstep^*) =  \trv-\epsilon^* -\tstep^*$) then from what is given, it holds that
	$\forall \tstep\in [0, \trv-\epsilon^*)$, $\theta^{\pm}_{\formula}(\sstraj,t\pm \tstep) > \trv-\epsilon^*-\tstep$.
	
	\begin{enumerate}
		\item First we show that $\tstep^* \in [0, \trv-\epsilon^*)$. 
		We know that $\tstep^* \in  [0, \trv)$, it follows that $\tstep^*-\trv<0$. We also know that $\theta^{\pm}_\varphi(\sstraj, t\pm \tstep^*) \geq 0$. Combining the two we get $\tstep^*-\trv< \theta^{\pm}_\varphi(\sstraj, t\pm \tstep^*)$. Since 
		$\theta^{\pm}_{\formula}(\sstraj,t\pm \tstep^*) =  \trv-2\epsilon^* -\tstep^*$ then $\tstep^*-\trv< \trv-2\epsilon^* -\tstep^*$ which leads to $\tstep^*-\trv < \epsilon^*$, i.e. $\tstep^* \in [0, \trv-\epsilon^*)$.
		
		\item Now since  $\tstep^* \in [0, \trv-\epsilon^*)$ then due to the above, 
		$\theta^{\pm}_{\formula}(\sstraj,t \pm \tstep^*) > \trv-\epsilon^*-\tstep^*$. 
		Since $\epsilon^*+\theta^{\pm}_{\formula}(\sstraj,t\pm \tstep^*) =  \trv-\epsilon^* -\tstep^*$, then   $\theta^{\pm}_{\formula}(\sstraj,t \pm \tstep^*)  > \epsilon^* +\theta^{\pm}_{\formula}(\sstraj,t \pm \tstep^*)$
		which means that $\epsilon^* <0$. 
		which is a contradiction. 
	\end{enumerate} 
}, it holds that
	$\forall \tstep\in [0, \trv)$, $\theta^{\pm}_{\formula}(\sstraj,t \pm \tstep) \geq \trv-\tstep$.
	
%		We showed, that $	\forall \tstep\in [0, i)$, $\theta^{\pm}_{\formula}(\sstraj,t\pm \tstep) \geq i-\tstep $.

	\textbf{1.2.} Let $\trv=\infty$. Then \eqref{eq:ref3} holds. 
Using the definition of an infimum (unbounded set) together with an induction hypothesis we can conclude that $\forall M\in\Re$, $\forall \tstep\in[0,M)$, $\theta^\pm_{\varphi}(\sstraj,t \pm \tstep) > M-\tstep$,  thus\footnote{\label{ft:infty} Assume the opposite, assume that $\exists \tstep^*\in[0,\infty)$, $\theta^\pm_{\varphi}(\sstraj,t \pm \tstep^*)<\infty$. Since $	\theta^{\pm}_\formula(\sstraj,t)=\infty>0$ then due to Thms.~\ref{thm:chi_async} and \ref{thm:sat_theta} we know that $\theta^\pm_{\varphi}(\sstraj,t \pm \tstep^*) \geq 0$. Thus, 
	%$\exists \tstep^*\in[0,\infty)$, $\exists j \geq 0$, $\theta^\pm_{\varphi}(\sstraj,t \pm \tstep^*) =j$. This means 
	$\exists \tstep^*\in[0,\infty)$, $\exists j > 0$, $\theta^\pm_{\varphi}(\sstraj,t \pm \tstep^*) <j$. Let $M=j+\tstep^*$ then $\exists \tstep^*\in[0,\infty)$, $\exists M>\tstep^*$, $\theta^\pm_{\varphi}(\sstraj,t \pm \tstep^*) <M - \tstep^*$ but we are given that $\forall M\in\Re$, $\forall \tstep\in[0,M)$, $\theta^\pm_{\varphi}(\sstraj,t \pm \tstep) > M-\tstep$ which is a contradiction.}, $\forall \tstep\in[0,\infty)$, $\theta^\pm_{\varphi}(\sstraj,t \pm \tstep)=\infty$.

\textbf{2.} Let $\chi_{\varphi}(\sstraj,t)= -1$. Since 
$\chi_{\varphi}(\sstraj,t)= -1$ then due to Thm.~\ref{thm:chi_async}, $\forall \tstep\in [0, \trv)$, $\chi_\formula(\sstraj,t \pm \tstep)=-1$ and thus, due to Thm.~\ref{thm:sat_theta}, 
$\forall \tstep\in [0, \trv)$, $\theta^{\pm}_\formula(\sstraj,t \pm \tstep) \leq 0$.
Same as in Section \ref{sec:chi_async} to prove the rest we distinguish between the cases of $\trv=\infty$ and $\trv<\infty$ as following:

	\textbf{2.1.} Let $\trv=\infty$. Then \eqref{eq:ref15} holds.
By the induction hypothesis and Lemma \ref{lemma:quant_laws_set}(\ref{i:quant_set}),(\ref{i:forall_set}),(\ref{i:quant_scope_set}), it holds that $\forall M\in\Re_+$, $\forall \tstep\in[0,M)$, $\theta^{\pm}_{\varphi}(\sstraj,t \pm \tstep) \leq -M+\tstep$,  and thus\footref{ft:infty}, $\forall \tstep\in[0,\infty)$, $\theta^{\pm}_{\varphi}(\sstraj,t \pm \tstep) = -\infty$, 
which means 
$\forall \tstep\in[0, \trv)$, 
$|\theta^{\pm}_{\varphi}(\sstraj,t \pm \tstep)| \geq \trv-\tstep$. 

\textbf{2.2.} Let $\trv<\infty$. Then \eqref{eq:j} holds.
Take any $t'\in t+I$. Same as in Section \ref{sec:chi_async}, we will work with each $j_{t'}$ separately. We will distinguish between the cases of $j_{t'}<\infty$ and $j_{t'}=\infty$ as following:		
\begin{enumerate}
	\item Let $j_{t'} <\infty$. Then \eqref{eq:ref4} holds.
	Note, that $|\theta^{\pm}_{\formula_2}(\sstraj,t')| = -\theta^{\pm}_{\formula_2}(\sstraj,t')= j_{t'} \geq \trv> \trv-\epsilon$ and $|\theta^{\pm}_{\formula_1}(\sstraj,t'')| =-\theta^{\pm}_{\formula_1}(\sstraj,t'') > j_{t'}-\epsilon \geq \trv-\epsilon$. 
	Then according to Thm.~\ref{thm:chi_async} and Thm.~\ref{thm:sat_theta}, item~\ref{i4},
	$\forall \tstep\in [0, -\theta^{\pm}_{\formula_2}(\sstraj,t'))$, 
	$\theta^{\pm}_{\varphi_2}(\sstraj, t' \pm \tstep) \leq 0$ and
	$\forall \tstep\in [0, -\theta^{\pm}_{\formula_1}(\sstraj,t''))$, 
	$\theta^{\pm}_{\varphi_1}(\sstraj, t'' \pm \tstep) \leq 0$.
	Therefore, by the induction hypothesis and Lemma \ref{lemma:quant_laws_set}(\ref{i:quant_set}),(\ref{i:forall_set}),
	$\forall \epsilon>0$, $\forall \tstep\in [0,\ \trv-\epsilon)$,
	$\theta^{\pm}_{\formula_2}(\sstraj,t' \pm \tstep)\ \sqcap
	\bigsqcap_{t'' \in [t,t')}   
	\theta^{\pm}_{\formula_1}(\sstraj,t'' \pm \tstep)< -\trv+\epsilon+\tstep$, 
%	Applying the induction hypothesis we get:
%	\begin{equation*}
%		\begin{aligned}
%			&\forall \epsilon>0,\ \forall \tstep\in [0,\trv-\epsilon),\quad  -\theta^{\pm}_{\formula_2}(\sstraj,t' \pm \tstep)> \trv-\epsilon-\tstep	\ \vee\\
%			&\forall \epsilon>0,\ 
%			\exists t'' \in [t,t'), \forall \tstep\in [0,\trv-\epsilon),\quad  -\theta^{\pm}_{\varphi_1}(\sstraj,t'' \pm \tstep)> \trv-\epsilon- \tstep
%		\end{aligned}			
%	\end{equation*}
%	Due to \ref{lemma:quant_laws_set}, items \ref{i:quant_set} and \ref{i:forall_set}
%it follows that
%\begin{equation*}
%	\forall \epsilon>0,\ \forall \tstep\in [0,\ \trv-\epsilon)\quad
%	\theta^{\pm}_{\formula_2}(\sstraj,t' \pm \tstep)\ \sqcap
%	\bigsqcap_{t'' \in [t,t')}   
%	\theta^{\pm}_{\formula_1}(\sstraj,t'' \pm \tstep)< -\trv+\epsilon+\tstep. 
%\end{equation*}
thus~\footref{footnote}, it holds that
\begin{equation}
	\label{eq:fin_j}
	\forall \tstep\in [0,\ \trv)\quad
	\theta^{\pm}_{\formula_2}(\sstraj,t' \pm \tstep)\ \sqcap
	\bigsqcap_{t'' \in [t,t')}   
	\theta^{\pm}_{\formula_1}(\sstraj,t'' \pm \tstep) \leq-\trv+\tstep. 
\end{equation}

\item Let $j_{t'} =\infty$.  Then \eqref{eq:ref5} holds. 
According to Thm.~\ref{thm:chi_async} and Thm.~\ref{thm:sat_theta}(\ref{i4}),
$\forall \tstep\in [0, -\theta^{\pm}_{\formula_2}(\sstraj,t'))$, 
$\theta^{\pm}_{\varphi_2}(\sstraj, t' \pm \tstep) \leq 0$ and
$\forall \tstep\in [0, -\theta^{\pm}_{\formula_1}(\sstraj,t''))$, 
$\theta^{\pm}_{\varphi_1}(\sstraj, t'' \pm \tstep) \leq 0$. Therefore, 
$\forall \tstep\in [0, \infty)$, 
$\theta^{\pm}_{\varphi_2}(\sstraj, t' \pm \tstep) \leq 0$ and 
$\forall \tstep\in [0, M)$, 
$\theta^{\pm}_{\varphi_1}(\sstraj, t'' \pm \tstep) \leq 0$.
The induction hypothesis leads to:
\begin{align*}
	&\forall \tstep\in [0,\ \infty),\ -\theta^{\pm}_{\formula_2}(\sstraj,t' \pm \tstep) = \infty \qquad\vee\quad\\
	&\forall M\in\Re_+,\ \exists t''\in [t,t'),\  
	\forall \tstep\in [0,\ M),\ 
	-\theta^{\pm}_{\formula_1}(\sstraj,t'' \pm \tstep) \geq M- \tstep. 
\end{align*}
Since $\trv$ is finite, $[0, \trv)\subset [0,\infty)$ and above property holds for $M\geq \trv$ and thus, 
%\begin{align*}
%	&\forall \tstep\in [0,\ \trv),\ \theta^{\pm}_{\formula_2}(\sstraj,t' \pm \tstep)< -\trv+\tstep\\&\qquad\vee\quad
%	\exists t''\in [t,t'),\  
%	\forall \tstep\in [0,\ \trv),\ 
%	\theta^{\pm}_{\formula_1}(\sstraj,t'' \pm \tstep)  \leq -\trv +\tstep. 
%\end{align*}
due to Lemma \ref{lemma:quant_laws_set}(\ref{i:quant_set}),(\ref{i:forall_set}) it follows that 
\begin{equation}
	\label{eq:inf_j}
	\forall \tstep\in[0, \trv) \qquad 
	\theta^{\pm}_{\formula_2}(\sstraj,t' \pm \tstep)\ \sqcap
	\bigsqcap_{t'' \in [t,t')}   
	\theta^{\pm}_{\formula_1}(\sstraj,t'' \pm \tstep) \leq -\trv+\tstep. 
\end{equation}

\end{enumerate}

Note that \eqref{eq:fin_j} and \eqref{eq:inf_j} are equivalent, therefore, they hold regardless of each $j_{t'}$ value. 
Thus, it holds that	$\forall t'\in t+I$, 
	$\forall \tstep\in[0, \trv)$,  
	$\theta^{\pm}_{\formula_2}(\sstraj,t' \pm \tstep)\ \sqcap
	\bigsqcap_{t'' \in [t,t')}   
	\theta^{\pm}_{\formula_1}(\sstraj,t'' \pm \tstep) \leq -\trv+\tstep$, 
which due to Lemma \ref{lemma:quant_laws_set}(\ref{i:quant_scope_set}) leads to $\forall \tstep\in[0, \trv)$, 
$\theta^{\pm}_{\varphi}(\sstraj,t \pm \tstep) \leq -\trv+\tstep$, thus, 
$\forall \tstep\in[0, \trv)$, 
$|\theta^{\pm}_{\varphi}(\sstraj,t \pm \tstep)| \geq \trv-\tstep$.

\end{enumerate}

%\subsection{Proof of Lemma \ref{lemma:theta_bound_disc}}
\subsection{Proof of Eq.\eqref{eq:theta_decrease_disc} (Theorem \ref{thm:theta_bound} for discrete-time)}

Let $\varphi$ be an STL formula, $\sstraj:\Ze \rightarrow X$ be a discrete-time signal and $t\in\Ze$ be a time point. For any value $\trv\in\Ze_{\geq 0}$, we want to prove that
if 
$|\theta^{\pm}_\varphi(\sstraj,t)| = \trv$  then $\forall \tau\in [0,\trv]$, $|\theta^{\pm}_{\varphi}(\sstraj,t \pm \tstep)| \geq \trv -\tstep$.
Note that the result is trivial for $\trv=0$. 
Therefore, let $|\theta^{\pm}_\varphi(\sstraj, t)|= \trv>0$. 
The rest of the proof is by induction on the structure of the formula $\varphi$ and follows the similar steps as Section \ref{sec:theta_bound}, therefore, we only present the base case.

\textbf{Base case} $\varphi=p$. Since $|\theta^{\pm}_p(\sstraj, t)|= \trv$ then due to Corollary \ref{rem:eq}, $|\eta^{\pm}_p(\sstraj,t)|=\trv$. From \eqref{eq:eta_decrease_disc}, i.e., Theorem \ref{thm:eta_decrease} for discrete-time, we get that
$\forall \tstep\in [0, \trv]$, $|\eta^{\pm}_{\varphi}(\sstraj,t \pm \tstep)| = \trv -\tstep$.
Which again due to Corollary \ref{rem:eq} leads to  $\forall \tstep\in [0, \trv]$, $|\theta^{\pm}_{p}(\sstraj,t \pm \tstep)| = \trv -\tstep$.

\subsection{Proof of Theorem \ref{thm:theta_bounded}}

%$|\theta^{\pm}_\varphi(\sstraj,t) | \leq |\eta^{\pm}_\varphi(\sstraj,t)|$	

Given an STL formula $\varphi$ and a signal $\sstraj:\TDom \rightarrow X$, we want to show that $|\theta^{\pm}_\varphi(\sstraj,t) | \leq |\eta^{\pm}_\varphi(\sstraj,t)|$ 
for any $t\in\TDom$.
Let $|\theta^{\pm}_\varphi(\sstraj,t) |=\trv\geq 0$. 
According to Thm.~\ref{thm:chi_async}, $\forall t'\in t\pm[0,\trv),\  \chi_\varphi(\sstraj,t')=\chi_\varphi(\sstraj,t)$.
To prove the rest we distinguish between the cases of $\trv=\infty$ and $\trv<\infty$ as following.
If $\trv=\infty$ then due to Thm.~\ref{thm:chi_sync_stronger}, $|\eta^{\pm}_\varphi(\sstraj,t) | =\infty=|\theta^{\pm}_\varphi(\sstraj,t)|$. 
Now consider the case when $\trv<\infty$.
There exist two separate cases: either it holds that (a) $\forall \epsilon>0$, $\exists \tstep\in[\trv, \trv+\epsilon)$, $\chi_{\varphi}(\sstraj,t \pm \tstep)\not=\chi_\varphi(\sstraj,t)$ or the opposite holds, i.e., it holds that (b) $\exists \epsilon>0$, $\forall \tstep\in [\trv, \trv+\epsilon)$,
$\chi_{\varphi}(\sstraj,t \pm \tstep)=\chi_\varphi(\sstraj,t)$.
In case of condition (a), due to Thm.~\ref{thm:chi_sync_stronger} we can conclude that $|\eta^{\pm}_\varphi(\sstraj,t) | =\trv$. Now consider the case (b). By assumption we know that 
$\forall t'\in t\pm[0,\trv),\  \chi_\varphi(\sstraj,t')=\chi_\varphi(\sstraj,t)$, which in combination with condition (b) leads to $\exists \epsilon>0$, $\forall t'\in t\pm[0,\trv+\epsilon),\  \chi_\varphi(\sstraj,t')=\chi_\varphi(\sstraj,t)$. 
Therefore, $\forall t'\in t\pm[0,\trv]$, $\chi_\varphi(\sstraj,t')=\chi_\varphi(\sstraj,t)$ or in other words, 
$\trv\in\varUpsilon=\{\tau\geq0 \ :\ \forall t'\in t\pm [0,\tau],\ \chi_\varphi(\sstraj,t')=\chi_\varphi(\sstraj,t)\}$. 
Therefore, $\trv\leq \sup\varUpsilon= |\eta^{\pm}_\varphi(\sstraj,t) |$.

\subsection{Proof of Lemma \ref{lemma:p1_eq}}

	Consider a formula $\varphi\in\stlnnf(\wedge,\always_I)$ and a signal $\sstraj:\TDom \rightarrow X$. We want to show that for any $t\in\TDom$ such that $\chi_\varphi(\sstraj,t)=\po$ it follows that $\eta^{\pm}_\varphi(\sstraj,t)=\theta^{\pm}_\varphi(\sstraj,t)$.
The proof is by induction on the structure of the formula $\varphi\in\stlnnf(\wedge,\always_I)$.	
\begin{enumerate}[label={\textbf{Case}}]
	\item $\varphi=p$ \textbf{and} $\varphi=\neg p$. 
	Immediately follows from Corollary~\ref{rem:eq}, e.g.,
	$\quad\eta^{\pm}_{\neg p}(\sstraj,t) 
		\overset{\text{Lemma}~\ref{rem:eta_neg}}{=}
		-\eta^{\pm}_{p}(\sstraj,t)
		\overset{\text{Cor.}\ref{rem:eq}}{=}
		-\theta^{\pm}_{p}(\sstraj,t)
		\overset{\eqref{eq:t_neg}}{=}
		\theta^{\pm}_{\neg p}(\sstraj,t)$.

	\item $\varphi=\varphi_1\wedge\varphi_2$.
	First note that since $\chi_{\varphi}(\sstraj,t)=\po$ then due to Thm.~\ref{thm:sat_eta}, $\theta^{\pm}_{\varphi}(\sstraj,t)\geq 0$ and $\eta^{\pm}_{\varphi}(\sstraj,t)\geq 0$. Also by Def.~\ref{def:char_func}, for $k\in\{1,2\}$, $\chi_{\varphi_k}(\sstraj,t)=\po$ and $\chi_{\varphi_k}(\sstraj,t)=\po$ and thus, $\theta^{\pm}_{\varphi_k}(\sstraj,t)\geq 0$ and $\eta^{\pm}_{\varphi_k}(\sstraj,t)\geq 0$. 
	WLOG assume that $\theta^{\pm}_{\varphi_1}(\sstraj,t) \leq \theta^{\pm}_{\varphi_2}(\sstraj,t)$.		
	From the induction hypothesis for $\varphi_k$ for $k\in\{1,2\}$ we get that 
	$\eta^{\pm}_{\varphi_1}(\sstraj,t)=\theta^{\pm}_{\varphi_1}(\sstraj,t)\leq
	\theta^{\pm}_{\varphi_2}(\sstraj,t)=\eta^{\pm}_{\varphi_2}(\sstraj,t)$.
	Then it holds that
	$\theta^{\pm}_{\varphi}(\sstraj,t)=
	\theta^{\pm}_{\formula_1}(\sstraj,t) \ \sqcap\ \theta^{\pm}_{\formula_2}(\sstraj,t)=
	\theta^{\pm}_{\varphi_1}(\sstraj,t)\leq
	\theta^{\pm}_{\varphi_2}(\sstraj,t)=\eta^{\pm}_{\varphi_2}(\sstraj,t)$.
	Let $\theta^{\pm}_{\varphi}(\sstraj,t) = \trv$ for some $\trv\in\CoRe_{\geq 0}$ or $\trv\in\CoZe_{\geq 0}$ depending on $\TDom=\{\Re,\Ze\}$.
	Thus, $\eta^{\pm}_{\varphi_1}(\sstraj,t)=\trv$ and $\eta^{\pm}_{\varphi_2}(\sstraj,t)\geq \trv$.
		Then according to Thm.~\ref{thm:chi_sync_stronger}, $\forall t'\in t\pm [0, \trv),\ \chi_{\varphi_k}(\sstraj,t')=\po$ for both $k\in\{1,2\}$, which by Def.~\ref{def:char_func} leads to
		$\forall t'\in t\pm [0, \trv),\ \chi_{\varphi}(\sstraj,t')=\po$.
	We distinguish between the cases of $\trv=\infty$ and $\trv<\infty$ as following.
	
	\textbf{1.} Let $\trv=\infty$. Since $\forall t'\in t\pm [0, \infty),\ \chi_{\varphi}(\sstraj,t')=\chi_{\varphi}(\sstraj,t)$, by Thm.~\ref{thm:chi_sync_stronger} we get that $|\eta_{\varphi}^{\pm}(\sstraj,t)|=\infty$.
	
	\textbf{2.} Let $\trv<\infty$. Since $\eta^{\pm}_{\varphi_1}(\sstraj, t)=\trv<\infty$ then
	from Thm.~\ref{thm:chi_sync_stronger}, we get that  
	$\forall\epsilon>0,\ \exists \tstep\in[\trv, \trv+\epsilon),\ \chi_{\varphi_1}(\sstraj,t \pm \tstep)=-1$. Which means that $\forall\epsilon>0,\ \exists \tstep\in[\trv, \trv+\epsilon),\ \chi_{\varphi}(\sstraj,t \pm \tstep)=-1 \sqcap \chi_{\varphi_2}(\sstraj,t \pm \tstep)=-1$ and in combination with obtained 
	$\forall t'\in t\pm [0, \trv),\ \chi_{\varphi}(\sstraj,t')=\po$ by Thm.\ref{thm:chi_sync_stronger} we get that $|\eta_{\varphi}^{\pm}(\sstraj,t)|=\trv$.
	
	In both cases, since $\eta_{\varphi}^{\pm}(\sstraj,t)\geq 0$ then $\eta_{\varphi}^{\pm}(\sstraj,t)=|\eta_{\varphi}^{\pm}(\sstraj,t)|=\trv=\theta_{\varphi}^{\pm}(\sstraj,t)$.

	\item $\varphi=\always_I \varphi_1$.
	First note that since $\chi_{\varphi}(\sstraj,t)=\po$ then due to Def.~\ref{def:char_func}, 
	$\forall t'\in t+ I$,  
	$\chi_{\varphi_1}(\sstraj,t')=\po$, due to Thm.~\ref{thm:sat_eta}, $|\eta^{\pm}_{\varphi}(\sstraj,t)|=\eta^{\pm}_{\varphi}(\sstraj,t)$ and due to Thm.~\ref{thm:sat_theta},
	$|\theta^{\pm}_{\varphi}(\sstraj,t)|=\theta^{\pm}_{\varphi}(\sstraj,t)$.	
From the induction hypothesis we get that $\forall t'\in t+I$, $\theta^{\pm}_{\varphi_1}(\sstraj,t)=\eta^{\pm}_{\varphi_1}(\sstraj,t)$.
	Let $\theta^{\pm}_{\varphi}(\sstraj,t) = \trv$. 
	Then by Def.~\ref{def:time_rob_rec},
	$\bigsqcap_{t'\in t+ I} 
		\theta^{\pm}_{\formula_1}(\sstraj,t') =\trv$.
			Next we consider the cases of $\trv=\infty$ and $\trv<\infty$.
	
	\textbf{1.} Let $\trv=\infty$. Since $\bigsqcap_{t'\in t+ I} 
	\theta^{\pm}_{\formula_1}(\sstraj,t') =\infty$, then it holds that $\forall t'\in t+I$, $\theta^{\pm}_{\formula_1}(\sstraj,t') =\infty$ which by the induction hypothesis leads to $\forall t'\in t+I$, $\eta^{\pm}_{\formula_1}(\sstraj,t') =\infty$. According to Thm.~\ref{thm:chi_sync_stronger}, 
			$\forall t'\in t+I$, $\forall \tstep\in [0, \infty),\  \chi_{\varphi_1}(\sstraj, t' \pm \tstep)=\po$ which by Thm.~\ref{thm:chi_sync_stronger} leads to $\eta^{\pm}_{\varphi}(\sstraj,t)=|\eta^{\pm}_{\varphi}(\sstraj,t)|=\infty=\theta^{\pm}_{\varphi}(\sstraj,t)$.

	\textbf{2.} 	Let $\trv<\infty$. Then by the definition of infimum,
			$\forall t'\in t+I$, $\theta^{\pm}_{\formula_1}(\sstraj,t') \geq \trv$ and $\forall \epsilon>0$, $\exists t''\in t+I$ such that $\trv\leq \theta^{\pm}_{\formula_1}(\sstraj,t'') < \trv +\epsilon$.
		The first condition due to Thm.~\ref{thm:chi_async} leads to $\forall t'\in t +I$,
$\forall \tstep\in [0,\trv)$, $\chi_{\varphi_1}(\sstraj,t'\pm \tstep)=\po$ which by Def.~\ref{def:char_func} leads to $\forall t'\in t\pm [0, \trv)$, $\chi_\varphi(\sstraj, t')=\po$.		
		
			The second condition due to an induction hypothesis can be rewritten as
			 $\forall \epsilon>0$, $\exists t''\in t+I$, $\exists j\in [\trv, \trv+\epsilon)$, $\theta^{\pm}_{\formula_1}(\sstraj,t'')=\eta^{\pm}_{\formula_1}(\sstraj,t'') = j$.
Since $\trv<\infty$ then due to Thm.~\ref{thm:chi_sync_stronger} we get that 
$\forall \epsilon>0$, $\exists t''\in t+I$, $\exists j\in [\trv, \trv+\epsilon)$,
$\forall\epsilon'>0$, $\exists \tstep\in[j, j+\epsilon')$, $\chi_{\varphi_1}(\sstraj,t'' \pm \tstep)=-1$.
This due to Lemma~\ref{lemma:quant_laws_set}(\ref{i:quant_set}),(\ref{i:quant_scope_set}) leads to $\forall \epsilon>0$, $\exists \tstep\in [\trv, \trv+\epsilon)$, 
$\exists t''\in t+I$, $\chi_{\varphi_1}(\sstraj,t'' \pm \tstep)=-1$ and 
by Def.~\ref{def:char_func} we get that $\forall \epsilon>0$, $\exists \tstep\in [\trv, \trv+\epsilon)$, $\chi_{\varphi}(\sstraj,t \pm \tstep)=-1$.
Thus, we obtained that $\forall t'\in t\pm [0, \trv)$, $\chi_\varphi(\sstraj, t')=\po$ and
$\forall \epsilon>0$, $\exists \tstep\in [\trv, \trv+\epsilon)$, $\chi_{\varphi}(\sstraj,t \pm \tstep)=-1$ which by Thm.~\ref{thm:chi_sync_stronger} leads to $|\eta^{\pm}_{\varphi}(\sstraj,t)|=\trv$. Since $\chi_{\varphi}(\sstraj,t)=\po$, we obtain that $\eta^{\pm}_{\varphi}(\sstraj,t)=|\eta^{\pm}_{\varphi}(\sstraj,t)|=\trv=\theta^{\pm}_{\varphi}(\sstraj,t)$.
\end{enumerate}

%%%%%%%%%%%%%%%%%%%%%%%%%%%%%%%%%%%%%%%%%%%%%%%%%%%
\subsection{Proof of Lemma \ref{lemma:m1_eq}}
	Consider a formula $\varphi\in\stlnnf(\vee,\eventually_I)$ and a signal $\sstraj:\TDom \rightarrow X$. We want to show that for any $t\in\TDom$ such that $\chi_\varphi(\sstraj,t)=\-1$ it follows that $\eta^{\pm}_\varphi(\sstraj,t)=\theta^{\pm}_\varphi(\sstraj,t)$.
Since $\varphi\in\stlnnf(\vee,\eventually_I)$ then $\psi\defeq\textbf{nnf}(\neg\varphi)\in \stlnnf(\wedge,\always_I)$. Also since we know that $\chi_{\formula}(\sstraj,t)=-1$ then $\chi_{\psi}(\sstraj,t)=\po$. 
Therefore, the following sequence hold
$\eta^{\pm}_\varphi(\sstraj,t)
	\overset{\text{Lemma}~\ref{rem:eta_neg}}{=}
	-\eta^{\pm}_{\neg\varphi}(\sstraj,t)
	\overset{\psi=\textbf{nnf}(\neg\varphi)}{=}
	-\eta^{\pm}_{\psi}(\sstraj,t) \overset{\text{Lemma}~\ref{lemma:p1_eq}}{=}
	-\theta^{\pm}_\psi(\sstraj,t)  
	%-\theta^{\pm}_{\neg \varphi}(\sstraj,t) 
	\overset{\eqref{eq:t_and}}{=}
	\theta^{\pm}_{\varphi}(\sstraj,t)$.

%%%%%%%%%%%%%%%%%%%%%%%%%%%%%%%%%%%%%%%%%%%%%%%%%%%%%%%%%%%%%%%%%%%%%%%%%%%%%%%%%%%%%
% Section: SHIFTS
\section{Proofs of Section~\ref{sec:robustness_shifts_time}}
\label{appndx:shifts}
\subsection{Proof of Remark \ref{rem:shift_sync}}
We want to show that for a signal $\mathbf{s}=\sstraj^{\leftrightarrows \tstep}$ it holds that $\forall p_k\in AP$ and $\forall t\in\TDom$,  $\chi_{p_k}^{}(\mathbf{s}, t)=
\chi_{p_k}(\sstraj, t\pm \tstep)$.
By Definition \ref{def:char_func}, $\forall p_k\in AP$, $\forall t\in\TDom$, 
	$\chi_{p_k}(\mathbf{s}, t)\overset{\text{Def.}\ref{def:char_func}}{=}
	\sign \mu_k(\mathbf{s}_t) = 
	\sign \mu_k(\sstraj_{t \pm \tstep}) \overset{\text{Def.}\ref{def:char_func}}{=} 
	\chi_{p_k}(\sstraj, t\pm \tstep)$.
\subsection{Proof of Corollary \ref{cor:sync_h}}
Given a formula $\varphi$ built upon the predicate set $AP$, $\tstep\in\TDom_{\geq 0}$ and signal $\sstraj:\TDom \rightarrow X$, we want to show that $\chi_\varphi(\sstraj^{\leftrightarrows \tstep}, t)=\chi_\varphi(\sstraj, t\pm \tstep)$,  $\forall t\in\TDom$.
The proof is by induction on the structure of the formula $\varphi$.

\begin{enumerate}[label={\textbf{Case}}]
	\item $\varphi =p_k$, $\forall p_k\in AP$. Then by Definition \ref{def:set_sync}, for any $p_k\in AP$, $\forall t\in\TDom$, $\chi_{p_k}^{}(\sstraj^{\leftrightarrows \tstep}, t)=\chi_{p_k}(\sstraj, t \pm \tstep)$. 
		
	\item $\varphi=\neg\varphi_1$. 
	Then due to the induction hypothesis for $\varphi_1$ and Def.~\ref{def:char_func}, it holds that
	$\chi_\varphi(\sstraj^{\leftrightarrows \tstep},t )
	=-\chi_{\varphi_1}(\sstraj^{\leftrightarrows \tstep},t )
	\overset{\text{Ind.H.}}{=}-\chi_{\varphi_1}(\sstraj, t\pm \tstep)
	=
	\chi_\varphi(\sstraj, t\pm \tstep)$, $\forall t\in\TDom$.
	\item $\varphi=\varphi_1 \wedge\varphi_2$.
	Then due to the induction hypothesis for $\varphi_k$, $k\in\{1,2\}$ and Def.~\ref{def:char_func}, it holds that
	%$\forall t\in\TDom$, 
	$\chi_\varphi(\sstraj^{\leftrightarrows \tstep},t )
	\overset{\text{Def.\ref{def:char_func}}}{=}\chi_{\varphi_1}(\sstraj^{\leftrightarrows \tstep},t ) \sqcap
	\chi_{\varphi_2}(\sstraj^{\leftrightarrows \tstep},t )
	\overset{\text{Ind.Hyp.}}{=}\chi_{\varphi_1}(\sstraj, t\pm \tstep)\sqcap
	\chi_{\varphi_2}(\sstraj, t\pm \tstep)
	\overset{\text{Def.\ref{def:char_func}}}{=}
	\chi_\varphi(\sstraj, t\pm \tstep)$.
	\item $\varphi=\varphi_1\until_I \varphi_2$. Follows directly from the induction hypothesis and Def.~\ref{def:char_func} as follows:
	$\forall t\in\TDom$, 
	$\chi_\varphi(\sstraj^{\leftrightarrows \tstep},t )
	\overset{\text{Def.~\ref{def:char_func}}}{=} \bigsqcup_{t'\in t+ I} \left( \chi_{\varphi_2}(\sstraj^{\leftrightarrows \tstep},t') \sqcap \bigsqcap_{t'' \in [t,t')}
	\chi_{\varphi_1}(\sstraj^{\leftrightarrows \tstep},t'')\right)
	\overset{\text{Ind.Hyp.}}{=}\\
	\bigsqcup_{t'\in t+ I} \left( \chi_{\varphi_2}(\mathbf{x},t'\pm \tstep) \sqcap \bigsqcap_{t'' \in [t,t')}
	\chi_{\varphi_1}(\mathbf{x},t''\pm \tstep)\right)=\\
	\bigsqcup_{t'\in t\pm \tstep+ I} \left( \chi_{\varphi_2}(\mathbf{x},t') \sqcap \bigsqcap_{t'' \in [t\pm \tstep,t')}
	\chi_{\varphi_1}(\mathbf{x},t'')\right)
	\overset{\text{Def.~\ref{def:char_func}}}{=}
	\chi_\varphi(\sstraj, t\pm \tstep)$.
\end{enumerate}

%\subsection{Proof of Theorem \ref{thm:shift_sync}}

%Denote $|\eta^{\pm}_\varphi(\sstraj,t)|=i$. 
%Note that due to Def.~\ref{def:uni}, $h < i \ \Longrightarrow\ 
%\chi_\varphi(\sstraj^{\leftrightarrows h},t )=\chi_\varphi(\sstraj,t)$
%is equivalent to $\forall \tstep\in [0, i)$, $\chi_\varphi(\sstraj^{\leftrightarrows h},t )=\chi_\varphi(\sstraj,t)$.
%\begin{align*}
%	|\eta^{\pm}_\varphi(\sstraj,t)|=i \quad\overset{\text{Thm.}~\ref{thm:chi_sync_stronger}}{\Longrightarrow}\quad 
%	&\forall \tstep\in [0, i),\ \chi_\varphi(\sstraj, t \pm \tstep)=\chi_\varphi(\sstraj,t)\\
%	\quad\overset{\text{Lemma}~\ref{lemma:sync_h}}{\Longrightarrow}\quad
%	&\forall \tstep\in [0, i),\ 
%	\chi_\varphi(\sstraj^{\leftrightarrows h},t )=\chi_\varphi(\sstraj,t).
%\end{align*} 

\subsection{Proof of Theorem \ref{thm:shift_sync_stronger}}
\label{sec:shift_sync_stronger}
	
	Let $\varphi$ be an STL formula, $\sstraj:\Re \rightarrow X$ be a continuous-time signal and $t\in\Re$ be a time point. For any value $\trv\in\CoRe_{\geq 0}$, we want to prove that the synchronous temporal robustness $|\eta^{\pm}_\varphi(\sstraj, t)|=\trv$ if and only if $\forall \tstep\in[0,\trv)$, 
	$\chi_\varphi(\sstraj^{\leftrightarrows \tstep}, t )=\chi_\varphi(\sstraj, t )$ and if $\trv<\infty$ then $\forall\epsilon>0$, $\exists \tstep\in[\trv, \trv+\epsilon)$, 
	$\chi_\varphi(\sstraj^{\leftrightarrows \tstep}, t )\not=\chi_\varphi(\sstraj, t )$.	
	Let $|\eta^{\pm}_\varphi(\sstraj, t)|=\trv$. Then the result follows directly from Theorem \ref{thm:chi_sync_stronger} and Corollary \ref{cor:sync_h}.
%\begin{align*}
%	|\eta^{\pm}_\varphi(\sstraj,t)|=\trv \quad\overset{\text{Thm.}~\ref{thm:chi_sync_stronger}}{\Longleftrightarrow}\quad 
%	&\forall \tstep\in [0, \trv),\ \chi_{\varphi}(\sstraj,t\pm \tstep)=\chi_\varphi(\sstraj,t) \qquad\text{and}\\
%	[&
%	\forall\epsilon>0,\ \exists \tstep\in[\trv, \trv+\epsilon),\ \chi_{\varphi}(\sstraj,t \pm \tstep)\not=\chi_\varphi(\sstraj,t)
%	\quad\text{or}\quad 
%	%	&\chi_{\varphi}(\sstraj,t \pm i)\not=\chi_\varphi(\sstraj,t)\quad \vee\\ 
%	\trv=\infty
%	]\\
%	\quad\overset{\text{Corollary}~\ref{cor:sync_h}}{\Longleftrightarrow}\quad
%	&\forall \tstep\in [0, \trv),\ \chi_{\varphi}(\sstraj^{\leftrightarrows \tstep},t)=\chi_\varphi(\sstraj,t) \qquad\text{and}\\
%	[&
%	\forall\epsilon>0,\ \exists \tstep\in[\trv, \trv+\epsilon),\ \chi_{\varphi}(\sstraj^{\leftrightarrows h},t)\not=\chi_\varphi(\sstraj,t)
%	\quad\text{or}\quad 
%	%	&\chi_{\varphi}(\sstraj,t \pm i)\not=\chi_\varphi(\sstraj,t)\quad \vee\\ 
%	\trv=\infty
%	].
%\end{align*} 
\subsection{Proof of Eq.\eqref{eq:shift_eta_disc} (Theorem \ref{thm:shift_sync_stronger} for discrete-time)}

Let $\varphi$ be an STL formula, $\sstraj:\Ze \rightarrow X$ be a discrete-time signal and $t\in\Ze$ be a time point. For any finite value $\trv\in\Ze_{\geq 0}$, we want to prove that the synchronous temporal robustness $|\eta^{\pm}_\varphi(\sstraj, t)|=\trv$ if and only if $\forall t'\in t\pm [0, \trv]$, $\chi_\varphi(\sstraj^{\leftrightarrows \tstep}, t )=\chi_\varphi(\sstraj, t )$ and
$\chi_\varphi(\sstraj^{\leftrightarrows \trv+1}, t )\not=\chi_\varphi(\sstraj, t )$.

Let $|\eta^{\pm}_\varphi(\sstraj,t)| =\trv$. 
 Then the result follows directly from equation \eqref{eq:chi_eta_disc} and Corollary \ref{cor:sync_h}.
 
%\begin{align*}
%	|\eta^{\pm}_\varphi(\sstraj,t)|=\trv \ \overset{\eqref{eq:chi_eta_disc}}{\Longleftrightarrow}\  
%	&\forall \tstep\in [0, \trv],\ \chi_{\varphi}(\sstraj,t\pm \tstep)=\chi_\varphi(\sstraj,t) \ \text{and}
%	\ \chi_{\varphi}(\sstraj,t \pm (\trv+1))\not=\chi_\varphi(\sstraj,t)\\
%	\quad\overset{\text{Corollary}~\ref{cor:sync_h}}{\Longleftrightarrow}\quad
%	&\forall \tstep\in [0, \trv],\ \chi_{\varphi}(\sstraj^{\leftrightarrows \tstep},t)=\chi_\varphi(\sstraj,t) \ \text{and}\
%	\chi_{\varphi}(\sstraj^{\leftrightarrows \trv+1},t)\not=\chi_\varphi(\sstraj,t).
%\end{align*} 

\subsection{Proof of Theorem \ref{thm:shift_async}}
\label{sec:shift_async}

Let $\varphi$ be an STL formula built upon the predicate set $AP\defeq\{p_1,\ldots,p_L\}$, $\sstraj:\Re \rightarrow X$ be a continuous-time signal and $t\in\Re$ be a time point. For any value $\trv\in\CoRe_{\geq 0}$, we want to prove that
if the asynchronous temporal robustness 
$|\theta^{\pm}_\varphi(\sstraj,t)| = \trv$  then 
$\forall \tstep_1,\ldots,\tstep_L\in[0, \trv)$,
$\chi_\varphi(\sstraj^{\leftrightarrows  \bar{\tstep}},t )=\chi_\varphi(\sstraj, t)$.

Note that the result is trivial for $\trv=0$. 
Therefore, let $|\theta^{\pm}_\varphi(\sstraj, t)|= \trv>0$. 
The rest of the proof is by induction on the structure of the formula $\varphi$.  	

\begin{enumerate}[label={\textbf{Case}}]
	\item $\varphi =p_k\in AP$. 
	Since $|\theta^{\pm}_{p_k}(\sstraj, t)|=\trv$ then due to Thm.~\ref{thm:chi_async},
	$\forall \tstep\in [0, \trv),\ \chi_{p_k}(\sstraj, t\pm \tstep)=\chi_{p_k}(\sstraj,t)$.
Using Def.~\ref{def:set_async} it leads to
	$\forall \tstep_1,\ldots,\tstep_L\in [0, \trv)$, 
	$\chi_{p_k}(\sstraj^{\leftrightarrows \bar{\tstep}}, t)\overset{\text{Def.}~\ref{def:set_async}}{=}
	\chi_{p_k}(\sstraj, t\pm \tstep_k) 
	\overset{\tstep_k< \trv}{=}\chi_{p_k}(\sstraj, t)$. 
	
	\item $\varphi=\neg\varphi_1$. 
		Since $|\theta^{\pm}_{\neg\varphi_1}(\sstraj,t)| =r$ then due to \eqref{eq:t_neg} it holds that $|\theta^{\pm}_{\varphi_1}(\sstraj,t)|=r$. By the induction hypothesis for $\varphi_1$ we get that
	$\forall \tstep_1,\ldots,\tstep_L\in[0, \trv)$,
	$\chi_{\varphi_1}(\sstraj^{\leftrightarrows \bar{\tstep}},t )=\chi_{\varphi_1}(\sstraj, t)$. Therefore, due to Def.~\ref{def:char_func}, it holds that
	$\forall \tstep_1,\ldots,\tstep_L\in[0, \trv)$, $\chi_{\varphi}(\sstraj^{\leftrightarrows \bar{\tstep}},t)
=
	-\chi_{\varphi_1}(\sstraj^{\leftrightarrows \bar{\tstep}},t) 
	=
	-\chi_{\varphi_1}(\sstraj,t) 
=
	\chi_{\varphi}(\sstraj,t)$.
	
	\item $\varphi=\varphi_1 \wedge \varphi_2$.
		We consider the two separate cases of $\chi_{\varphi}(\sstraj,t)= 1$ and $\chi_{\varphi}(\sstraj,t)= -1$ as follows:			
	
	\textbf{1.} Let $\chi_{\varphi}(\sstraj,t)= 1$. 
	Since $|\theta^{\pm}_\varphi(\sstraj, t)|= \trv>0$
	then due to Thm.~\ref{thm:sat_theta}(\ref{i3}), $\theta^{\pm}_{\formula}(\sstraj,t) = \trv> 0$. WLOG assume that $\theta^{\pm}_{\varphi_1}(\sstraj,t) \leq \theta^{\pm}_{\varphi_2}(\sstraj,t)$. Therefore,
	$	|\theta^{\pm}_\varphi(\sstraj,t)|=
	\theta^{\pm}_\varphi(\sstraj,t)\overset{\eqref{eq:t_and}}{=}
	\theta^{\pm}_{\formula_1}(\sstraj,t) \ \sqcap\ \theta^{\pm}_{\formula_2}(\sstraj,t)=
	\theta^{\pm}_{\varphi_1}(\sstraj,t)=\trv$.
	If we denote $\theta^{\pm}_{\formula_2}(\sstraj,t)=j$ then 
	$0<\trv=\theta^{\pm}_{\varphi_1}(\sstraj,t) \leq \theta^{\pm}_{\varphi_2}(\sstraj,t)=j$.
	From $\trv\leq j$ and by the induction hypothesis for $\varphi_k$ for $k\in\{1,2\}$ it holds that
			$\forall \tstep_1,\ldots,\tstep_L\in[0, \trv)$, 
			$\chi_{\varphi_k}(\sstraj^{\leftrightarrows \bar{\tstep}},t )=\chi_{\varphi_k}(\sstraj, t)$.
		Thus, using Def.~\ref{def:char_func}, $\forall \tstep_1,\ldots,\tstep_L\in[0, \trv)$ it holds that
		$\chi_{\varphi}(\sstraj^{\leftrightarrows \bar{\tstep}},t) =
			\chi_{\varphi_1}(\sstraj^{\leftrightarrows \bar{\tstep}},t) \sqcap 
			\chi_{\varphi_2}(\sstraj^{\leftrightarrows \bar{\tstep}},t)=
			\chi_{\varphi_1}(\sstraj,t) \sqcap
			\chi_{\varphi_2}(\sstraj,t)=
			\chi_\varphi(\sstraj,t)$.
		
		\textbf{2.} Let $\chi_{\varphi}(\sstraj,t)= -1$.
		Since $|\theta^{\pm}_\varphi(\sstraj, t)|= \trv>0$
		then due to Thm.~\ref{thm:sat_theta}(\ref{i4}), $\theta^{\pm}_{\formula}(\sstraj,t) = -\trv< 0$.
		WLOG assume that $\theta^{\pm}_{\varphi_1}(\sstraj,t) \leq \theta^{\pm}_{\varphi_2}(\sstraj,t)$. Therefore,
		$
		-|\theta^{\pm}_\varphi(\sstraj,t)|=
		\theta^{\pm}_\varphi(\sstraj,t)\overset{\eqref{eq:t_and}}{=}
		\theta^{\pm}_{\formula_1}(\sstraj,t) \ \sqcap\ \theta^{\pm}_{\formula_2}(\sstraj,t)=
		\theta^{\pm}_{\varphi_1}(\sstraj,t)=-\trv<0
		$.
		Since $\theta^{\pm}_{\varphi_1}(\sstraj,t)<0$ then $\chi_{\varphi_1}(\sstraj,t)=-1$.
		From the induction hypothesis for $\varphi_1$ we get that
		$\forall \tstep_1,\ldots,\tstep_L\in[0, \trv)$,  $\chi_{\varphi_1}(\sstraj^{\leftrightarrows \bar{\tstep}},t)=\chi_{\varphi_1}(\sstraj,t)=-1$.
	Thus, $\forall \tstep_1,\ldots,\tstep_L\in[0, \trv)$,
		$\chi_{\varphi}(\sstraj^{\leftrightarrows \bar{\tstep}},t) =
			\chi_{\varphi_1}(\sstraj^{\leftrightarrows \bar{\tstep}},t) \sqcap 
			\chi_{\varphi_2}(\sstraj^{\leftrightarrows \bar{\tstep}},t)
		=
			-1 \sqcap 
			\chi_{\varphi_2}(\sstraj^{\leftrightarrows \bar{\tstep}},t)
			=
			-1=
			\chi_\varphi(\sstraj,t)$.
	%%%%%%%%%%%%%%%%%%%%%%%%%%%%%%%%%%%%%%%%%%%%%
	\item $\varphi=\varphi_1\until_I \varphi_2$.
	The proof is analogous to Section \ref{sec:chi_async}.
	We consider the two separate cases of $\chi_{\varphi}(\sstraj,t)= 1$ and $\chi_{\varphi}(\sstraj,t)= -1$ as follows:	
	
	\textbf{1.} Let $\chi_{\varphi}(\sstraj,t)= \po$. 	Since $|\theta^{\pm}_\varphi(\sstraj, t)|= \trv>0$
	then due to Thm.~\ref{thm:sat_theta}(\ref{i3}), $\theta^{\pm}_{\formula}(\sstraj,t) = \trv > 0$.
	Next, same as in Section \ref{sec:chi_async} we distinguish between the cases of $\trv<\infty$ and $\trv=\infty$ as following:
	
	%	Therefore, 
	%	\begin{equation}
		%	\etam_\varphi(\sstraj,t) =
		%	\min\{\tau\geq0 \ :\ \forall t'\in[t,t+\tau],\ \chi_{\varphi_1 \until_I \varphi_2}(\sstraj,t')=\po\} \geq 0.	
		%	\end{equation}

	\textbf{1.1.} Let $\trv<\infty$. Then \eqref{eq:ref2} holds. Note that since $\trv>0$ then using Thm.~\ref{thm:sat_theta}, $\chi_{\varphi_2}(\sstraj,t')=\chi_{\varphi_1}(\sstraj,t'')=\po$. By the induction hypothesis we get that
		\begin{equation*}
			\begin{aligned}
				\forall \epsilon>0,\ &\exists t'\in t+I\ ( \forall \tstep_1,\ldots,\tstep_L\in [0,\trv-\epsilon),\  
				\chi_{\varphi_2}(\sstraj^{\leftrightarrows \bar{\tstep}},t')=\po	\ \wedge\\
				&\forall t'' \in [t,t'), \forall \tstep_1,\ldots,\tstep_L\in [0,\trv-\epsilon),\  \chi_{\varphi_1}(\sstraj^{\leftrightarrows \bar{\tstep}},t'')=\po
				)
			\end{aligned}			
		\end{equation*}
	
		Thus, due to Lemma \ref{lemma:quant_laws_set}(\ref{i:quant_scope_set}), (\ref{i:forall_exists_set}) and (\ref{i:quant_set}) it holds that
	$\forall \epsilon>0$, $\forall \tstep_1,\ldots,\tstep_L\in[0,\trv-\epsilon)$, $\exists t'\in t+I$,
		\begin{equation*}
		 \chi_{\varphi_2}(\sstraj^{\leftrightarrows \bar{\tstep}},t')=\po	\ \wedge\ 
				\forall t'' \in [t,t'),\  \chi_{\varphi_1}(\sstraj^{\leftrightarrows \bar{\tstep}},t'')=\po
		\end{equation*}
	Therefore, by Def.~\ref{def:char_func},
		$\forall \epsilon>0,\ \forall \tstep_1,\ldots,\tstep_L\in[0,\trv-\epsilon)$,
		$\chi_{\varphi}(\sstraj^{\leftrightarrows \bar{\tstep}},t)=1$, 
		and similarly to \footref{footnote_eps}, we can conclude that
%		\footnote{Assume the opposite, assume $\exists \tstep_1^*,\ldots,\tstep_L^*\in[0,\trv)$, $\chi_{\varphi}(\sstraj^{\leftrightarrows \bar{\tstep^*}},t)=-1$. This means $\exists \epsilon^* > 0$ ($\epsilon^* \leq \trv$) such that $\max\{\tstep_1^*,\ldots,\tstep_L^*\}=\trv-\epsilon^*$ 
%			%(means $\tstep_1^*,\ldots,\tstep_L^* \leq i-\epsilon^*$) 
%			for which $\chi_{\varphi}(\sstraj^{\leftrightarrows \bar{\tstep^*}},t)=-1$.
%			But since $\epsilon=\frac{\epsilon^*}{2}>0$ then 
%			$\tstep_1^*,\ldots,\tstep_L^*\leq \trv-\epsilon^*\in [0,\trv-\frac{\epsilon^*}{2})$ therefore, due to the above,
%			$\chi_{\varphi}(\sstraj^{\leftrightarrows \bar{\tstep^*}},t)=\po$ which is a contradiction with 
%		$\chi_{\varphi}(\sstraj^{\leftrightarrows \bar{\tstep^*}},t)=-1$.}, 
$\forall \tstep_1,\ldots,\tstep_L\in[0,\trv)$,  $\chi_{\varphi}(\sstraj^{\leftrightarrows \bar{\tstep}},t)=\po=\chi_{\varphi}(\sstraj,t)$.
		
		\textbf{1.2.} 
		 Let $\trv=\infty$. Then \eqref{eq:ref3} holds. 
		Using the definition of an infimum (unbounded set) together with an induction hypothesis we can conclude that $\forall M\in\Re$, $\forall \tstep_1,\ldots,\tstep_L\in[0,M)$, $\chi_{\varphi}(\sstraj^{\leftrightarrows \bar{\tstep}},t)=\po$,  and thus, $\forall \tstep_1,\ldots,\tstep_L\in[0,\infty)$,  $\chi_{\varphi}(\sstraj^{\leftrightarrows \bar{\tstep}},t)=\po$.
		
	\textbf{2.} Let $\chi_{\varphi}(\sstraj,t)= -1$. 	
		Since $|\theta^{\pm}_\varphi(\sstraj, t)|= \trv>0$
		then due to Thm.~\ref{thm:sat_theta}(\ref{i4}), $\theta^{\pm}_{\formula}(\sstraj,t) = -\trv < 0$. 
		Next, same as in Section \ref{sec:chi_async} we distinguish between the cases of $\trv=\infty$ and $\trv<\infty$ as following:		

	\textbf{2.1.} Let $\trv=\infty$. Then \eqref{eq:ref15} holds.
By the induction hypothesis and Lemma \ref{lemma:quant_laws_set}(\ref{i:quant_set}),(\ref{i:forall_set}),(\ref{i:quant_scope_set}), it holds that 		
$\forall M\in\Re_+$, $\forall \tstep_1,\ldots,\tstep_L\in[0,M)$, $\chi_{\varphi}(\sstraj^{\leftrightarrows \bar{\tstep}},t)=-1$,  i.e. $\forall \tstep_1,\ldots,\tstep_L\in[0,\infty)$, $\chi_{\varphi}(\sstraj^{\leftrightarrows \bar{\tstep}},t)=-1$.  	

\textbf{2.2.} Let $\trv<\infty$. Then \eqref{eq:j} holds.
Take any $t'\in t+I$. Same as in Section \ref{sec:chi_async}, we will work with each $j_{t'}$ separately. We will distinguish between the cases of $j_{t'}<\infty$ and $j_{t'}=\infty$ as following:		

\begin{enumerate}
	\item Let $j_{t'} <\infty$. Then \eqref{eq:ref4} hold.
		Note, that $|\theta^{\pm}_{\formula_2}(\sstraj,t')| = -\theta^{\pm}_{\formula_2}(\sstraj,t')= j_{t'} \geq \trv> \trv-\epsilon$ and $|\theta^{\pm}_{\formula_1}(\sstraj,t'')| =-\theta^{\pm}_{\formula_1}(\sstraj,t'') > j_{t'}-\epsilon \geq \trv-\epsilon$. 
	Therefore, by the induction hypothesis and Lemma \ref{lemma:quant_laws_set}(\ref{i:quant_set}),(\ref{i:forall_set}) 
	it holds that $\forall \epsilon>0$, $\forall \tstep_1,\ldots,\tstep_L\in [0,\ \trv-\epsilon)$, $\chi_{\varphi_2}(\sstraj^{\leftrightarrows \bar{\tstep}},t' )\ \sqcap
	\bigsqcap_{t'' \in [t,t')}   
	\chi_{\formula_1}(\sstraj^{\leftrightarrows \bar{\tstep}},t'' )=-1$, and thus, similarly to \footref{footnote_eps}, we can conclude that 
	\begin{equation}
		\label{eq:j_finite2}
		\forall \tstep_1,\ldots,\tstep_L\in [0,\ \trv)\quad
		\chi_{\varphi_2}(\sstraj^{\leftrightarrows \bar{\tstep}},t' )\ \sqcap
		\bigsqcap_{t'' \in [t,t')}   
		\chi_{\formula_1}(\sstraj^{\leftrightarrows \bar{\tstep}},t'' )=-1. 
	\end{equation}
	
	\item Let $j_{t'} =\infty$.  Then \eqref{eq:ref5} holds. 
	Since $|\theta^{\pm}_{\formula_1}(\sstraj,t'')| = -\theta^{\pm}_{\formula_1}(\sstraj,t'')>M$ then applying the induction hypothesis 
%	leads to:
%		\begin{align*}
%		&\forall \tstep_1,\ldots,\tstep_L\in [0,\ \infty),\ \chi_{\formula_2}(\sstraj^{\leftrightarrows \bar{\tstep}},t' )= -1 \\&\qquad\vee\quad
%		\forall M\in\Re_+,\ \exists t''\in [t,t'),\  
%		\forall \tstep_1,\ldots,\tstep_L\in \in [0,\ M),\ 
%		\chi_{\formula_1}(\sstraj^{\leftrightarrows \bar{\tstep}},t'' ) =-1. 
%	\end{align*}
	and using $M \geq \trv$ leads to:
	\begin{align*}
		&\forall \tstep_1,\ldots,\tstep_L\in [0,\ \trv),\ \chi_{\varphi_2}(\sstraj^{\leftrightarrows \bar{\tstep}},t' )= -1 \\&\qquad\vee\quad
		\exists t''\in [t,t'),\  
		\forall \tstep_1,\ldots,\tstep_L\in [0,\ \trv),\ 
		\chi_{\varphi_1}(\sstraj^{\leftrightarrows \bar{\tstep}},t'' ) =-1. 
	\end{align*}
	Due to Lemma \ref{lemma:quant_laws_set}(\ref{i:quant_set}),(\ref{i:forall_set}) it follows that 
	\begin{equation}
		\label{eq:j_finite3}
		\forall \tstep_1,\ldots,\tstep_L\in [0,\ \trv)\quad
		\chi_{\varphi_2}(\sstraj^{\leftrightarrows \bar{\tstep}},t' )\ \sqcap
		\bigsqcap_{t'' \in [t,t')}   
		\chi_{\formula_1}(\sstraj^{\leftrightarrows \bar{\tstep}},t'' )=-1. 
	\end{equation}
\end{enumerate}
Note that \eqref{eq:j_finite2} and \eqref{eq:j_finite3} are equivalent, therefore, they hold regardless of each $j_{t'}$ value. 
Thus, it holds that 
$\forall t'\in t+I$, 
$\forall \tstep_1,\ldots,\tstep_L\in [0,\ \trv)$,  
$\chi_{\varphi_2}(\sstraj^{\leftrightarrows \bar{\tstep}},t' )\ \sqcap
\bigsqcap_{t'' \in [t,t')}   
\chi_{\formula_1}(\sstraj^{\leftrightarrows \bar{\tstep}},t'' )=-1$,
which due to Lemma \ref{lemma:quant_laws_set}(\ref{i:quant_scope_set}) leads to $\forall \tstep_1,\ldots,\tstep_L\in [0,\ \trv)$, $\chi_{\varphi}(\sstraj^{\leftrightarrows \bar{\tstep}},t)=-1$.
\end{enumerate}
\subsection{Proof of Eq.\eqref{eq:shift_theta_disc} (Theorem \ref{thm:shift_async} for discrete-time)}

Let $\varphi$ be an STL formula built upon the predicate set $AP\defeq\{p_1,\ldots,p_L\}$, $\sstraj:\Ze \rightarrow X$ be a discrete-time signal and $t\in\Ze$ be a time point. For any value $\trv\in\Ze_{\geq 0}$, we want to prove that
if the asynchronous temporal robustness 
$|\theta^{\pm}_\varphi(\sstraj,t)| = \trv$  then 
$\forall \tstep_1,\ldots,\tstep_L\in[0, \trv]$, $\chi_\varphi(\sstraj^{\leftrightarrows  \bar{\tstep}},t )=\chi_\varphi(\sstraj, t)$.
Note that the result is trivial for $\trv=0$. 
Therefore, let $|\theta^{\pm}_\varphi(\sstraj, t)|= \trv>0$. 
The rest of the proof is by induction on the structure of the formula $\varphi$ and follows the similar steps as Section \ref{sec:shift_async}, therefore, we only present the base case.

\textbf{Base case} $\varphi =p_k$. 
	Since $|\theta^{\pm}_{p_k}(\sstraj, t)|=\trv$ then due to \eqref{eq:chi_theta_disc}, i.e., Theorem~\ref{thm:chi_async} for discrete-time, we get that
$\forall \tstep\in [0, \trv],\ \chi_{p_k}(\sstraj, t\pm \tstep)=\chi_{p_k}(\sstraj,t)$.
Using Def.~\ref{def:set_async} it leads to
$\forall \tstep_1,\ldots,\tstep_L\in [0, \trv]$, 
$\chi_{p_k}(\sstraj^{\leftrightarrows \bar{\tstep}}, t)\overset{\text{Def.}~\ref{def:set_async}}{=}
\chi_{p_k}(\sstraj, t\pm \tstep_k) 
\overset{\tstep_k\leq \trv}{=}\chi_{p_k}(\sstraj, t)$.

\subsection{Proof of Lemma \ref{lemma:shift_async}}

Let $x_t\defeq(\cluster\coordup{1}_t,\ldots,\cluster\coordup{c}_t)$ where $c$ the number of groups in which $x_t$ is clustered.
First, let every predicate  $p_k\in\{p_1,\ldots,p_L\}$ in $\varphi$ be defined over only a single $\cluster\coordup{d}$ for some $d$, 
i.e., $\exists d\in\{1,\ldots,c\}$ and a real-valued function $\nu_k$, such that 
$p_k
%\defeq\mu_k(x_t)\geq 0
\defeq\nu_k(\cluster\coordup{d}_t)\geq 0$.
Second, let $\mathbf{s}$ be defined as 
$s_t\defeq(\cluster\coordup{1}_{t \pm \kappa_1}, \ldots, \cluster\coordup{c}_{t\pm \kappa_c})$.
Then we want to show that $\mathbf{s}=\sstraj^{\leftrightarrows \bar{\tstep}}$.

Due to the above two assumptions, the predicate $p_k$ over the sate $s_t$ is defined through the real-valued function 
$\mu_k(s_t)=
\nu_k(\zeta\coordup{d}_{t \pm \kappa_d}) = 
\mu_k(x_{t \pm \kappa_d})$.
Then by Def.~\ref{def:char_func}, we get that $\chi_{p_k}(\mathbf{s}, t)=\sign (\mu_k(s_t))=
\sign (x_{t \pm \kappa_d})=\chi_{p_k}(\sstraj, t\pm \kappa_d)$.
Therefore, for $\bar{\tstep} = (\tstep_1,\ldots,\tstep_L)$ where each $\tstep_k\in\{\kappa_1,\ldots,\kappa_c\}$, we get that $\mathbf{s}=\sstraj^{\leftrightarrows \bar{\tstep}}$.

\section{Proofs of Section~\ref{sec:control}}
\label{appndx:proof_sec_control}
\subsection{Proof of Proposition \ref{thm:milp_sync_encoding}}
\begin{enumerate}
	\item The proof is by construction, following Section~\ref{sec:milp_eta}.
	\item 	The Boolean encoding of STL constraints \eqref{eq:milp_zphi} introduces $O(H\cdot|AP|)$ binary and $O(H\cdot|\varphi|)$ continuous variables according to \cite{raman2014model}. The encoding of the synchronous \textbf{temporal robustness} then introduces integer counter variables $\cophi_t$, $\czphi_t$ and adjusted counters $\acophi_t$, $\aczphi_t$. Since the algorithm introduces a constant amount of counters per time step, it leads to $O(H)$ additional integer variables, which does not increase the above mentioned $O(H\cdot|\varphi|)$ number of continuous variables.
	
	\textbf{Note.} There exist a purely binary version of the Boolean encoding of STL constraints \eqref{eq:milp_zphi} that introduces $O(H\cdot(|AP|+|\varphi|))$ binary variables (no integers). Together with the introduced counters it leads to $O(H\cdot(|AP|+|\varphi|))$ binary and $O(H)$ continuous variables. Though such encoding is possible and depending on the problem might lead to a faster implementation due to the heuristic solvers and the order of linear relaxations, the worst-case complexity of solving an MILP is exponential in the number of binary variables \cite{wolff2014optimization}, therefore, for the generic implementation, the former  encoding that requires  $O(H\cdot|AP|)$ binary variables is beneficial.
\end{enumerate}

\subsection{Proof of Proposition \ref{thm:milp_asyn_encoding}}
\begin{enumerate}
	\item The proof is by construction, following Section~\ref{sec:milp_async}.
	\item 	
Due to Corollary~\ref{rem:eq}, $\theta^{\pm}_p(\sstraj,t)=\eta^{\pm}_p(\sstraj,t)$ for any $t\in\TDom$. Therefore, using Proposition \ref{thm:milp_sync_encoding}, in order to encode each predicate, we introduce $O(H)$ binary and continuous variables. Therefore, we introduce $O(H\cdot |AP|)$ binary and continuous variables for the encoding of STL predicates.

The second step of the encoding is the encoding of STL operators, where for each operator $\psi$ within $\varphi$ and every $t\in\TDom$, we introduce continuous $\thetap_\psi(\sstraj, t)$ and a finite set of boolean variables $b_j$, see Section~\ref{sec:milp_async}. Therefore, to encode all operators within $\varphi$ we require $O(H\cdot|\varphi|)$ additional integer and boolean variables.
	
	Thus, the MILP encoding of $\bm{\theta}^+_{\varphi}(\sstraj)$ requires $O(H\cdot(|AP|+|\varphi|))$ binary and integer variables. 
\end{enumerate} 
%\section{Laws of Quantifier Distribution}
\section{Auxiliary Lemmas}

\begin{lemma}[Quantifier Laws \cite{hammond2007introduction}]
	\label{lemma:quant_laws_set}
	The following quantifier laws hold:
	\begin{enumerate}
		\item\label{i:quant_scope_set}
		$\forall x\in X,\ \forall y\in Y.\ P(x,y)\  \Longleftrightarrow\ 
		\forall y\in Y,\ \forall x\in X.\  P(x,y)$.
		\item\label{i:quant_set}
		$	\exists x\in X,\ \forall y\in Y.\  Q(x,y)\ \Longrightarrow\ 
		\forall y\in Y,\ \exists x\in X.\  Q(x,y)$.
		\item\label{i:forall_exists_set}
		$\forall x\in X.\ P(x) \ \wedge\ \forall x\in X.\ Q(x)\  \Longleftrightarrow\ 
		\forall x\in X\ (P(x) \wedge Q(x))$.
		\item\label{i:forall_set}
		$\forall x\in X.\ P(x) \ \vee\ \forall x\in X.\ Q(x)\ \, \Longrightarrow\ \,\forall x\in X\  (P(x) \vee Q(x))$.
	\end{enumerate}
\end{lemma}

\begin{lemma}[{If-then-else product construct~\cite{bemporad2001discrete}}]
	\label{lemma:product}
	Let $b\in\{0,1\}$ be a Boolean variable and let $x$ be an integer variable such that lower and upper bounds are known constants, $x_l\leq x\leq x_{u}$. The expression 
	$y=b\cdot x$ can be equivalently expressed as a set of mixed-integer linear constraints as follows:
	\begin{equation*}
		\label{eq:inq}
		\begin{aligned}
			x_lb\leq\  &y\leq x_{u} b \\
			x - x_{u}(1-b) \leq &y \leq x- x_l(1-b).
		\end{aligned}
	\end{equation*}
\end{lemma}

\begin{lemma}
	\label{rem:eta_neg}
		For any STL formula $\varphi$, signal $\sstraj:\TDom \rightarrow X$ and any time $t\in\TDom$, 
	$\eta^{\pm}_{\neg\varphi}(\sstraj,t) = -\eta^{\pm}_\varphi(\sstraj,t)$.
\end{lemma}
\begin{proof}
	The proof is derived directly from Definitions \ref{def:time_rob_all} and \ref{def:char_func} as following:
\begin{align*}
	\eta^{\pm}_{\neg\varphi}(\sstraj,t) &\overset{\text{Def.}\ref{def:time_rob_all}}{=} \chi_{\neg\varphi}(\sstraj, t)\cdot
	\sup\{\tau\geq0 \ :\ \forall t'\in t\pm [0,\tau],\ \chi_{\neg\varphi}(\sstraj,t')=\chi_{\neg\varphi}(\sstraj,t)\}\\
	&\overset{\text{Def.}\ref{def:char_func}}{=}
	-\chi_{\varphi}(\sstraj, t)\cdot
	\sup\{\tau\geq0 \ :\ \forall t'\in t\pm [0,\tau],\ \chi_{\varphi}(\sstraj,t')=\chi_{\varphi}(\sstraj,t)\}\\
	&\overset{\text{Def.}\ref{def:time_rob_all}}{=} -\eta^{\pm}_\varphi(\sstraj,t).
\end{align*}
\end{proof}

\end{document}